\documentclass[11pt,a4paper]{article}
\pdfoutput=1

\usepackage{jheppub}
\usepackage{slashbox}
\usepackage{mathrsfs}
\usepackage{bbold}
\usepackage{subfigure}
\usepackage{enumerate}
\usepackage{color}
\usepackage{cancel}

\newcommand{\be}{\beta}

\newcommand{\La}{\Lambda}
\newcommand{\om}{\omega}

 \def\cK{{\cal K}}
\def\cL{{\cal L}} \def\cM{{\cal M}}

\def\cW{{\cal W}} 

\newcommand{\LL}{\mscr{L}}
\newcommand{\DD}{\mscr{D}}

\newcommand{\mscr}[1]{\mathscr{#1}}

\newcommand{\sign}{\mathrm{sign}}
\newcommand{\hc}{\text{h.c.}}
\newcommand{\mean}[1]{\left\langle#1\right\rangle}
\newcommand{\derp}{\partial}
\newcommand{\unity}{\mathbb{1}}
\newcommand{\diag}{\mathrm{diag}}

\def\thh{\theta^2}
\def\thhb{\overline{\theta}^2}
\newcommand{\msusy}{m_{0}}

\newcommand{\phit}{\varphi_T}
\newcommand{\phis}{\varphi_S}
\newcommand{\xit}{\tilde{\xi}}

\newcommand{\xipp}{\xi^{\prime\prime}}
\newcommand{\sphenod}{{\tt SPheno-2.2.3}}
\newcommand{\spheno}{{\tt SPheno} }

\newcommand\dd{\displaystyle}
\newcommand\nn{\nonumber}
\newcommand{\ov}{\overline}
\def\be{\begin{equation}}
\def\ee{\end{equation}}
\def\bea{\begin{eqnarray}}
\def\eea{\end{eqnarray}}
\def\ba{\begin{array}}
\def\ea{\end{array}}
\newcommand{\bac}{\be\begin{array}}
\newcommand{\eac}{\end{array}\ee}
\newcommand{\bal}{\begin{align}}
\newcommand{\eal}{\end{align}}
\newcommand{\bald}{\be\begin{aligned}}
\newcommand{\eald}{\end{aligned}\ee}
\newcommand{\bgad}{\be\begin{gathered}}
\newcommand{\egad}{\end{gathered}\ee}
\def\raw{\rightarrow}
\def\meg{$\mu \raw e \gamma$~}
\def\bsg{$b \raw s \gamma$~}

\leftline{TUM-HEP-812/11\hspace{3cm}
DFPD-11/TH/06 \hspace{3cm}
IFT-UAM/CSIC-11-60}
\bigskip

\title{\boldmath Flavour violation in a supersymmetric $T'$ model}

 \author[a,b]{Luca Merlo,} 
 \author[c]{Stefano Rigolin}
 \author[d]{and Bryan Zald\'ivar} 

\affiliation[a]{Physik-Department, Technische Universit\"at M\"unchen, \\
  	       James-Franck-Strasse, D-85748 Garching, Germany}
\affiliation[b]{TUM Institute for Advanced Study, Technische Universit\"at M\"unchen, \\
  	       Lichtenbergstrasse 2a, D-85748 Garching, Germany}
\affiliation[c]{Dipartimento di Fisica ``G.~Galilei'', Universit\`a di Padova and \\
  	       INFN, Sezione di Padova, Via Marzolo~8, I-35131 Padua, Italy}
\affiliation[d]{Instituto de F\'isica Te\'orica, IFT-UAM/CSIC, \\
	       Nicol\'as Cabrera 15, UAM Cantoblanco 28049, Madrid}

\emailAdd{luca.merlo@ph.tum.de}
\emailAdd{rigolin@pd.infn.it}
\emailAdd{bryan.zaldivar@uam.es}

\abstract{We describe the phenomenology of the flavour changing neutral current sector of a supersymmetric model, 
invariant under the $T'$ discrete flavour group. This model has been proposed in Ref.~\cite{FHLM:Tp} and describes 
realistic leptonic and hadronic masses and mixings and predicts the amount of flavour changing in terms of the 
small flavour breaking parameter $u\in[0.007,\,0.05]$. 
For small values of $u$, the model almost reduces to the MSUGRA framework, while sizable deviations from MSUGRA can be, 
instead, observed when larger values of $u$ and $\tan\beta$ are considered. We analyse in detail $T'$ \meg prediction, 
concerning the leptonic sector, while for the hadronic sector we concentrate on \bsg and neutral B meson mass differences 
$\Delta M_{B_{s,d}}$. Moreover, for the first time a comparative study on leptonic and hadronic observables for the 
$T'$ model is performed. The experimental data on FCNC observables severely constrain the model in the 
small $m_0$ region. Conversely for larger $m_0$, the $T'$ model satisfies all the bounds.}


\begin{document} 

\maketitle

\flushbottom

%
%

\section{Introduction}
\label{sec:Intro}

In the Standard Model (SM) of particle physics an explanation of the origin of the fermionic mass and mixing patterns 
is missing. The mass spectrum is highly hierarchical, spreading over several orders of magnitude. The quark mixing matrix 
$V_{CKM}$ is close to the unity matrix and the deviations from unity are usually parametrised in powers of the (sin of the)
Cabibbo angle $\lambda$:
\be
V_{CKM} = \left(
\begin{array}{ccc}
	1-\lambda^2/2               & \lambda        & A \lambda^3 (\bar\rho -i \bar\eta) \\
	-\lambda                     & 1-\lambda^2/2  & A \lambda^2                \\
	A \lambda^3 (1-\bar\rho -i\bar\eta)  & - A \lambda^2  & 1                          \\
	\end{array}\right) \; .
\label{CKMWolf}
\ee
Recent global fits on the $V_{CKM}$ elements \cite{PDG2010} give the following values (at $1\sigma$):
\bald
&\lambda = 0.2253 \pm0.0007\,, &&\qquad\qquad 
A = 0.808^{+0.022}_{-0.015}\,, \\
&\bar\rho = 0.132^{+0.022}_{-0.014}\,, &&\qquad\qquad 
\bar\eta = 0.341\pm 0.013 \,. 
\eald
A fundamental theory that explains all the SM flavour parameters is missing and its search constitutes the so-called 
{\em SM flavour problem}.

While already present in the original version of the SM, the flavour problem has worsened even more after the discovery 
that neutrinos oscillate and consequently have both a non-vanishing mass and a non-vanishing leptonic flavour mixings. 
The global fit in neutrino oscillation experiments presented in Ref.~\cite{PDG2010} gives the following bounds at 
$99.73\%$ of C.L.: 
\bgad
7.05\times 10^{-5}\,{\rm eV}^2 \leq \Delta\,m^2_{\rm sol} \leq 8.34\times 10^{-5}\,{\rm eV}^2 \\
2.07\times 10^{-3}\,{\rm eV}^2 \leq \Delta\,m^2_{\rm atm} \leq 2.75\times 10^{-3}\,{\rm eV}^2
\egad
with the electron neutrino mass $m_{\nu_e} \le 2$ eV from Tritium experiments \cite{PDG2010}. Moreover the 
leptonic mixing matrix, $U_{PMNS}$ has a completely different structure than the $V_{CKM}$, with two very large 
angles and the third very small. The solar, atmospheric and reactor angle (namely $\theta_{12}, \theta_{24}$ 
and $\theta_{13}$) bounds read (at $99.73\%$ of C.L.):
\bgad
0.25 \leq \sin^2\theta_{12} \leq 0.37 \\
0.36 \leq \sin^2\theta_{23} \leq 0.67 \\ 
\sin^2\theta_{13} \leq 0.056\,.
\egad
Recently new data from T2K collaboration \cite{Abe:2011sj} and corresponding fits \cite{Newfit13,Schwetz:2011qt} 
indicate a $3\sigma$ evidence of a non--vanishing $\theta_{13}$ with a relatively ``large'' central value.

A very good approximation for the lepton mixing matrix is provided by the so-called Tri-Bimaximal (TB) pattern 
\cite{HPS:TBM,HS:TBM,Xing:TBM}, which corresponds to the mixing angles $\sin^2\theta^{TB}_{12}=1/3$, 
$\sin^2\theta^{TB}_{23}=1/2$, $\sin\theta^{TB}_{13}=0$, and agrees at about the $1\sigma$ level with the data. The 
appeal of the TB scheme is also due to the independence of the mixing angles from any mass parameter: indeed they are 
defined only by simple numerical factors. Other mixing schemes, like the Bimaximal \cite{Vissani:BM,BPWW:BM,AF:BM,MS:BM,Stancu:1999ct} 
and the Golden Ratio \cite{KRS:GR,Rodejohann:GR} ones, exhibit analogous features, but the agreement with the 
experimental data is less promising compared to the TB one\footnote{Typical TB models predict $\sin^2\theta_{13}\approx 
0.003$ with a tension with the present T2K central value: $\sin^2\theta_{13}=0.029$.}.

The Supersymmetric extension of the Standard Model in its most straightforward (Minimal) description (MSSM), 
while stabilizing the Higgs scale, severely worsens the flavour problem as order of one hundred extra parameters 
in the flavour sector are introduced. Most of these parameters are already highly constrained (``fine-tuned'') by present 
experimental data. A fundamental theory that explains the flavour mass patterns is missing in the MSSM too.

Following the idea that the fermion mass structures could arise from a symmetry principle, flavour symmetries have 
been introduced, both in the context of the SM and its supersymmetric (SUSY) extensions. However a commonly 
accepted approach is still missing and many different examples have been proposed in the literature based on 
a large variety of symmetries: either abelian on non-abelian, local or global, continuous or discrete. 

The largest flavour symmetry allowed in the SM in the limit of vanishing Yukawas couplings is $U(3)^5$, i.e. the 
symmetry of the kinetic terms. Such a symmetry is implemented in the so-called Minimal Flavour Violation (MFV) ansatz 
\cite{DGIS:MFV,CGIW:MLFV,DP:MLFV,AIMMN:VMLFV}: a very concise and predictive framework built on the assumption that, 
at low energy, the Yukawa couplings are the only sources of flavour 
and CP violation both in the SM and in any possible extension beyond it. The technical realisation of the MFV ansatz 
promotes the Yukawa matrices to {\em spurion} fields, thus making the Lagrangian manifestly invariant under the full 
flavour symmetry. Any model beyond the SM, implemented with the MFV ansatz, is successful in suppressing flavour changing 
neutral current (FCNC) contributions with a typical new physics flavour scale in the TeV region. When the symmetry is 
gauged \cite{GRV:SU3gauged,Feldmann:SU5gauged,GMS:SU3gaugedLR}, new effects could arise as it has been recently pointed 
out in \cite{BMS:NeutralBSG}. 

On the other hand, a natural mechanism to explain fermions masses and mixings in the MFV framework is still missing 
\cite{FJM:MFVScalarPotentialBiFund,AGMR:ScalarPotentialMFV}. Some improvements in this 
direction are possible when considering smaller symmetry groups. However, any deviation from the MFV scheme, usually 
allows the appearance of dangerous flavour violating contributions. Interesting attempts, in which FCNC processes are 
under control and the fermion mass origin is discussed, are the one adopting as continuous symmetries $U(2)$ 
\cite{PT:U2,BDH:U2,BHRR:U2}, $U(2)^3$\cite{BIJLS:U2MFVSusy, Crivellin:2011sj}, $SU(3)$ and $SO(3)$ \cite{KR:SU3QuarkLepton,
CFN:SU3QuarkLepton,RV:SU3QuarkLepton,KR:SU3QuarkLeptonGUT,MR:SU3TB}. 
Notice that only in Refs.~\cite{KR:SU3QuarkLepton,CFN:SU3QuarkLepton,RV:SU3QuarkLepton,KR:SU3QuarkLeptonGUT,MR:SU3TB} 
both the fermion sectors are fully discussed; however, in such papers, an \textit{ad hoc} dynamics is adopted 
to achieve the Yukawa textures.

Discrete symmetries represent an alternative solution to the flavour problem, exhibiting their power especially 
in the leptonic sector. In \cite{MR:A4EWscale,BMV:A4TBM,AF:Extra,AF:Modular,AFL:Orbifold} it has been pointed out 
that a (spontaneously broken) flavour symmetry based on the discrete group $A_4$ appears to be particularly suitable 
to reproduce the TB lepton mixing pattern as a first approximation. Many other solutions have also been considered 
to generate the TB scheme, based on larger discrete flavour groups\footnote{As a review on the possible discrete 
groups see for example \cite{AF:ReviewDiscreteSyms}.}, such as $S_4$, $T'$ and $\Delta(27)$. Discrete symmetries 
have been successfully used to describe also the Bimaximal \cite{AFM:BMS4,ABM:PSS4} and the Golden Ratio 
\cite{ES:GRA5,FP:GRA5} schemes. 

Models based on discrete symmetries, however, usually highly deviate from the MFV 
framework and potentially large flavour violating contributions are expected. In spite of the large effort devoted 
in describing flavour models predicting mass patterns in agreement with the data, only few studies analysed possible 
contributions to lepton flavour violating (LFV) and FCNC observables \cite{HKY:LFVS3, MMP:LFVS3, Kifune:2007fj,IKOT:LFVD4,
IKOT:LFVA41,IKOST:LFVDelta54, Ho:2010yp,Frampton:2010uw,ABMP:Constraining1,ABMP:Constraining2}. In particular, the phenomenology of the 
Altarelli-Feruglio (AF) lepton model \cite{AF:Extra,AF:Modular,AFL:Orbifold}, based on $A_4$, has been studied 
in details both in the SM \cite{FHLM:LFVinA4,FHLM:LFVA4proceeding} and in the SUSY \cite{FHM:Vacuum,
LMP:RunningA4,FHLM:LFVinSUSYA4,FP:RareDecaysA4} scenario. The main result of these analyses is that the 
flavour symmetry not only governs the structure of the fermion masses and mixings, but also constrains the relative 
contribution to flavour violating observables, such as for example the $\ell_i\to\ell_j\gamma$ decays. 
In this context, in fact, the predicted rates of FCNC processes are more suppressed than in a general effective operator 
approach \cite{INP:FP_Review}, due to peculiar cancellations, allowing for a new physics scale close to the TeV range, 
without conflicting with the present bounds. This type of analysis is therefore essential for testing a specific model 
and for providing predictions which could allow to discriminate among all the available proposals. 

In Ref.~\cite{FHLM:Tp} a SUSY model based on the discrete group $T'$ has been constructed. This model accounts both for 
leptons, predicting the TB mixing pattern, and for quarks, providing a realistic CKM matrix. In particular, the lepton 
sector corresponds to that of the AF model and therefore this $T'$ model can be considered as an extension of the SUSY 
AF model to the quark sector. However in Ref.~\cite{FHLM:Tp} no phenomenological analysis of the hadronic sector 
has been presented. Consequently, the main goal of the present paper is to fill this gap and to present a complete 
description of FCNC observables both for leptons and hadrons. 

The paper is organized has follows: in Sect.~\ref{GenFeatures} the general features of the $T'$ model are 
briefly reminded. In Sect.~\ref{KinMass} the kinetic terms and the mass matrices for fermions and sfermions 
are derived in a particularly convenient basis, and then in Sect.~\ref{sec:PhysicalBasis} the results are 
presented in a more ``physical'' basis. Most of the details regarding the derivation of the soft mass matrices 
and the connection between the two basis are deferred to the Appendices. The main phenomenological results 
of the $T'$ model are described in Sect.~\ref{sec:Phen}. Finally we conclude in Sect.~\ref{sec:Conclusions}. 

%
%

\mathversion{bold}
\section{General features of the supersymmetric $T'$ model}
\label{GenFeatures}
\mathversion{normal}

The $T'$ model, we are interested in, has been proposed in Ref.~\cite{FHLM:Tp}. We summarise here the notation 
and the main results, pointing out, when necessary, the few slight differences introduced here. 

The full flavour group, $G_f$, is a product of various terms,
\be
G_f=T'\times Z_3\times U(1)_{FN}\times U(1)_{R} \,,
\ee
each of them playing a different role. The spontaneous breaking of the $T'$ group (see Appendix \ref{AppA} and 
Ref.~\cite{FHLM:Tp}) is mainly responsible for the fermion mixings, while the correct fermion mass hierarchies 
mainly originate from a combined effect of the $Z_3$ and $U(1)_{FN}$ groups. Indeed the $Z_3$ symmetry is used to 
forbid unwanted couplings between the SM fields and the extra fields added in the model that will be specified 
in the following; the continuous $U(1)_{FN}$ provides, according to the original Froggatt-Nielsen mechanism \cite{FN}, 
the suppressions necessary to reproduce the mass hierarchies. Finally, the $U(1)_R$ is a common ingredient of SUSY 
constructions, containing the usual $R$-parity as a subgroup. 

\begin{table}[!t]
\centering
\begin{tabular}{|c||c|c|c|c||c|c|c|c|c|c||c|}
\hline
&&&&&&&&&&&\\[-4mm]
& $\ell$ & $e^c$ & $\mu^c$ & $\tau^c$ &  $D_q$ & $D_u^c$ & $D_d^c$ & $q_3$ & $t^c$ & $b^c$  &  
    $H_{u,d}$ \\[2mm]
\hline
&&&&&&&&&&&\\[-4mm]
$T'$ & \bf3 & \bf1 & $\bf1''$ & $\bf1'$ & $\bf2''$ & $\bf2''$ & $\bf2''$ & \bf1 & \bf1 & \bf1 & \bf1 \\[2mm]
$Z_3$ & $\om$ & $\om^2$ & $\om^2$ & $\om^2$ & $\om$ & $\om^2$ & $\om^2$ & $\om$ & $\om^2$ & 
    $\om^2$ & $1$ \\[2mm]
$U(1)_{FN}$ & 0 & 2 & 1 & 0 & 0 & 1 & 1 & 0 & 0 & 1 & 0 \\[2mm]
$U(1)_R$ & 1 & 1 & 1 & 1 & 1 & 1 & 1 & 1 & 1 & 1 & 0 \\[2mm]
\hline
\end{tabular}
\caption{\it The transformation properties of matter fields and Higgses under the flavour group $G_f$.}
\label{table:TransformationsMatter}
\end{table}

The matter fields of the model, together with their transformation 
properties under the flavour group, are listed in Tab.~\ref{table:TransformationsMatter} adopting, both for fields 
and their superpartners, the following notation: $\ell=(\ell_1,\ell_2,\ell_3)$, where $\ell_1=(\nu_e,e)$, 
$\ell_2=(\nu_\mu,\mu)$ and $\ell_3=(\nu_\tau,\tau)$, are the $SU(2)_L$-doublet (s)leptons; $e^c$, $\mu^c$ and 
$\tau^c$ are the $SU(2)_L$-singlet (s)leptons; $D_q=(q_1,q_2)$, where $q_1=(u,d)$ and $q_2=(c,s)$, are the 
$SU(2)_L$-doublet (s)quarks of the first two generations; $D_u^c=(u^c,c^c)$ and $D_d^c=(d^c,s^c)$ are the 
$SU(2)_L$-singlet (s)quarks of the first two generations; $q_3=(t,b)$ is the $SU(2)_L$-doublet (s)quark of the 
third generation while $t^c$ and $b^c$ are the $SU(2)$-singlet (s)quarks of the third generation. $H_{u,d}$ are 
the two usual SUSY Higgs doublets. Under the discrete symmetry group $T'$, $\ell$ transforms as a triplet, 
$D_q$, $D_u^c$ and $D_d^c$ as doublets, $e^c$, $\mu^c$, $\tau^c$, $q_3$, $t^c$, $b^c$ and $H_{u,d}$ are singlets. 

\begin{table}[!ht]
\centering
\begin{tabular}{|c||c||c|c|c|c|c|}
\hline
&&&&&&\\[-4mm]
& $\theta$ & $\phit$ & $\phis$ & $\xi$, $\xit$ & $\eta$ & $\xipp$ \\[2mm]
\hline
&&&&&&\\[-4mm]
$T'$ & {\bf1} & {\bf3} & {\bf3} & {\bf1} &  $\bf2'$ & $\bf1''$ \\[2mm]
$Z_3$ & $1$ & $1$ & $\om$ & $\om$  & 1 & 1 \\[2mm]
$U(1)_{FN}$ & $-1$ & 0 & 0 & 0 & 0 & 0 \\[2mm]
$U(1)_R$ & 0 & 0 & 0 & 0 & 0 & 0 \\[2mm]
\hline
\end{tabular}
\caption{\it The transformation properties of flavon fields under the flavour group $G_f$.}
\label{table:TransformationsFlavons}
\end{table}

Apart from matter superfields the spectrum of the model accounts for several scalar fields that are neutral 
under the SM gauge group. Their transformation properties under the flavour group $G_f$ are shown in 
Tab.~\ref{table:TransformationsFlavons}. This new set of scalar fields is responsible for the spontaneous 
breaking of the flavour symmetry and are usually called {\it flavons}. As described in details in 
\cite{FHLM:Tp}, it is possible to construct a scalar potential in such a way that flavons develop vacuum 
expectation values (vevs) along the following (flavour) directions:
\bgad
\dfrac{\mean{\varphi_T}}{\Lambda_f}=(u,\,0,\,0)+(c_{T1}\, u^2,\,c_{T2}\, t\,u^2,\,c_{T3}\, u^2)\,,\\
\dfrac{\mean{\varphi_S}}{\Lambda_f}=c_b(u,\,u,\,u)+ {\cal O}(u^2)\,, \\
\dfrac{\mean{\eta}}{\Lambda_f}=c'(u,\,0)+(c_{\eta1}\,u^2,\,c_{\eta2}\,u^2)\,,\\
\dfrac{\mean{\xi}}{\Lambda_f}=c_a\, u \, +{\cal O}(u^2)\,,\qquad
\dfrac{\mean{\xit}}{\Lambda_f}=c_c\, u^2\,,\qquad
\dfrac{\mean{\xi''}}{\Lambda_f}=c''\,u^2\,,\qquad
\dfrac{\mean{\theta_{FN}}}{\Lambda_f}=t \,.
\label{VEVs}
\egad
All the coefficients $c_i$, appearing in Eqs.~(\ref{VEVs}), are complex numbers with absolute value of order 
one, while $u$ and $t$ are two, small, $T'$ symmetry breaking parameters. In Eqs.~(\ref{VEVs}) we explicitly 
considered all contributions up to second order in $u$, when relevant. Notice that both $u$ and $t$ can, in general, 
be  complex, but one can show that, through fields redefinitions, they can be taken real and positive. The cutoff 
scale $\Lambda_f$ is associated to the flavour dynamics and it is expected close to the grand unification scale, 
$\Lambda_f \lesssim 10^{16}$ GeV.

In Ref.~\cite{FHLM:Tp} the most general superpotential invariant under the SM gauge group and under $G_f$ has 
been constructed, according to the transformation properties listed in Tabs.~\ref{table:TransformationsMatter}
and \ref{table:TransformationsFlavons}. Considering only the leading-order (LO) terms in the $u$ expansion of 
such a superpotential and the corresponding LO approximation of the expressions in Eq.~(\ref{VEVs}), the 
charged leptons mass matrix reads:
\be
M_\ell \propto \left(
\begin{array}{ccc}
	y_e\, t^2 & 0& 0\\
	0& y_\mu\, t& 0\\
	0& 0& y_\tau
	\end{array}\right) u\; ,
\label{mlLO}
\ee
with $y_e$, $y_\mu$ and $y_\tau$ being complex numbers with absolute values of order one. The parameter $t$ 
governs the relative hierarchy among the charged lepton masses. Through a comparison with the experimental 
data one infers $t\approx 0.05$.
A lower bound on $u$ can be obtained from the relation that connects the $T'$ breaking parameter to the SM electroweak
vev ($v$), the $\tau$ Yukawa coupling and pole mass ($y_\tau$ and $m_\tau$) and the MSSM neutral Higgs vevs. 
ratio ($\tan\beta =v_u/v_d$):
\be
u=\dfrac{1}{|y_\tau|} \dfrac{\sqrt{2}\, m_\tau}{v\, \cos\beta} \approx 0.01 \dfrac{\tan\beta}{|y_\tau|}\,.
\label{tanb&u&yt}
\ee
By requiring a $\tau$--Yukawa coupling consistent with the perturbative regime, $|y_\tau|<3$, one gets $u \geq 
0.05 ~(0.007)$ for $\tan\beta=15~(2)$ respectively.

Following Ref.~\cite{FHLM:Tp}, at the LO the light neutrino mass matrix is given by
\be
M_\nu \propto \left(
\begin{array}{ccc}
	3a+2 b & -b & -b\\
	-b & 2b & 3a-b\\
	-b & 3a-b & 2 b
	\end{array}\right) u\,,
\label{mnuLO}
\ee
and it can be diagonalised by the TB mixing matrix, defined, up to phases, by
\be
U_{TB}=\left(
\begin{array}{ccc}
	\sqrt{2/3}& 1/\sqrt{3}& 0\\
	-1/\sqrt{6}& 1/\sqrt{3}& -1/\sqrt{2}\\
	-1/\sqrt{6}& 1/\sqrt{3}& +1/\sqrt{2}\\
	\end{array}\right)\,.
\label{UTB}
\ee

When considering the sub-leading contributions to the matter superpotential in the $u$ expansion (see 
Ref.~\cite{FHLM:Tp} for details) and the corresponding sub-leading terms in Eq.~(\ref{VEVs}), the mass matrices 
in Eqs.~(\ref{mlLO}) and (\ref{mnuLO}) get modified and consequently the diagonalisation matrix in Eq.~(\ref{UTB}) 
slightly deviates from the TB pattern. An upper bound on $u$ of about $0.05$ is obtained by the requirement that 
these corrections do not perturb excessively the TB values of the neutrino mixing angles: the strongest constraint 
provided by the solar mixing angle. As a consequence, for $\tan\beta=2$ only values for $u$ in the range 
\be
0.007 \lesssim u \lesssim 0.05 
\label{ubound}
\ee
are allowed, while for $\tan\beta=15$ only $u=0.05$ is permitted.

Considering the quark sector, the particular choice of the transformation properties for the quark fields and 
the specific alignment of the flavon vevs. determine the characteristic ``shell'' filling of the mass matrices. 
In terms of the $T'$ breaking parameters $u$ and $t$, they read:
\be
M_u \sim \left(
        \begin{array}{ccc}
           t^2\,u^2  & t\,u^2 & t\,u^2 \\
            t\,u^2 & t\,u & t\,u \\
            u^2 & u & 1 \\
        \end{array}
\right)\,,\qquad\qquad
M_d \sim \left(
        \begin{array}{ccc}
            t^2\,u^2  & \sqrt{t}\,u^2 & t\,u^2 \\
            \sqrt{t}\,u^2 & t\,u & t\,u \\
            t\,u^2 & t\,u & t \\
        \end{array}
\right)\,.
\label{Qmassmatrices}
\ee
In the previous mass matrices we didn't explicitly write the unknown, order one, coefficients entering in each 
of the entries. In the largest part of the allowed range for $u$, realistic quark masses and CKM mixings follow 
from Eqs.~(\ref{Qmassmatrices}). Only when very small values for $u$ are considered, it is necessary to compensate 
with coefficients slightly larger than one. We will discuss this issue more in detail in the following sections.

In order to break the flavour group along the required directions and to allow the flavons to get vevs. as in 
Eqs.~(\ref{VEVs}),a new set of fields, called {\it driving} superfields, has to be introduced. These fields 
transform only under the flavour group $G_f$ and couple only to the flavons.
The driving fields, together with their transformation properties under $G_f$, are listed in 
Tab.~\ref{table:TransformationsDriving}. 

\begin{table}[ht!]
\centering
\begin{tabular}{|c||c|c|c|c|c|}
\hline
&&&&&\\[-9pt]
Field & $\varphi^0_T$ & $\varphi^0_S$ & $\eta^0$ & $\xi^0$ & $ \xi^{\prime0}$ \\
&&&&&\\[-9pt]
\hline
&&&&&\\[-9pt]
$T^{\prime}$ & $\bf3$ & $\bf3$ & $\bf2''$ & $\bf1$ & $\bf1'$ \\[3pt]
$Z_3$ & $1$ & $\omega$ & $1$ & $\omega$ & $1$ \\ [3pt]
$U(1)_{FN}$ & $0$ & $0$ & $0$ & $0$ & $0$ \\ [3pt]
$U(1)_R$ & $2$ & $2$ & $2$ & $2$ & $2$ \\ [3pt]
\hline
\end{tabular}
\caption{\it The transformation properties of the driving fields under the flavour group $G_f$.}
\label{table:TransformationsDriving}
\end{table}

As it has been discussed in \cite{FHM:Vacuum,FHLM:LFVinSUSYA4}, in the context of the $A_4$ model, the driving fields 
develop non-vanishing vevs. only once soft SUSY breaking terms are introduced.
Extending such an analysis to the case of the $T'$ model, the following vevs. for the driving superfields are obtained: 
to first order in $u$ they are given by
\bgad
\dfrac{\mean{\varphi_T^0}}{\msusy}=c_T^0(1,\,0,\,0)+(c^0_{T1}\, u,\,c^0_{T2} u,\,c^0_{T3}\, u)\,,\\
\dfrac{\mean{\varphi_S^0}}{\msusy}=c_b^0(1,\,1,\,1)+(c^0_{S1}\,u,\,c^0_{S2}\,u,\,c^0_{S3}\,u)\,,\\
\dfrac{\mean{\eta^0}}{\msusy}=(0,\,c^0_\eta)+(c^0_{\eta1}\,u,\,c^0_{\eta2}\,u)\,,\\
\dfrac{\mean{\xi^0}}{\msusy}=c^0_a+c^0_c\, u\,,\qquad\qquad
\dfrac{\mean{\xi^{\prime 0}}}{\msusy}=c^0_{\xi'}\,u\,.
\label{VEVsDriving}
\egad
All the coefficients $c^0_i$, appearing in the previous equations, are complex numbers with absolute value 
of order one. Here, $m_0$ represents the common soft SUSY breaking scalar mass. Starting from Eqs.~(\ref{VEVs}) 
and (\ref{VEVsDriving}), the vevs. of the flavon $F$-terms which non-trivially contribute to the sfermion 
mass matrices can be derived:
\bgad
\dfrac{1}{\Lambda_f}\mean{\dfrac{\derp w}{\derp \varphi_T}}=
   \msusy\left[c^F(u,\,0,\,0)+(c^F_{T1}\, u^2,\,c^F_{T2}\, u^2,\,c^F_{T3}\, u^2)\right]\,,\\
\dfrac{1}{\Lambda_f}\mean{\dfrac{\derp w}{\derp \varphi_S}}=
   \msusy\left[c^F_b(u,\,u,\,u)+(c^F_{S1}\,u^2,\,c^F_{S2}\,u^2,\,c^F_{S3}\,u^2)\right]\,,\\
\dfrac{1}{\Lambda_f}\mean{\dfrac{\derp w}{\derp \eta}}=
   \msusy\left[c^{F\prime}(u,\,0)+(c^F_{\eta1}\,u^2,\,c^F_{\eta2}\,u^2)\right]\,,\\
\dfrac{1}{\Lambda_f}\mean{\dfrac{\derp w}{\derp \xi}}=
   \msusy c^F_a\, u\,,\quad
\dfrac{1}{\Lambda_f}\mean{\dfrac{\derp w}{\derp \xit}}=
   \msusy c^F_c\, u^2\,,\quad
\dfrac{1}{\Lambda_f}\mean{\dfrac{\derp w}{\derp \xi''}}=
   \msusy\, c^{F\prime\prime}\,u^2\,.
\label{VEVsFTerms}
\egad
Again all the coefficients $c^F_i$ are complex numbers with absolute value of order one. It is interesting to note 
that the LO terms in Eq.~(\ref{VEVsFTerms}) are proportional to those in Eq.~(\ref{VEVs}), confirming what 
found in \cite{FHM:Vacuum,FHLM:Tp,RV:Supergravity}.

%
%

\section{Kinetic terms and mass matrices}
\label{KinMass}

The Lagrangian of the SUSY $T'$ model from which both fermion and sfermion masses are obtained, is given by
\be
\LL=\int d^2\theta d^2\overline{\theta}\, \cK(\ov{z}, e^{2 V} z)+\left[\int d^2 \theta\, w(z)+\hc\right]+
     \dfrac{1}{4}\left[\int d^2\theta\, f(z) \,\cW\,\cW+\hc\right]\,,
\label{leel}
\ee
where $\cK(\ov{z},z)$ is the K\"ahler potential, $w(z)$ is the superpotential and $f(z)$ is the gauge kinetic 
function. $V$ is the Lie-algebra valued vector supermultiplet, describing the gauge fields and their superpartners 
while $\cW$ is the chiral superfield describing, together with the function $f(z)$, the kinetic terms of gauge 
bosons and their superpartners. Notice that  the fermion sector has already been discussed in Ref.~\cite{FHLM:Tp}, 
but assuming canonical kinetic terms.

It is assumed that the breaking of SUSY occurs at a scale higher than or comparable to the flavour scale, so that 
at energies close to the cutoff scale we have non-universal boundary conditions for the soft SUSY breaking terms, 
dictated by the flavour symmetry.
The soft SUSY breaking terms are generated from the SUSY Lagrangian by promoting all coupling constants 
(such as the Yukawa couplings, the couplings in the flavon superpotential and the couplings in the K\"ahler 
potential) to superfields with constant $\thh$ and $\thh\thhb$ components \cite{Luty:TASISusyBreaking}. 
The analysis of the lepton sector is going to be the same as the one presented in Ref.~\cite{FHLM:LFVinSUSYA4} 
for the AF model, but for the presence in the $T'$ case of two additional flavons: $\eta$ and $\xi''$.

%
%
\subsection{K\"ahler potential}

For non-vanishing values of the $T'$ breaking parameters $u$ and $t$, the K\"ahler potential deviates 
from the canonical form, $\cK(\ov{z},z)=\ov{z} z$, due to the contributions of non-renormalizable terms, 
invariant under both the gauge and the flavour symmetries, containing the flavon fields. Such terms are 
suitably suppressed by the flavour scale $\Lambda_f$, but cannot be neglected as they affect the fermionic mass matrices, through redefinitions needed to move into the basis of canonically normalised kinetic terms.
In the following we are going to include all the terms up to the second 
order in the expansion in $u$ and additional contributions proportional to $t$, when needed in order not to have vanishing elements in the sfermion mass matrices.

The K\"ahler potential can be written as $\cK=\cK_\ell+\cK_q+\cK_r$ where the three terms correspond 
to lepton, quark and remaining field contributions, respectively.
Let us start considering the leptonic contribution:
\be
\cK_\ell=\cK_\ell^{(0)}+\cK_\ell^{(1)}+\cK_\ell^{(2)}+\ldots
\ee
where $\cK_\ell^{(i)}$ represents the $i^{th}$ term in the $u$ expansion. The LO term reads:
\be
\cK_\ell^{(0)}=k^\ell_0\sum_{i=1}^3\ov{\ell}_i \ell_i + \sum_{i=1}^3 \left[(k_0^e)_i +({\hat k}^e_0)_i 
              \dfrac{\vert\theta_{FN}\vert^2}{\Lambda_f^2}\right]\ov{\ell}^c_i  \ell^c_i
\label{kappa0Lep}
\ee
with $\ell^c=(e^c,\mu^c,\tau^c)$. 
Notice that the contributions from the superfield $\theta_{FN}$ can be neglected, with the exception 
of the right-handed sector, where we consider terms up to second order in $t$. The next order term 
in the $u$ expansion is given by
\be
\begin{split}
\cK_\ell^{(1)}=&\quad\dfrac{k^\ell_S}{\Lambda_f}  (\varphi_T (\ov{\ell}\,\ell)_S) + \dfrac{k^\ell_A}{\Lambda_f} 
  (\varphi_T (\ov{\ell}\,\ell)_A)+\dfrac{k^\ell_{\xi''}}{\Lambda_f}  \xi''(\ov{\ell}\,\ell)'+\\
&+\dfrac{k^{\ell\prime}_S}{\Lambda_f}  (\ov{\varphi}_T (\ov{\ell}\,\ell)_S)+\dfrac{k^{\ell\prime}_A}{\Lambda_f} 
  (\ov{\varphi}_T (\ov{\ell}\,\ell)_A)+\dfrac{k^{\ell\prime}_{\xi''}}{\Lambda_f}  \ov{\xi}''(\ov{\ell}\,\ell)''+\hc
\end{split}
\label{kappa1Lep}
\ee
where $(\ldots)$, $(\ldots)'$ and $(\ldots)''$ denote the $\bf 1$, $\bf 1'$ and $\bf 1''$ singlets of $T'$, while 
$(\ldots)_S$ and  $(\ldots)_A$ the symmetric and antisymmetric triplets originating from the contraction of two 
$\bf 3$ representations (see the Appendix \ref{AppA} for details). Notice that the $SU(2)_L$ singlet superfields $\ell^c$ are not affected at this order in the expansion. To discuss the next term it is useful to distinguish the contributions to the left-handed doublets 
and the right-handed singlets: 
\be
\cK_\ell^{(2)}=\cK_{\ell\,L}^{(2)}+\cK_{\ell\,R}^{(2)}\,.
\ee
For lepton doublets one finds:
\be
\cK_{\ell\,L}^{(2)}= \sum_{i=1}^{17} \dfrac{k^\ell_i}{\Lambda_f^2}  (X^\ell_i\, \ov{\ell}\,\ell)
\label{kappa2LLep}
\ee
where $X^\ell$ is a list of $Z_3$-invariant operators, bilinear in the flavon superfields and their conjugates,
\be
\begin{split}
X^\ell=&\Big\{\ov{\xi}\, \xi,\,\varphi_T^2,\,\ov{\varphi}_T^2,\,{\ov \varphi_T}\,\varphi_T,\,\ov{\varphi}_S\, 
        \varphi_S,\,\ov{\xi}''\, \xi'',\,\ov{\eta}\, \eta,\,\ov{\xi}\,\varphi_S,\,\ov{\varphi}_S\,\xi,\,\\
&\quad\xi^{\prime\prime2},\,\ov{\xi}^{\prime\prime2},\,\eta^2,\,\ov{\eta}^2,\,\varphi_T\,\xi'',\,
        \ov{\varphi}_T\,\xi'',\,\varphi_T\,\ov{\xi}'',\,\ov{\varphi}_T\,\ov{\xi}'' \Big\}\,,
\end{split}
\label{XlistLep}
\ee
and each quantity $k^\ell_i$ represents a list of parameters since there can be different non-equivalent ways 
of combining $X^\ell_i$ with $\ov{\ell}\,\ell$ to form a $T'$-invariant. There are also obvious relations 
among the coefficients $k^\ell_i$  to guarantee that $\cK_{\ell\,L}^{(2)}$ is real. Notice that in $X^\ell$ 
we do not consider the possible combinations with the flavon $\xit$, because it has exactly the same quantum 
numbers as $\xi$ and therefore their contributions can be identified by the same coupling constant. Furthermore, 
notice that some of the terms in the sum of Eq.~(\ref{kappa2LLep}) are vanishing once we consider specific 
vevs. of the flavons in Eq.~(\ref{VEVs}).

For lepton singlets, we can distinguish a diagonal contribution and a non-diagonal one:
\be
\cK_{\ell\,R}^{(2)}=\left[\cK_{\ell\,R}^{(2)}\right]_{d}+\left[\cK_{\ell\,R}^{(2)}\right]_{nd} \nn
\ee
respectively given by:
\bea
&&\left[\cK_{\ell\,R}^{(2)}\right]_d =\dfrac{1}{\Lambda_f^2}\sum_{i=1}^{7} \left[ (k^e_e)_i\, (X^\ell_i)\ov{e}^c e^c 
    +(k^e_\mu)_i\,(X^\ell_i)\ov{\mu}^c \mu^c+(k^e_\tau)_i\,(X^\ell_i) \ov{\tau}^c \tau^c\right] \,,\label{kappa2R1Lep} \\
&&\begin{split}
\left[\cK_{\ell\,R}^{(2)}\right]_{nd}=&\quad
\dfrac{\ov{\theta}_{FN}}{\Lambda_f^2}(k_{e\mu}^e)_1\,\ov{\xi}''\ov{e}^c\mu^c+ 
\dfrac{\ov{\theta}^2_{FN}}{\Lambda_f^3}(k_{e\tau}^e)_1\,\xi''\ov{e}^c\tau^c+ 
\dfrac{\ov{\theta}_{FN}}{\Lambda_f^2}(k_{\mu\tau}^e)_1\,\ov{\xi}''\ov{\mu}^c\tau^c+ \\
&+\dfrac{\ov{\theta}_{FN}}{\Lambda_f^3}\left[\sum_{i=2}^5(k_{e\mu}^e)_i\,(X^\ell_i)' + 
    (k_{e\mu}^e)_6\,X^\ell_{10} +(k_{e\mu}^e)_7\,X^\ell_{13} \right]\ov{e}^c\mu^c+ \\
&+\dfrac{\ov{\theta}_{FN}^2 }{\Lambda_f^4} \left[\sum_{i=2}^5(k_{e\tau}^e)_i\,(X^\ell_i)'' + 
    (k_{e\tau}^e)_6\, X^\ell_{11}+(k_{e\tau}^e)_7\, X^\ell_{12} \right]\ov{e}^c \tau^c+ \\
&+\dfrac{\ov{\theta}_{FN}}{\Lambda_f^3}\left[\sum_{i=2}^5(k_{\mu\tau}^e)_i\,(X^\ell_i)' + 
    (k_{\mu\tau}^e)_6\, X^\ell_{10} +(k_{\mu\tau}^e)_7\, X^\ell_{13}\right]\ov{\mu^c}  \tau^c + \,\hc\,.
\end{split}
\label{kappa2R2Lep}
\eea
Due to the structure of the flavon vevs. at LO, the first row in Eq.~(\ref{kappa2R2Lep}) vanish. If also the 
next-to-leading order (NLO) is considered, these terms contribute at the same level as the operators in the 
subsequent rows. The expressions in Eqs.~(\ref{kappa0Lep})-(\ref{kappa2R2Lep}) reduce to the ones of 
\cite{FHLM:LFVinSUSYA4} when the contributions related to the flavons $\eta$ and $\xi''$ are singled out.

The same expansion in $u$ can be considered similarly for the K\"ahler potential of the quark sector, $\cK_q$: 
\be
\cK_q=\cK_q^{(0)}+\cK_q^{(1)}+\cK_q^{(2)}+\ldots \, .
\ee
The LO terms in the $u$ expansion are given by:
\be
\begin{split}
\cK_q^{(0)}=&\quad k^Q_0\sum_{i=1}^2{\ov{D}_q}_{i} {D_q}_{i}+\left[k^{q_3}_0+\hat{k}^{q_3}_0\dfrac{|\theta_{FN}|^2}{\La_f}\right]
   \ov{q}_3q_3+\\
&+k^U_0\sum_{i=1}^2{\ov{D}^c_u}_i {D^c_u}_i+\left[k^t_0+\hat{k}^t_0\dfrac{|\theta_{FN}|^2}{\La_f}\right]
   \ov{t}^c t^c+ \\
&+k^D_0\sum_{i=1}^2{\ov{D}^c_d}_i {D^c_d}_i+\left[k^b_0+\hat{k}^b_0\dfrac{|\theta_{FN}|^2}{\La_f}\right]\ov{b}^c b^c\,.
\label{kappa0Quark}
\end{split}
\ee
Notice again that the 
contributions from the superfield $\theta_{FN}$ can be neglected, with the exception of the third families, 
where we consider terms up to second order in $t$. The next terms in the $u$ expansion are given by:
\be
\begin{split}
\cK_q^{(1)}=&\quad\dfrac{k^Q_T}{\Lambda_f}  (\varphi_T (\ov{D}_q D_q)_3)+\dfrac{k^{Q\prime}_T}{\Lambda_f}  
     (\ov{\varphi}_T (\ov{D}_q D_q)_3)+\dfrac{k^Q_\eta}{\Lambda_f}  (\ov{\eta}\,\ov{D}_q)q_3+\\
&+\dfrac{k^U_T}{\Lambda_f}  (\varphi_T (\ov{D}^c_u D^c_u)_3)+\dfrac{k^{U\prime}_T}{\Lambda_f}  
     (\ov{\varphi}_T (\ov{D}^c_u D^c_u)_3)+
\dfrac{k^U_\eta}{\La_f^2}\ov{\theta}_{FN}(\ov{\eta}\,\ov{D}_u^c)t^c+\\
&+\dfrac{k^D_T}{\Lambda_f}  (\varphi_T (\ov{D}^c_d D^c_d)_3)+\dfrac{k^{D\prime}_T}{\Lambda_f}  
     (\ov{\varphi}_T (\ov{D}^c_d D^c_d)_3)+ 
\dfrac{k^D_\eta}{\Lambda_f}  (\ov{\eta}\,\ov{D}^c_d)b^c+\hc
\end{split}
\label{kappa1Quark}
\ee
\be
\begin{split}
\cK_q^{(2)}=&\quad
\sum_{i=1}^{15} \dfrac{k^Q_i}{\Lambda_f^2}  (X^q_i \ov{D}_qD_q) +  
    \sum_{i=1}^{7} \dfrac{k^{q_3}_i}{\Lambda_f^2}  (X^q_i)\, \ov{q}_3q_3+ 
    \sum_{i=1}^{5} \dfrac{k^{Qq_3}_i}{\Lambda_f^2}  (X^q_{i+15} \ov{D}_q)\,q_3+\\
&+\sum_{i=1}^{15}\dfrac{k^U_i}{\La_f^2}(X_i^q\ov{D}_u^cD_u^c)+ 
    \sum_{i=1}^{7}\dfrac{k^t_i}{\La_f^2}(X_i^q)\,\ov{t}^ct^c+ 
    \sum_{i=1}^{5}\dfrac{k^{Ut}_i}{\La_f^3}\ov{\theta}_{FN}(X_{i+15}^q\ov{D}_u^c)\,t^c+\\
&+\sum_{i=1}^{15}\dfrac{k^D_i}{\La_f^2}(X_i^q\ov{D}_d^cD_d^c)+ 
    \sum_{i=1}^{7}\dfrac{k^b_i}{\La_f^2}(X_i^q)\,\ov{b}^cb^c+ 
    \sum_{i=1}^{5}\dfrac{k^{Db}_i}{\La_f^2}(X_{i+15}^q\ov{D}_d^c)\,b^c+\hc
\end{split}
\label{kappa2Quark}
\ee
where $X^q$ is a list of $Z_3$-invariant operators, bilinear in the flavon superfields and their conjugates,
\bea
X^q&=&\Big\{\ov{\xi}\, \xi,\,\varphi_T^2,\,(\ov{\varphi}_T)^2,\,{\ov \varphi}_T\,\varphi_T,\, \ov{\varphi}_S\,
  \varphi_S,\,\ov{\xi}'' \xi'',\,\ov{\eta}\, \eta,\,\ov{\xi}\,\varphi_S,\,\ov{\varphi}_S\,\xi,\varphi_T\,\xi'', \,
  \ov{\varphi}_T\,\xi'', \nn \\
&&\quad \varphi_T\,\ov{\xi}'',\,\ov{\varphi}_T\,\ov{\xi}'',\,\eta^2,\,\ov{\eta}^2,\,\ov{\xi}''\,\eta,\,
  \varphi_T\,\eta,\,\ov{\varphi}_T\,\eta,\,\varphi_T\,\ov{\eta},\,\ov{\varphi}_T\,\ov{\eta} \Big\}\,,
\label{XlistQuark}
\eea
and each $k_i$ represents a list of parameters since there can be different non-equivalent ways of 
combining $X^q_i$ to quark superfields to form a $T'$-invariant. There are also obvious relations among 
the coefficients $k_i$ in order to guarantee the realness of $\cK_{q\,L}^{(2)}$. For the same reason as 
for $X^\ell$, also in this case we do not consider in $X^q$ the possible combinations with the flavon $\xit$.

Finally, the K\"ahler potential containing the Higgs and FN fields is given by:
\be
\cK_r=k_u \vert H_u\vert^2+k_d \vert H_d\vert^2+ k_{FN} \vert  \theta_{FN} \vert^2+\ldots
\ee
where dots stand for additional contributions related to the flavon sector. 

In order to generate a set of SUSY breaking soft mass terms, each coupling constant $k_I$ is treated as a 
superfield with non-vanishing constant $\thh\thhb$ component: 
\be
k_I=p_I+s_I\, \thh\thhb \msusy^2
\label{deck}
\ee
where $p_I$ and $s_I$ are complex numbers with absolute value of order one. 
A more general expression could be considered, including also terms proportional to $\thh$ and to $\thhb$ 
\cite{Luty:TASISusyBreaking}, but these contributions can be absorbed by a suitable reparametrisation.
In the following analysis it is useful to choose a simplified notation for some of the $k$ parameters:
\bald
k^\ell_0=1+s^\ell_0~ \thh\thhb m_0^2\,,\qquad\qquad
(k_0^e)_i=1+(s_0^e)_i~ \thh\thhb m_0^2\,,\\
k_{u,d}=1+s_{u,d}~ \thh\thhb m_0^2\,,\qquad\qquad
k_{FN}=1+s_{FN}~ \thh\thhb m_0^2\,.
\label{kudexp}
\eald 
When the flavons develop vevs, the K\"ahler potential gives rise to non-canonical kinetic terms:
\be
\begin{split}
\LL_{kin}=&\quad 
i~ K^\ell_{ij}\,\ov{\ell}_i\, \ov{\sigma}^\mu\, \DD_\mu\, \ell_j\,+\,i~ K^e_{ij}\,\ov{\ell}^c_i\, \ov{\sigma}^\mu\, \DD_\mu\, \ell^c_j\,+\\ 
&+\, i~ K^q_{ij}\,\ov{Q}_i\, \ov{\sigma}^\mu\, \DD_\mu\, Q_j\,+\,i~ K^u_{ij}\,\ov{U}^c_i\, \ov{\sigma}^\mu\, \DD_\mu\, U^c_j\,+ \,
   i~ K^d_{ij}\,\ov{D}^c_i\, \ov{\sigma}^\mu\, \DD_\mu\, D^c_j\,+\\ 
&+\, K^\ell_{ij}\,\ov{\DD^\mu\, \tilde{\ell}}_i\, \DD_\mu\, \tilde{\ell}_j\,+\,K^e_{ij}\, \ov{\DD^\mu\, \tilde{\ell}^c}_i 
  \,\DD_\mu\, \tilde{\ell}^c_j\,+\\
&+\, K^q_{ij}\,\ov{\DD^\mu\, \tilde{Q}}_i \,\DD_\mu\, \tilde{Q}_j\,+\,K^u_{ij}\, \ov{\DD^\mu \,\tilde{U}^c}_i  
\DD_\mu\, \tilde{U}^c_j\,+ \,K^d_{ij}\, \ov{\DD^\mu\, \tilde{D}^c}_i\,  \DD_\mu \,\tilde{D}^c_j\,,
\end{split}
\label{KineticL}
\ee
with $Q\equiv\{q_1,\,q_2,\,q_3\}$, $U^c\equiv\{u^c,\,c^c,\,t^c\}$, $D^c\equiv\{d^c,\,s^c,\,b^c\}$, and
the matrices $K^i$ given by
\be
K^\ell=
\left(
	\begin{array}{ccc}
		1+2 t^\ell_1~ u& t^\ell_4~ u^2& t^\ell_5~ u^2\\
		\ov{t}^\ell_4~ u^2& 1-(t^\ell_1+t^\ell_2)~ u& t^\ell_6~ u^2\\
		\ov{t}^\ell_5~u^2& \ov{t}^\ell_6~ u^2& 1-(t^\ell_1-t^\ell_2)~ u
	\end{array}
\right)\,,
\label{Tl}
\ee
\be
K^e=
\left(
	\begin{array}{ccc}
		1+t^e_1~ u^2+{t'}^e_1~ t^2& t^e_4~ u^2\, t& t^e_5~ u^2\, t^2\\
		\ov{t^e}_4~ u^2\, t& 1+t^e_2~ u^2+{t'}^e_2~ t^2& t^e_6~ u^2\, t\\
		\ov{t^e}_5~ u^2\, t^{2}& \ov{t^e}_6~ u^2\, t & 1+t^e_3~ u^2+{t'}^e_3~ t^2
	\end{array}
\right)\,,
\label{Te}
\ee
\be
K^q=
\left(
	\begin{array}{ccc}
		t_d^q- t^q_1~ u & t^q_5~ u^2 & t^q_6~ u^2 \\
		\ov{t}^q_5~ u^2 & t_d^q+t^q_1~ u & t^q_4~ u \\
		\ov{t}^q_6~u^2 & \ov{t}^q_4~ u & t_s^q+t^q_2~u^2+ t^q_3~t^2 \\ 
	\end{array}
\right)\,,
\label{Tq}
\ee
\be
K^u=
\left(
	\begin{array}{ccc}
		t_d^u- t^u_1~ u & t^u_5~ u^2 & t^u_6~ u^2\,t \\
		\ov{t}^u_5~ u^2 & t_d^u+t^u_1~ u & t^u_4~ u\,t \\
		\ov{t}^u_6~u^2\,t & \ov{t}^u_4~ u\,t & t_s^u+t^u_2~u^2+ t^u_3~t^2 \\ 
	\end{array}
\right)\,,
\label{Tu}
\ee
\be
K^d=
\left(
	\begin{array}{ccc}
		t_d^d- t^d_1~ u & t^d_5~ u^2 & t^d_6~ u^2 \\
		\ov{t}^d_5~ u^2 & t_d^d+t^d_1~ u & t^d_4~ u \\
		\ov{t}^d_6~u^2 & \ov{t}^d_4~ u & t_s^d+t^d_2~u^2+ t^d_3~t^2 \\ 
	\end{array}
\right)\,.
\label{Td}
\ee
All these coefficients are complex, apart from $t^\ell_{1,2}$, $t^e_{1,2,3}$, ${t'}^e_{1,2,3}$ and 
$t_{d,s,1,2,3}^{q,u,d}$ that are real. Furthermore, all the $t_I$ parameters are linearly related to 
the parameters $p_I$ introduced before in Eq.~(\ref{deck}), but such relations are not particularly 
significant and in the rest of this paper we will treat the $t_I$ coefficients as input parameters. 
Notice that $K^\ell$ and $K^e$ have exactly the same structure of the corresponding matrices in Ref.~\cite{FHLM:LFVinSUSYA4}. The only difference from Ref.~\cite{FHLM:LFVinSUSYA4} is hidden in different relations 
between the parameters $t_I$ and $p_I$ in the matrices $K^\ell$ and $K^e$, due to the presence of the 
flavons $\eta$ and $\xi''$. The matrices for the quark sector are, instead, derived for the first time 
and extend the analysis presented in Ref.~\cite{FHLM:LFVinSUSYA4}.

%
%
\subsection{Superpotential}

As for the K\"ahler potential, also the superpotential can be separated in several parts:
\be
w=w_\ell+w_\nu+w_q+w_d+w_h+...
\ee
representing respectively the charged lepton, the neutrino, the quark, the driving field and the Higgs terms.  
Again, each term can be written as an expansion, up to the second order, in the parameter $u$. 

The part responsible for the charged lepton masses has a LO and NLO terms:
\be
w_\ell=w_\ell^{(1)}+w_\ell^{(2)}+...\,,
\ee
respectively given by
\bea
&&w_\ell^{(1)}=\dfrac{x_e}{\Lambda_f}\, \dfrac{\theta_{FN}^2}{\La_f^2}\,e^c\, H_d\, \left(\varphi_T \ell\right)
+\dfrac{x_\mu}{\Lambda_f}\, \dfrac{\theta_{FN}}{\La_f}\,\mu^c\, H_d\, \left(\varphi_T \ell\right)'
+\dfrac{x_\tau}{\Lambda_f}\, \tau^c\, H_d\, \left(\varphi_T \ell\right)''\,,
\label{wLep1}\\[2mm]
&&\begin{split}
w_\ell^{(2)}=&\quad
\dfrac{1}{\Lambda_f^2}\, \dfrac{\theta_{FN}^2}{\Lambda_f^2}\, e^c\, H_d\, \left[x^T_e\left(\varphi_T^2 \ell\right) + 
  x^\eta_e\left(\eta^2 \ell\right) + x^{\xi''}_e\xi''\left(\varphi_T \ell\right)'\right] + \\
&+\dfrac{1}{\Lambda_f^2}\, \dfrac{\theta_{FN}}{\Lambda_f}\, \mu^c\, H_d\,\left[x^T_\mu\left(\varphi_T^2 \ell\right)' + 
  x^\eta_\mu\left(\eta^2 \ell\right)'+ x^{\xi''}_\mu\xi''\left(\varphi_T \ell\right)''\right] + \\
&+\dfrac{1}{\Lambda_f^2}\, \tau^c\, H_d\,\left[x^T_\tau\left(\varphi_T^2 \ell\right)'' + 
  x^\eta_\tau\left(\eta^2 \ell\right)''+ x^{\xi''}_\tau\xi''\left(\varphi_T \ell\right)\right] \,.\\
\end{split}
\label{wLep2}
\eea
To generate both the Yukawa interactions and the soft mass contributions of the RL type, we promote the 
quantities $x_f$ and $x^{T,\eta,\xi''}_f$ to constant superfields, of the type
\be
x_f=y_f-z_f\, \thh\, m_0\,,\qquad x^{T,\eta,\xi''}_f=y^{T,\eta,\xi''}_f-z^{T,\eta,\xi''}_f\, \thh\, m_0 
\quad(f=e,\mu,\tau)\,,
\label{xdecompLep}
\ee
where the coefficients $y_f$, $z_f$, $y^{T,\eta,\xi''}_f$ and $z^{T,\eta,\xi''}_f$ are complex numbers 
of order one. From Eqs.~(\ref{wLep1}) and (\ref{wLep2}), after the flavour and the electroweak symmetry 
breakings, the following mass matrix for the charged leptons are obtained:
\begin{equation}
M_\ell = \left( \
	\begin{array}{ccc}
		y_e\, t^2\, u+ y'_e\, t^2\, u^2 & c_{T3}\,y_e\, t^2\,u^2 & c_{T2}\, y_e\, t^3\, u^2 \\
		c_{T2}\, y_\mu\, t^2\, u^2 & y_\mu\, t\, u +y'_\mu\, t\, u^2 & c_{T3}\, y_\mu\, t\, u^2 \\
		c_{T3}\, y_\tau\, u^2 & c_{T2}\, y_\tau\, t\, u^2 & y_\tau\, u+y'_\tau\, u^2
	\end{array}
\right) 
\dfrac{v \cos\beta}{\sqrt{2}}\,.
\label{yl_subleading}
\end{equation}
with $y'_f\equiv y^T_f+c^{\prime2}~y^\eta_f+c_{T1}~y_f$. The matrix $M_\ell$ is shown in the basis in which 
the kinetic terms are non-canonical. Notice that $M_\ell$ slightly differs from the corresponding matrix 
derived in Ref.~\cite{FHLM:LFVinSUSYA4} due to the different NLO $\varphi_T$ vev structure between the AF and 
the  $T'$ models.

Similarly, the neutrino superpotential can be expanded in powers of $u$: 
\be
w_\nu=w_\nu^{(1)}+w_\nu^{(2)}+...
\ee
The LO term is given by:
\be
w_\nu^{(1)}=
\dfrac{x_a}{\Lambda_f\Lambda_L} \xi (H_u \ell H_u \ell) +\dfrac{x_b}{\Lambda_f\Lambda_L} 
 (\varphi_S H_u \ell H_u \ell)\,,
\label{wNu}
\ee
with $\Lambda_L$ referring to the lepton number violation scale. Also in this case the constants $x_{a,b}$ are 
promoted to superfields of the form:
\be
x_{a,b}=y_{a,b}+z_{a,b}\, \thh\, m_0\,.
\label{xdecompNu}
\ee
From this expression for $w^{(1)}_\nu$ one can recover the neutrino mass matrix $m_\nu$ of Eq.~(\ref{mnuLO}) 
simply setting $a=y_a\, c_a$ and $b=y_b\, c_b$. The NLO term in the expansion, $w_{\nu} ^{(2)}$, gives rise to 
order $u$ deviations from the TB mixing scheme. Being not relevant for the subsequent discussion its analytical 
form will be left out.

The LO and NLO superpotential terms in the $u$ expansion, responsible for the quark masses,
\be
w_q=w_q^{(1)}+w_q^{(2)}+...
\ee
are respectively given by
\bea
&&\begin{split}
w_q^{(1)}=&\quad 
x_t\left(t^c q_3\right)H_u+\dfrac{x_b}{\Lambda_f}\theta_{FN}\left(b^c q_3\right)H_d+\\[3mm]
&+\dfrac{x_1}{\Lambda_f^2}\theta_{FN}(\phit D_u^c D_q) H_u +\dfrac{x_5}{\La_f^2}\theta_{FN}(\phit D_d^c D_q) H_d+\\[3mm]
&+\dfrac{x_2}{\La_f^2}\theta_{FN}\xipp(D_u^c D_q)' H_u + \dfrac{x_6}{\La_f^2}\theta_{FN}\xipp(D_d^c D_q)' H_d+\\[3mm]
&+\dfrac{1}{\La_f}\left[x_3\, t^c (\eta D_q) + \dfrac{x_4}{\La_f}\theta_{FN}(D_u^c \eta)q_3\right] H_u+\\[3mm]
&+\dfrac{1}{\Lambda_f^2}\theta_{FN}\Big[x_7\, b^c (\eta D_q) + x_8\, (D_d^c \eta) q_3 \Big] H_d\,, 
\end{split}
\label{wQuark1}\\[2mm]
&&\begin{split}
w_q^{(2)}=&\quad
\dfrac{x_9}{\La_f^4}\theta_{FN}\xi^2(\varphi_S D_u^c D_q)H_u+ \dfrac{x_{11}}{\La_f^4}\theta_{FN}\xi^2 
   (\varphi_S D_d^c D_q)H_d+ \\
&+\dfrac{x_{10}}{\La_f^4}\theta_{FN}(\varphi^3_S D_u^c D_q)H_u+ \dfrac{x_{12}}{\La_f^4}\theta_{FN}
   (\varphi^3_S D_d^c D_q)H_d+\ldots
\end{split}
\label{wQuark2}
\eea
Here dots stand terms which do not introduce new flavour structures in the mass matrices and therefore can be 
re-absorbed in the parametrisation, as discussed in Ref.~\cite{FHLM:Tp}. The quantities $x_i$ are taken to be constant 
superfields with
\be
x_f=y_f-z_f\, \thh\, m_0.
\label{xdecompQuark}
\ee

The up and down quark mass matrices obtained from Eqs.~(\ref{wQuark1}) and (\ref{wQuark2}), after the 
breaking of the flavour and the electroweak symmetries, are given by:
\be
\begin{aligned}
&M_u=\left(\begin{array}{ccc}
   i\,c_{T2}\,y_1\,t^2\,u^2+i\,c_b(c_a^2\,y_9+c_b^2\,y_{10})\,t\,u^3 & (\frac{(1-i)}{2}c_{T3}\,y_1+c''\,y_2)\,t\,u^2 &  
       -c_{\eta_2}\,y_4\,t\,u^2 \\
   (\frac{(1-i)}{2}c_{T3}\,y_1-c''\,y_2)\,t\,u^2 & y_1\,t\,u & c'\,y_4\, t\,u \\
       c_{\eta_2}\,y_3\,u^2 & c'\,y_3\,u & y_t \\
            \end{array}\right)\dfrac{v \sin\beta}{\sqrt{2}}\,,\\
&M_d=\left(\begin{array}{ccc}
   i\,c_{T2}\,y_5\,t^2\,u^2+i\,c_b(c_a^2\,y_{11}+c_b^2\,y_{12})\,t\,u^3 & (\frac{(1-i)}{2}c_{T3}\,y_5+c''\,y_6)\,t\,u^2 &  
        -c_{\eta_2}\,y_8\,t\,u^2 \\
   (\frac{(1-i)}{2}c_{T3}\,y_5-c''\,y_6)\,t\,u^2 & y_5\,t\,u & c'\,y_8\,t\,u \\
         -c_{\eta_2}\,y_7\,t\,u^2 & c'\,y_7\,t\,u & y_b\,t \\
            \end{array}\right)\dfrac{v \cos\beta}{\sqrt{2}}\,.
\label{yq_subleading}
\end{aligned}
\ee
where all the $y_i$ are complex number with absolute value of order one, apart for $y_6$ which is taken to be 
slightly larger than its natural value,
\be
y_6\equiv \tilde{y}_6\dfrac{1}{\sqrt{t}}\,,
\label{y6}
\ee 
in order to reproduce the correct value of the Cabibbo angle (see Ref.~\cite{FHLM:Tp} for more details). Notice that 
the matrices $M_{u,d}$ are shown in the basis in which the kinetic terms are non-canonical. 

The term $w_d$ is responsible for the vacuum alignment of the flavon fields and, at leading and renormalisable order, 
is given by
\be
\begin{split}
w_d\,=&\quad 
M \,(\varphi^0_T\,\varphi_T)+g\,(\varphi^0_T\,\varphi_T\,\varphi_T)+g_{7}\,\xi''\,(\varphi^0_T\,\varphi_T)^{\prime} + 
    g_{8}\,(\varphi^0_T\,\eta\,\eta)+\\
&+g_{1}\,(\varphi^0_S\,\varphi_S\,\varphi_S)+g_{2}\,\tilde{\xi}\,(\varphi^0_S\,\varphi_S)+\\
&+g_{3}\,\xi^0\,(\varphi_S\,\varphi_S)+g_{4}\,\xi^0\,\xi^2+g_{5}\,\xi^0\,\xi\,\tilde{\xi}+g_{6}\,\xi^0\,\tilde{\xi}^2+\\
&+M_{\eta}\,(\eta^0\,\eta)+g_9\,(\varphi_T\,\eta^0\,\eta)+\\
&+M_{\xi}\,\xi^{\prime\,0}\,\xi'' +g_{10}\,\xi^{\prime\,0}\,(\varphi_T\,\varphi_T)^{\prime\,\prime}+\ldots
\end{split}
\label{wDriving}
\ee
where dots denote sub-leading non-renormalizable corrections. In Ref.~\cite{FHLM:Tp} it has been shown how this 
superpotential produces the alignment of the flavon vevs. reported in Eq.~(\ref{VEVs}). 

The Froggat-Nielsen 
flavon $\theta_{FN}$ gets its vev through a $D$-term, once we assume that the symmetry $U(1)_{FN}$ is gauged. 
The corresponding scalar potential is:
\be
V_{D, FN}=\dfrac{1}{2}\left(M_{FI}^2- g_{FN}\vert\theta_{FN}\vert^2+...\right)^2
\label{DtermFN}
\ee
where $g_{FN}$ is the gauge coupling constant of $U(1)_{FN}$ and $M_{FI}^2$ denotes the contribution of the 
Fayet-Iliopoulos (FI) term. Dots represent other terms which are not relevant for the calculation of the FN field vev. 
In the SUSY limit one gets
\be
|\langle\theta_{FN}\rangle|^2= \dfrac{M_{FI}^2}{g_{FN}}\,
\label{VEVFN}
\ee
which defines the $T'$ symmetry breaking parameter $t$ via Eq.~(\ref{VEVs}).

Finally the term $w_h$ is associated with the $\mu$ parameter:
\be
w_h=\mu\, H_u\, H_d\,.
\label{muterm}
\ee
This term explicitly breaks the continuous $U(1)_{R}$ symmetry of the model, while preserving the usual $R$-parity. 
Furthermore, the soft SUSY breaking term $B\mu$ can then arise from the $\mu$ term once $\mu$ is promoted to a 
superfield $\mu + \theta^2\, B\mu$. 

%
%
\subsection{Sfermion masses}
\label{sec:SfermionMasses}

With the aid of the K\"ahler potential and of the superpotential written in the previous subsections one can 
derive the sfermion mass matrices for the $T'$ model. The matter Lagrangian reads:
\be
\begin{split}
-\cL_m\supset&\quad
\left(\ov{\tilde{e}}\quad\tilde{e}^c\right)
\cM_e^2
\left(
  \begin{array}{c}
    \tilde{e} \\
    \ov{\tilde{e}}^{c} \\
  \end{array}
\right)+
\ov{\tilde{\nu}}\, m^2_{\nu LL}\,\tilde{\nu}\,+\\
&+\left(\ov{\tilde{u}}\quad\tilde{u}^c\right)\cM_u^2
\left(
  \begin{array}{c}
    \tilde{u} \\
    \ov{\tilde{u}}^{c} \\
  \end{array}
\right)+
\left(\ov{\tilde{d}}\quad\tilde{d}^c\right)\cM_d^2
\left(
  \begin{array}{c}
    \tilde{d} \\
    \ov{\tilde{d}}^{c} \\
  \end{array}
\right)\,,
\end{split}
\ee
where
\be
\cM_f^2=\left(
  \begin{array}{cc}
    m^2_{fLL} & m_{fLR}^2 \\[1mm]
    m_{fRL}^2 & m^2_{fRR} \\
  \end{array}
\right)\,, \qquad f=e,u,d \, .
\ee
with $m^2_{(f,\nu)LL}$ and $m^2_{fRR}$ being $3\times 3$ hermitian matrices, while $m^2_{fLR} = 
\left(m^2_{fRL}\right)^\dag$. Each of these blocks receives contributions from different part of 
the SUSY Lagrangian:
\be
\begin{aligned}
m^2_{(f,\nu)LL}&=(m^2_{(f,\nu) LL})_K+(m^2_{(f,\nu) LL})_F+(m^2_{(f,\nu) LL})_D\,,\\
m^2_{fRR}&=(m^2_{fRR})_K+(m^2_{fRR})_F+(m^2_{fRR})_D+(m^2_{fRR})_{D,FN}\,,\\
m^2_{fRL}&=(m^2_{fRL})_1+(m^2_{fRL})_2 + (m^2_{fRL})_3 \,.
\end{aligned}
\ee
that are going to be described more in details in the following. Notice that in the sneutrino sector only 
the LL block is present and that any contribution to the sneutrino masses associated to $w_\nu$ has been neglected. 

The contributions to the sfermion masses from the SUSY breaking terms in the K\"ahler potential are given by two 
distinct sources: the first one originates from Eqs.~(\ref{kappa0Lep})--(\ref{kappa2Quark}) and is proportional
 to the parameters $s_I$. These contributions for the sleptons are given by
\be
(m^2_{(e,\nu)LL})_K=
\left(
\begin{array}{ccc}
n^\ell_0+2 n^\ell_1\, u& n^\ell_4\, u^2& n^\ell_5\, u^2\\[1mm]
\ov{n}^\ell_4\, u^2& n^\ell_0-(n^\ell_1+n^\ell_2)\, u& n^\ell_6\, u^2\\[1mm]
\ov{n}^\ell_5\,u^2& \ov{n}^\ell_6\, u^2& n^\ell_0-(n^\ell_1-n^\ell_2)\, u
\end{array}
\right) m^2_0\,,
\label{me2LL}
\ee
\be
(m^2_{eRR})_K=
\left(
\begin{array}{ccc}
n^e_1& n^e_4\, u^2\,t& n^e_5\, u^2\, t^2\\
\ov{n}^e_4\, u^2\,t&n^e_2& n^e_6\, u^2\, t\\
\ov{n}^e_5\, u^2\,t^2& \ov{n}^e_6\, u^2\, t &n^e_3
\end{array}
\right) m^2_0\,,
\label{me2RR}
\ee
while for the squarks are
\be
(m^2_{uLL})_K=(m^2_{dLL})_K=
	\left(\begin{array}{ccc}
		n^q_0-n^q_2\,u& n^q_5\, u^2& n^q_6\, u^2 \\[1mm]
		\ov{n}^q_5\, u^2& n^q_0+n^q_2\, u& n^q_4\, u \\[1mm]
		\ov{n}^q_6\,u^2& \ov{n}^q_4\, u & n^q_1+n^q_3\, t^2 \\
	\end{array}\right) m^2_0\,,
\label{mq2LL}
\ee
\be
(m^2_{uRR})_K=
	\left(\begin{array}{ccc}
		n^u_0-n^u_2\,u& n^u_5\, u^2& n^u_6\, u^2 \,t\\[1mm]
		\ov{n}^u_5\, u^2& n^u_0+n^u_2\, u& n^u_4\, u\,t \\[1mm]
		\ov{n}^u_6\,u^2\,t& \ov{n}^u_4\, u\,t& n^u_1+n^u_3\, t^2 \\\end{array}
	\right) m^2_0\,,
\label{mu2RR}
\ee
\be
(m^2_{dRR})_K=
	\left(\begin{array}{ccc}
		n^d_0-n^d_2\,u& n^d_5\, u^2& n^d_6\, u^2 \\[1mm]
		\ov{n}^d_5\, u^2& n^d_0+n^d_2\, u& n^d_4\, u \\[1mm]
		\ov{n}^d_6\,u^2& \ov{n}^d_4\, u & n^d_1+n^d_3\, t^2 \\\end{array}
	\right) m^2_0\,.
\label{md2RR}
\ee
The coefficients $n_i$ are linearly related to the parameters $s_I$ introduced in Eq.~(\ref{deck}). As such 
relations do not give any deeper insight, for the rest of the paper $n_i$ will be considered free input 
parameters, with absolute values of order one. All these coefficients are in general complex, with the obvious 
exception of the ones appearing in the diagonal, that, for hermiticity, are real. In addition, one assumes 
$n^\ell_0$, $n^e_{1,2,3}$ and $n^{q,u,d}_{0,1}$ positive, in order to have positive definite squared masses, 
to avoid electric-charge breaking minima and further sources of electroweak symmetry breaking.

The second contribution to sfermion masses coming from the K\"ahler potential is related to the auxiliary fields of the 
flavon supermultiplets, that acquire non-vanishing vevs, as shown in Eq.~(\ref{VEVsFTerms}), when soft SUSY 
breaking terms are included into the flavon potential. This type of contribution originates from a fourth derivative 
term of the K\"ahler potential:
\be
\left\langle \dfrac{\derp^4 \cK}{\derp f_i \derp \ov{f}_j \derp \Phi_k \derp \ov{\Phi}_l } \right\rangle 
\left\langle \frac{\partial w}{\partial \Phi_k} \right\rangle \overline{\left\langle \frac{\partial w}
 {\partial \Phi_l} \right\rangle} 
     \tilde f_i \ov{\tilde f}_j \,,
\label{FVEVsKahler}
\ee
where $f$ refers to the matter scalar fields and $\Phi$ to the flavons.
For completeness all these contributions are explicitly derived in Appendix \ref{AppB}, where it is also shown 
how these terms can be safely absorbed in a redefinition of the coefficients $n_i$. 

The SUSY contribution from the $F$ terms is completely negligible for sneutrinos, while for charged sfermions 
they read:
\be
\begin{aligned}
(m^2_{eLL})_F&=M_\ell^\dag~\left[(K^e)^{-1}\right]^T~M_\ell\,,\qquad(m^2_{eRR})_F=M_\ell~(K^\ell)^{-1}~M_\ell^\dag\,,\\
(m^2_{uLL})_F&=M_u^\dag~\left[(K^u)^{-1}\right]^T~M_u\,,\qquad(m^2_{uRR})_F=M_u~(K^q)^{-1}~M_u^\dag\,,\\
(m^2_{dLL})_F&=M_d^\dag~\left[(K^d)^{-1}\right]^T~M_d\,,\qquad(m^2_{dRR})_F=M_d~(K^q)^{-1}~M_d^\dag\,,
\end{aligned}
\ee
where $M_f$ is the fermion mass matrix and $K^f$ is the matrix specifying the kinetic terms, explicitly 
written in Eqs.~(\ref{Tl}--\ref{Td}).

The SUSY contribution from the $D$ terms are given by:
\be
\begin{aligned}
(m^2_{eLL})_D&=\left(-\dfrac{1}{2}+\sin^2\theta_W\right) \cos 2\beta ~m_Z^2 K^\ell\,,\\
(m^2_{eRR})_D&=(-1) \sin^2\theta_W \cos 2\beta~m_Z^2 (K^e)^T\,,\\
(m^2_{\nu LL})_D&=\left(+\dfrac{1}{2}\right) \cos 2\beta~m_Z^2 K^\ell\,,\\
(m^2_{uLL})_D&=\left(+\dfrac{1}{2}-\dfrac{2}{3}\sin^2\theta_W\right) \cos 2\beta ~m_Z^2 K^q\,,\\
(m^2_{uRR})_D&=\left(+\dfrac{2}{3}\right)\sin^2\theta_W \cos 2\beta~m_Z^2 (K^u)^T\,,\\
(m^2_{d LL})_D&=\left(-\dfrac{1}{2}+\dfrac{1}{3}\sin^2\theta_W\right) \cos 2\beta~m_Z^2 K^q\,,\\
(m^2_{dRR})_D&=\left(-\dfrac{1}{3}\right) \sin^2\theta_W \cos 2\beta~m_Z^2 (K^d)^T\,,
\end{aligned}
\ee
where $m_Z$ is the $Z$ mass and $\theta_W$ is the Weinberg angle. For all the right-handed fields but $t^c$ 
an additional $D$ term contribution, coming from the gauged $U(1)_{FN}$ sector, is present.
One can check that this contribution can be simply reabsorbed in the redefinition of the $n_i^{e,u,d}$ coefficients   
parametrizing $(m_{f RR}^2)_K$. 

Concerning the RL mass blocks, they are the results of three distinct terms: the first one originates from 
the superpotential, Eqs.~(\ref{wLep1}, \ref{wLep2}, \ref{wQuark1}, \ref{wQuark2}), and is proportional to 
the parameters $z_f$ and $z'_f$ of the decomposition in Eqs.~(\ref{xdecompLep}, \ref{xdecompQuark}):
\be
(m^2_{(e,d)RL})_1 = A^{(e,d)}_1 \dfrac{v \cos\beta}{\sqrt{2}} A_0 \,,\qquad\qquad 
(m^2_{uRL})_1 = A^u_1 \dfrac{v \sin\beta}{\sqrt{2}} A_0\,,
\label{m2RL1}
\ee
with $A_0$ a common (SUSY) trilinear mass term and where for the sleptons we have defined 
\be
A^e_1 = \left( \
	\begin{array}{ccc}
		z_e\, t^2\, u+ z'_e\, t^2\, u^2 & c_{T3}\,z_e\, t^2\, u^2 & c_{T2}\,z_e\, t^3\, u^2 \\
		c_{T2}\, z_\mu\, t^2\, u^2 & z_\mu\, t\, u +z'_\mu\, t\, u^2 & c_{T3}\,z_\mu\, t\, u^2 \\
		c_{T3}\, z_\tau\, u^2 & c_{T2}\, z_\tau\, t\, u^2 & z_\tau\, u+z'_\tau\, u^2
	\end{array}\right)\,,
\label{Ae1}
\ee
with $z'_i=z_f^T+c^{\prime2}\,z_i^\eta+c_{T1}\,z_i$ for $i=e,\mu,\tau$, while for the squarks we have 
\be
A^u_1=\left(\begin{array}{ccc}
       i\,c_{T2}\,z_1\,t^2\,u^2+i\,z_9\,t\,u^3 & \left(\frac{(1-i)}{2}c_{T3}\,z_1+c''\,z_2\right)\,t\,u^2 &  - 
                       c_{\eta_2}\,z_4\,t\,u^2 \\
                    \left(\frac{(1-i)}{2}c_{T3}\,z_1-c''\,z_2\right)\,t\,u^2 & z_1\,t\,u & c'\,z_4\, t\,u \\
                    -c_{\eta_2}\,z_3\,u^2 & c'\,z_3\,u & z_t \\
            \end{array}\right)\,,
\label{Au1}
\ee
\be
A^d_1=\left(\begin{array}{ccc}
        i\,c_{T2}\,z_5\,t^2\,u^2+i\,z_{11}\,t\,u^3 & \left(\frac{(1-i)}{2}c_{T3}\,z_5+c''\,z_6\right)\,t\,u^2 &  - 
                         c_{\eta_2}\,z_8\,t\,u^2 \\
                    \left(\frac{(1-i)}{2}c_{T3}\,z_5-c''\,z_6\right)\,t\,u^2 & z_5\,t\,u & c'\,z_8\,t\,u \\
                    -c_{\eta_2}\,z_7\,t\,u^2 & c'\,z_7\,t\,u & z_b\,t \\
            \end{array}\right)\,,
\label{Ad1}
\ee
with $z_9=c_b(c_a^2\,z_9+c_b^2\,z_{10})$ and $z_{11}=c_b(c_a^2\,z_{11}+c_b^2\,z_{12})$. Notice that the 
matrix $A^e_1$ in Eq.~(\ref{Ae1}) differs from the corresponding matrix in Ref.~\cite{FHLM:LFVinSUSYA4}, the reason 
being the, already mentioned, different choice of the flavon $\varphi_T$ vev.

The second term in the LR mass matrices has the same source of Eq.~(\ref{FVEVsKahler}) and it is related to 
the non-vanishing vev of the auxiliary fields of the flavon supermultiplets. This contribution can be written as
\be
\left\langle \dfrac{\derp^3 w}{\derp \Phi \derp f^c_i \derp f_j} \right\rangle 
\overline{\left\langle \frac{\partial w}{\partial \Phi} \right\rangle} 
\tilde f^c_i \tilde f_j  + \hc
\ee
so that the soft mass matrices $(m^2_{fRL})_2$ read as
\be
(m^2_{(e,d)RL})_2 = A^{(e,d)}_2 \dfrac{v \cos\beta}{\sqrt{2}} A_0\,,\qquad\qquad
(m^2_{uRL})_2 = A^u_2 \dfrac{v \sin\beta}{\sqrt{2}} A_0\,.
\label{m2RL2}
\ee
The matrices $A^{(e,u,d)}_2$ have exactly the same structure of $A^{(e,u,d)}_1$ in Eqs.~(\ref{Ae1})--(\ref{Ad1}) 
in terms of $u$ and $t$, apart from the $(3,3)$ entries of $A^{(u,d)}_2$ which are vanishing. Furthermore only 
the coefficients of order one are different. As it will be discussed in the next section, these two contributions 
cannot be absorbed into the same matrix, by a simple redefinition of the parameters. When moving to the physical 
basis, in the $A^{(e,u,d)}_1$ matrices several cancellations among the parameters happen, while this is not the case 
for the $A^{(e,u,d)}_2$ matrices. Consequently, these two contributions are kept distinct and the explicit form 
of $A^{(e,u,d)}_2$ follows from that of $A^{(e,u,d)}_1$ in Eqs.~(\ref{Ae1})--(\ref{Ad1}), simply substituting 
$\{z_f,\,z'_f,\,z_i,\,c_i\}$ with $\{\ov{c}^F\,y_f,\,y^F_f,\,y_i,\,\ov{c}^F_i\}$, where $f=e,\,\mu,\,\tau$ and 
$i=1,\ldots,11$. The new parameters in this list are 
\bald
&y^F_f=2 \ov{c}^F\,y^T_f + 2\ov{c}^{F\prime}\,y_f^\eta + \ov{c}^F_{T1}\, y_f\,,\\
&y^F_9=2\,c_a(\ov{c}^F_ac_b+c_a\ov{c}^F_b)y_{9}+3\,\ov{c}^F_bc_b^2\,y_{10}\,,\\
&y^F_{11}=2\,c_a(\ov{c}^F_ac_b+c_a\ov{c}^F_b)y_{11}+3\,\ov{c}^F_bc_b^2\,y_{12}\,.
\eald

Finally, the last contribution to the LR mass matrices is proportional to the fermion masses and explicitly 
depends on $\tan\beta$ through:
\be
(m^2_{(e,d)RL})_3=-\ov\mu\, \tan\beta\, M_{(\ell,d)}\,,\qquad\qquad
(m^2_{uRL})_3=-\ov\mu\, \dfrac{1}{\tan\beta}\, M_u\,.
\label{m2RL3}
\ee
The origin of this contribution is quite similar to the previous one, since it also arises from the auxiliary 
component of a superfield: the terms in Eq.~(\ref{m2RL2}) originate from the auxiliary component of the flavon superfields, 
while the terms in Eq.~(\ref{m2RL3}) from the auxiliary component associated to the Higgs doublets $H_d$ and $H_u$.

%
%
\section{The physical basis}
\label{sec:PhysicalBasis}

All the matrices of the previous section are written in a basis in which the kinetic terms of (s)fermion 
are non-canonical. To have a better phenomenological insight, it is useful to move to a {\em physical} basis 
defined as a basis in which the kinetic terms are in a canonical form and the charged lepton and down quark 
matrices are diagonal. All matrices in the {\em physical} basis are going to be denoted with a hat. 
Most of the details of this basis change are deferred to the Appendices \ref{AppB} and \ref{AppC}.

In the preferred allowed $T'$--breaking parameters space one has that $u\approx t\approx \lambda^2$.
In terms of powers of the flavour symmetry breaking parameters $u$ and $t$, the charged fermion mass matrices 
in the {\em physical} basis read
\be
\hat{M}_\ell \sim \diag(t^2\,u,\,t\,u,\,u)\,,\qquad
\hat{M}_d \sim \diag(u^3,\,t\,u,\,t)\,,\qquad
\hat{M}_u \sim \left(
        \begin{array}{ccc}
             \sqrt{t}\,u^3 & t\,u^2 & t\,u^2 \\
             \sqrt{t}\,u^2 & t\,u & t\,u \\
             \sqrt{t}\,u & u & 1 \\
        \end{array}
\right)\,.
\ee
As exemplification we can show here how it is possible to recover the correct $V_{CKM}$ structure and to obtain 
the experimental 
values just selecting appropriate order one value for the model unknown coefficients entering in the mixing matrix. 
The derivation follows straightforwardly from Ref.~\cite{FHLM:Tp}, inserting explicitly the substitution in Eq.~(\ref{y6}) 
and the vev alignment for $\varphi_T$ in Eq.~(\ref{VEVs}). From Eqs.~(51) of Ref.~\cite{FHLM:Tp} one obtains the following 
identification between the CKM parameters in the Wolfenstein parametrisation and the $T'$ model coefficients:
\be
\lambda = -\frac{\tilde{y}_6}{y_5} c'' \frac{u}{\sqrt{t}}\,, \qquad\qquad 
       A = \left(\frac{y_7}{y_b}-\frac{y_3}{y_t} \right) \frac{\tilde{y}_6^2}{y_5^2} \frac{c'}{{c''}^2} \frac{t}{u}\,.
\ee
These expressions automatically lead to a consistent identification of the leading 13 and 31 CKM elements,
\be
V_{td} \sim V_{ub} \approx A \lambda^3 = 
\left(\frac{y_7}{y_b}-\frac{y_3}{y_t} \right) \frac{\tilde{y}_6}{y_5} c' c'' \frac{u^2}{\sqrt{t}}\,,
\ee
while leave enough freedom for recovering the experimental (order one) values for the last two CKM parameters $\rho$ 
and $\eta$. The corresponding expressions in terms of the $T'$ order one coefficients can be easily derived but are 
not reported here not particularly suggestive.

Moving to the physical basis and performing the corresponding transformations to the sfermion sector give quite involved 
results. Here only the naive structure of the sfermion mass matrices in terms of the $T'$ breaking parameters $u$ and $t$ 
are showed, thus omitting all the order one unknown coefficients of the model. A complete description can be found in 
Appendix \ref{AppB}. Starting with the LL block, the K\"ahler potential contributions are given by
\be
(\hat{m}_{(e,\nu)LL}^2)_K \sim
\left( \begin{array}{ccc}
                1+u  & u^2 & u^2 \\
                u^2 & 1+u & u^2 \\
                u^2 & u^2 & 1+u \\
\end{array}\right) \, m_0^2
\label{meLL_hat}
\ee
for the sleptons and 
\be
(\hat{m}_{(u,d)LL}^2)_K \sim
\left( \begin{array}{ccc}
                1+u & u\,\sqrt{t} & u\,\sqrt{t}\\[3mm]
                u\,\sqrt{t} & 1+u & u \\[3mm]
                u\,\sqrt{t} & u & 1 \\
\end{array}\right) \, m_0^2
\label{mqLL_hat}
\ee
for the squarks. 
 
The supersymmetric $F$ and $D$ term contributions are respectively given by
\be
\begin{aligned}
(\hat{m}^2_{eLL})_F&=\hat{M}_\ell^T\, \hat{M}_\ell\,, \qquad& 
(\hat{m}_{\nu LL}^2)_F&=0\,,\\
(\hat{m}^2_{uLL})_F&=\hat{M}_u^T\, \hat{M}_u\,, \qquad& 
(\hat{m}^2_{dLL})_F&=\hat{M}_d^T\,\hat{M}_d\,, 
\end{aligned}
\ee
and by
\bea
(m^2_{eLL})_D &=& \left(-\dfrac{1}{2}+\sin^2\theta_W\right) \cos 2\beta ~m_Z^2\times\unity \,,\\
(m^2_{\nu LL})_D &=& \left(+\dfrac{1}{2}\right) \cos 2\beta~m_Z^2\times \unity\,,\\
(m^2_{uLL})_D &=& \left(+\dfrac{1}{2}-\dfrac{2}{3}\sin^2\theta_W\right) \cos 2\beta ~m_Z^2\times \unity\,,\\
(m^2_{d LL})_D &=& \left(-\dfrac{1}{2}+\dfrac{1}{3}\sin^2\theta_W\right) \cos 2\beta~m_Z^2\times \unity\,.
\eea
Both, $F$ and $D$ term contributions are suppressed by a factor of order $\hat{M}_f^T \hat{M}_f/m_0^2$ or 
order $m_Z^2/m_0^2$, respectively, compared to those coming from the K\"ahler potential and so numerically 
negligible for typical values of $m_0$ around 1 TeV. 

The K\"ahler potential contributions to the RR block of the slepton, up- and down-squark matrices are respectively
\bea
&
(\hat{m}_{eRR}^2)_K \sim \left( \begin{array}{ccc}
                1 & \dfrac{m_e}{m_\mu} u & \dfrac{m_e}{m_\tau} u\\[3mm]
                \dfrac{m_e}{m_\mu} u & 1 & \dfrac{m_\mu}{m_\tau} u\\[3mm]
                \dfrac{m_e}{m_\tau} u & \dd \frac{m_\mu}{m_\tau} u & 1\\
	\end{array}\right) \, m_0^2\,,
\label{meRR_hat}\\[2mm]
&
(\hat{m}_{uRR}^2)_K \sim \left(
        \begin{array}{ccc}
            1+u & u^2 & t\,u^2 \\
           u^2  & 1+u & t\,u \\
           t\,u^2 & t\,u & 1 \\
        \end{array}
\right)\,\msusy^2\,,
\label{muRR_hat}\\[2mm]
&
(\hat{m}_{dRR}^2)_K \sim \left(
        \begin{array}{ccc}
            1+u & u\,\sqrt{t} & u\,\sqrt{t}\\[3mm]
            u\,\sqrt{t} & 1+u & u \\[3mm]
            u\,\sqrt{t}& u & 1 \\
        \end{array}
\right)\,\msusy^2\,.
\label{mdRR_hat}
\eea

The supersymmetric F and D contributions read:
\be
\begin{aligned}
(\hat{m}^2_{eRR})_F&=\hat{M}_\ell\, \hat{M}_\ell^T\,,\\
(\hat{m}^2_{uRR})_F&=\hat{M}_u\, \hat{M}_u^T\,, \qquad& (\hat{m}^2_{dRR})_F&=\hat{M}_d\,\hat{M}_d^T\,, 
\end{aligned}
\ee
and
\be
\begin{aligned}
(m^2_{eRR})_D&=(-1) \sin^2\theta_W \cos 2\beta~m_Z^2\times\unity\,,\\
(m^2_{uRR})_D&=\left(+\dfrac{2}{3}\right)\sin^2\theta_W \cos 2\beta~m_Z^2\times\unity\,,\\
(m^2_{dRR})_D&=\left(-\dfrac{1}{3}\right) \sin^2\theta_W \cos 2\beta~m_Z^2\times\unity\,.
\end{aligned}
\ee
Also in this case the SUSY contributions are numerically negligible in most of the parameter space. 

Finally, the contributions to the RL block for charged sfermions read
\bea
&
(\hat{m}_{eRL}^2)_1 \sim \left( \begin{array}{ccc}
                m_e & m_e\,u & m_e\,u\\[3mm]
                m_\mu\,t\,u^2 & m_\mu & m_\mu\,u\\[3mm]
                m_\tau\,u^2 & m_\tau\,t\,u^2 & m_\tau\\
	\end{array}\right) \, A_0\,.
\label{meRL1_hat}\\[2mm]
&
(\hat{m}_{eRL}^2)_{2} \sim \left( \begin{array}{ccc}
               m_e & m_e\, u & m_e\, u\\[1mm]
               m_\mu\, u & m_\mu  & m_\mu\,u\\[1mm]
               m_\tau\,u & m_\tau\,u &m_\tau \\
\end{array}
\right) \, A_0 \,,
\label{meRL2_hat}\\[2mm]
&
(\hat{m}^2_{eRL})_3 = -\mu^\dag\,\tan\beta\,\hat{M}_\ell\,.
\label{meRL3_hat}
\eea
An important feature of $(\hat{m}_{eRL}^2)_1$ is that the elements below the diagonal are suppressed by a factor 
$\sim u$ compared to the corresponding elements of the matrix in the non--canonical basis. However, this fact does 
not happen for $(\hat{m}_{eRL}^2)_2$. As a result the elements of $(\hat{m}_{eRL}^2)_2$ dominate with respect to 
those of $(\hat{m}_{eRL}^2)_1$. 

The RL matrix for the up squarks sector is given by the three following contributions:
\bea
&
(\hat{m}^2_{uRL})_1 \sim \left( \begin{array}{ccc}
                \sqrt{t}\,u^3 & t\,u^2 & t\,u\\[3mm]
                \sqrt{t}\,u^2 & t\,u & u\\[3mm]
                u^2/\sqrt{t} & u & 1\\
	\end{array}\right) \dfrac{v \sin\beta}{\sqrt{2}} A_0\,,
\label{muRL1_hat}\\[2mm]
&
(\hat{m}^2_{uRL})_2 \sim \left( \begin{array}{ccc}
                t\,u^2 & t\,u^2 & t\,u^2\\[3mm]
                \sqrt{t}\,u^2 & t\,u & t\,u\\[3mm]
                u^2/\sqrt{t} & u & u^2\\
	\end{array}\right) \dfrac{v \sin\beta}{\sqrt{2}} A_0\,,
\label{muRL2_hat}\\[2mm]
&
(\hat{m}^2_{uRL})_3 =-\mu^\dag\, \dfrac{1}{\tan\beta}\,\hat{M}_u\,.
\label{muRL3_hat}
\eea
Comparing these results with the un--hatted matrices of Eqs.~(\ref{m2RL1}) and (\ref{m2RL2}), one can observe 
that the first column of $(\hat{m}^2_{uRL})_{1}$ is enhanced by a factor $t^{-1/2}$, while the entry $(13)$ 
of $(\hat{m}^2_{uRL})_1$ is enhanced by a factor $u^{-1}$. Also the first column of $(\hat{m}^2_{uRL})_2$ is 
increased by a factor $t^{-1/2}$ with respect to $(m^2_{uRL})_2$, except for the (11) entry which is, instead, 
enhanced by a factor $t^{-1}$. Moreover the entry $(33)$ of $(\hat{m}^2_{uRL})_2$ is not vanishing as in 
$(m^2_{uRL})_2$, but it is proportional to $u^2$.

Finally the down squarks LR contributions are the following:
\be
(\hat{m}^2_{dRL})_1 \sim \left( \begin{array}{ccc}
                u^3 & \sqrt{t}\,u^2 & \sqrt{t}\,u^2\\[3mm]
                \sqrt{t}\,u^2 & t\,u & t\,u\\[3mm]
                \sqrt{t}\,u^2 & t\,u & t\\
	\end{array}\right) \dfrac{v \cos\beta}{\sqrt{2}} A_0\,,
\label{mdRL1_hat}
\ee
\be
(\hat{m}^2_{dRL})_2 \sim \left( \begin{array}{ccc}
                u^3 & \sqrt{t}\,u^2 & \sqrt{t}\,u^2\\[3mm]
                \sqrt{t}\,u^2 & t\,u & t\,u\\[3mm]
                \sqrt{t}\,u^2 & t\,u & t\,u^2\\
	\end{array}\right) \dfrac{v \cos\beta}{\sqrt{2}} A_0\,,
\label{mdRL2_hat}
\ee
\be
(\hat{m}^2_{dRL})_3 =-\mu^\dag\,\tan\beta\,\hat{M}_d\,.
\label{mdRL3_hat}
\ee
The first column of $(\hat{m}^2_{dRL})_{1,2}$ shows the same enhancement, compared to the un--hatted 
matrices of Eqs.~(\ref{m2RL1}) and (\ref{m2RL2}), as for the up-type squarks, while the entry $(33)$ of 
$(\hat{m}^2_{dRL})_2$ is not vanishing as in $(m^2_{dRL})_2$, but scales as $t\,u^2$.   

%
%

%
%
\mathversion{bold}
\section{Phenomenology of the $T'$ FCNC sector }
\label{sec:Phen}
\mathversion{normal}

While several flavour models can account for the correct SM mass and mixing patterns, a discrimination between them 
can only be obtained through a detailed analysis of the associated phenomenology. This section will be dedicated to study 
the predictions of the $T'$ model to the most relevant, leptonic and hadronic, FCNC observables. The results for the 
FCNC leptonic sector are going to be a straightforward replica of those obtained in Ref.~\cite{FHLM:LFVinSUSYA4} for the 
$A_4$ model, as the two realisations (almost) coincide when restricted to leptons: only slight differences are expected 
from the extra flavons $\eta$ and $\xi''$ (see Eq.~(\ref{VEVs})) that are included when the embedding of $A_4$ in $T'$ 
is considered. No corresponding analysis, instead, has ever been performed on the $T'$ predictions for the FCNC hadronic 
sector. Therefore the main goal of this section will be twofold: from one side we will explore the $T'$ model constraints 
coming form the hadronic sector; from the other side, we will combine such constraints with those arising from the 
leptonic sector. This combined analysis will help in better identifying which cross--correlations can be most useful 
in discriminating between several (discrete) flavour models.

As discussed in the previous sections both the fermion and sfermion flavour structures are obtained in our model 
in terms of the two $T'$ symmetry breaking parameters $u$ and $t$. While $t$ governs essentially the charged fermion 
mass hierarchy, the amount of flavour changing is proportional to the $u$ parameter. The allowed range in which 
$u$ and $t$ can be varied is constrained by the experimental data on lepton masses and mixings. As stated in 
Sect.~\ref{GenFeatures} one gets $t \simeq 0.05$ and $0.007 \lesssim u \lesssim 0.05$. The allowed values of 
$\tan \beta$ are fixed by the requirement that the $\tau$ mass is consistent with experimental data (see 
Eq.~(\ref{tanb&u&yt})) giving a range $2 \lesssim \tan\beta \lesssim 15$ for $u = 0.05$ while for $u = 0.007$ 
only $\tan\beta \simeq 2$ is permitted. 

Although it is not possible to set consistently $u=0$, as viable fermion masses and mixings cannot be obtained, an 
interesting limit is found when $u=0$ is selected only in the sfermion sector. Indeed, at the high energy scale 
$\Lambda_f \approx M_{GUT}$, the SUSY scalar sector very much resembles the MSUGRA framework, with the scalar 
mass matrices function of the common, flavour universal, soft bilinear and trilinear parameters $(m_0, A_0)$.
This fact is inherited from the underlying assumption that all the SUSY breaking of the model is provided by a unique 
(hidden) sector. However, our construction deviates from plain MSUGRA due to the different $T'$ embedding 
of the  matter fields that permits, also in the $u=0$ limit, non--universal (order one) diagonal soft terms in the 
sfermion mass matrices, as it appears evident, for example, from Eqs.~(\ref{kappa0Lep}) and (\ref{kappa0Quark}).
The flavour symmetry $G_f$ does not predict the parameters involved either in the gaugino or in the Higgs(ino) sectors. 
Therefore, inspired by MSUGRA, a common gaugino mass $M_{1/2}$ and a common scalar mass $m_0$ are assumed at the high 
energy scale $\Lambda_f$. The $\mu$ parameter is fixed by imposing the requirement of a correct EW symmetry breaking: it 
can be expressed as a (one-loop) function of the other MSUGRA parameters, with the only freedom of the $\sign[\mu]$. 

Due to all these constraints, one expects our $T'$ model to (almost) approximate MSUGRA, in the small $u$ region, while 
significant deviations can be, instead, expected only in the large $u$ (and large $\tan\beta$) region. We will concentrate 
on the two most compelling cases: 
\begin{enumerate}[A)]
\item the (almost) universal ``limit'' $u=0.01$, with $\tan\beta=5$. In this scenario one does not expect any sizable 
difference between the $T'$ and the MSUGRA FCNC phenomenology\footnote{The lowest allowed value for $(u,\,\tan\beta)$ 
is $(0.007,\,2)$, for which it turns out to be statistically harder to found a physically allowed SUSY spectrum, 
due to the smallness of $\tan\beta$. No significant phenomenological difference can be seen with respect to the 
considered case.}. In the following we will refer to this case as {\em Reference Point A} ($RP_A$);
\item the non-universal ``limit'' $u=0.05$ with $\tan\beta=15$. In this scenario one has the strongest allowed 
deviation from the universal soft breaking term case. The $T'$ model might be experimentally 
distinguished from the MSUGRA assumption. We will refer to this case as {\em Reference Point B} ($RP_B$); 
\end{enumerate}
Regarding the scalar bilinear and trilinear parameters we will show the results for two different choices:
low ($A_0=2\,m_0=400$ GeV) and high ($A_0=2\,m_0=2000$ GeV) soft breaking scale, with the gaugino mass term 
spanning in the $100 < M_{1/2} < 1000$ GeV range.

%
%
\mathversion{bold}
\subsection{SUSY mass spectrum: MSUGRA vs. $T'$}
\mathversion{normal}

The SUSY spectrum at low energies has been obtained by means of the code {\sphenod} 
\cite{SPheno}. This program implements SUSY one-loop (and partially two loops) Renormalisation Group Equations 
(RGE) from the high energy scale where the model parameters are defined, down to the electroweak scale, where 
observables are measured. This procedure is performed in an iterative way until a stable (under RGE) SUSY 
low energy spectrum is obtained. Even if the default algorithm deals with universal boundary conditions, typical of the MSUGRA setting, it is already built in the possibility to account for non-diagonal Yukawas and soft terms, as in the MSSM. 

Using this general framework we have implemented in {\tt SPheno} our $T'$ model in a consistent way, i.e. by requiring that the iteration procedure does not spoil the agreement with the physical values of fermionic masses and mixing, also predictions of the model. In practice, the structures of the high energy fermion and sfermion mass matrices introduced in \spheno are the 
ones given in Sect.~\ref{sec:PhysicalBasis} in the {\em physical} basis. At the end of the iteration, the stable 
set of SUSY masses are used for calculating the corrections to the fermionic masses and mixings at the EW scale 
(see Ref.~\cite{SPheno}). Consistency at $1\sigma$ level with the known SM flavour (low energy) parameters 
\cite{PDG2010} is obtained by accordingly fixing the (order one) values of few of the $T'$ free parameters. 
All the other $T'$ coefficients are, instead, considered as free order-one parameters, with an absolute value 
chosen randomly in the $(1/2,\,2)$ range. 

This procedure is simultaneously ``natural'' and ``stable''. Natural 
because only order one values for all the $T'$ parameters (let fixed or free) are chosen. Stable because once the 
fixed $T'$ parameter are set, the effect on the SM parameters of choosing random (order one) free $T'$ parameters is 
negligible.
Notice, moreover, that fixing some of the $T'$ parameters does not introduce any fine-tuning in the model, 
as it has been exemplified in the previous section for the $V_{CKM}$ case.
This procedure in principle can restrict the SUSY spectrum obtained at low energy, as a consequence of the combined 
RGE evolution. However, this effect is (partially) compensated by the presence of the many free
(order one) parameters in the soft SUSY sector.

\begin{figure}
\centering
\vspace{-0.5cm}
\subfigure[\it MSUGRA (left) vs. $T'$ (right) spectra for $u=0.01$ and $\tan\beta=5$. In the upper (lower) plots 
$A_0 = 2 \, m_{0}=400$ GeV ($A_0 = 2 \, m_{0}=2000$ GeV). \label{Fig:scaRPA}]
{\includegraphics[width=15cm,height=10cm]{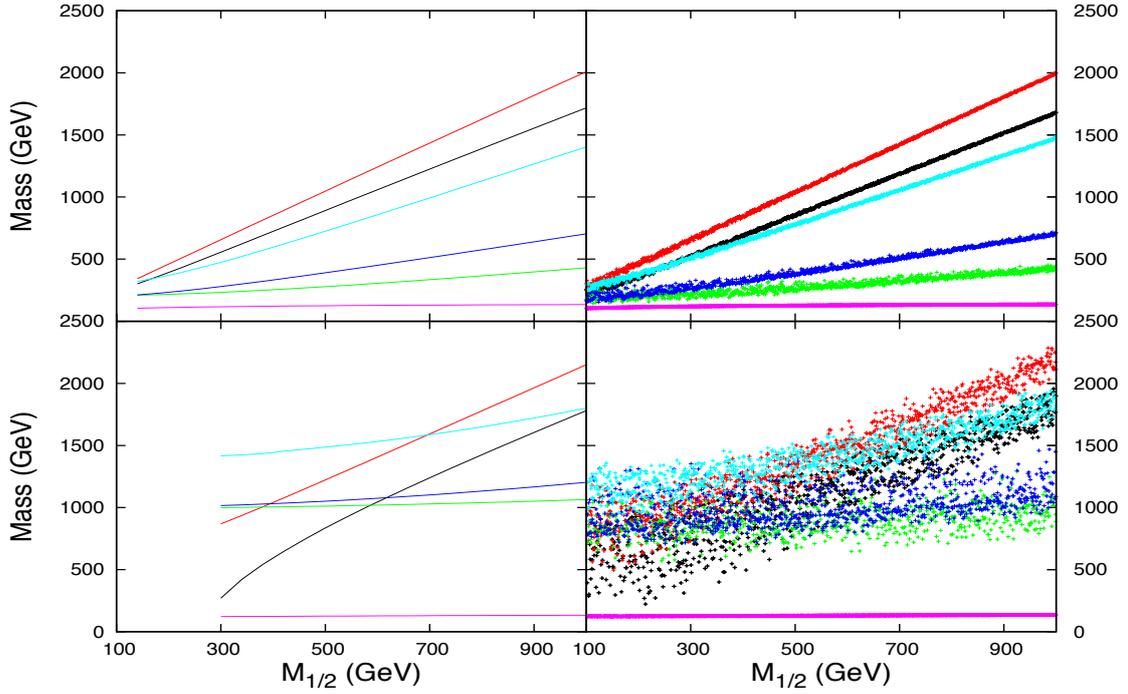}} \\ 
\vspace{-0.5cm}
\subfigure[\it MSUGRA (left) vs. $T'$ (right) spectra for $u=0.05$ and $\tan\beta=15$. In the upper (lower) plots 
$A_0 = 2 \, m_{0}=400$ GeV ($A_0 = 2 \, m_{0}=2000$ GeV). \label{Fig:scaRPB}]
{\includegraphics[width=15cm,height=10cm]{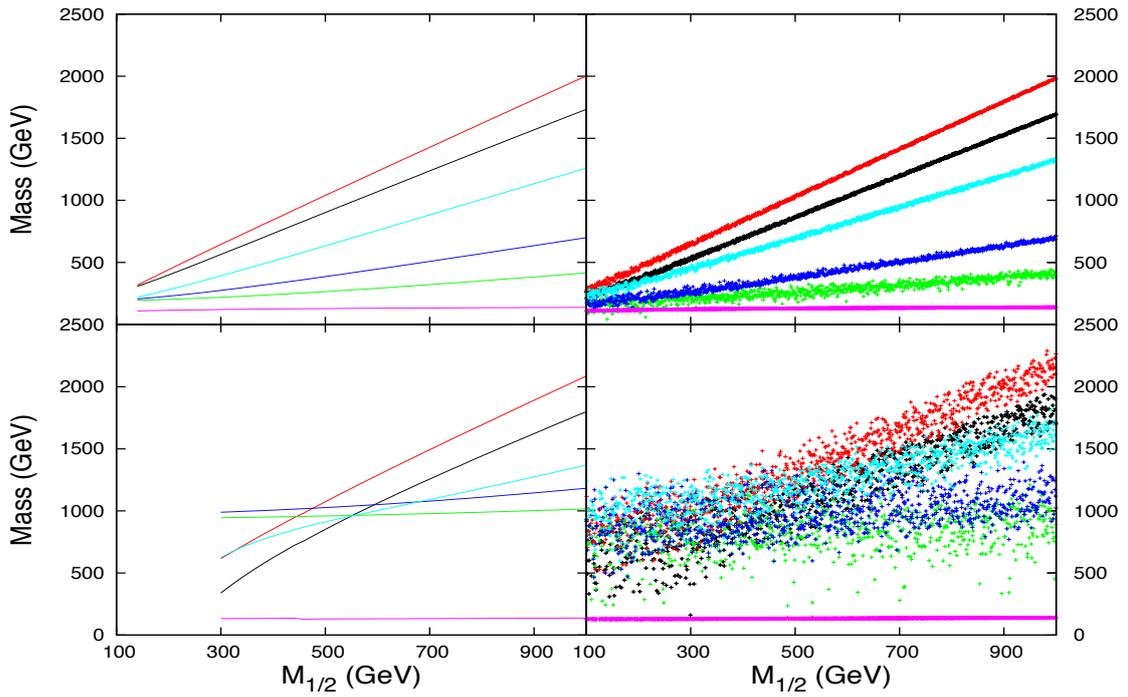}}
\caption{\it SUSY scalar spectra. The colors in the plots refer respectively to: Magenta for $h^0$, Green for $\tilde \ell$, 
Blue for $\tilde \nu$, Cyan for $H^\pm$, Black for $\tilde t$, and Red for $\tilde b$.} \label{Fig:scatot}
\end{figure}

In Fig.~\ref{Fig:scatot} we compare the MSUGRA scalar spectrum (left plots) with the $T'$ model one (right plots), 
as function of common gaugino mass $M_{1/2}$. The results in Fig.~\ref{Fig:scaRPA} refer to the $RP_A$ case 
(i.e $u=0.01$ and $\tan\beta=5$) while the plots in Fig.~\ref{Fig:scaRPB} cover the $RP_B$ case (i.e $u=0.05$ and 
$\tan\beta=15$). In each figure, the two upper plots are shown for a common bilinear and trilinear scalar SUSY scale 
$A_0=2\,m_0=400$ GeV, while for the two lower plots $A_0=2\,m_0=2000$ GeV has been chosen. For definiteness only the 
$\sign[\mu]>0$ results have been shown, as no significative dependence on it is expected, nor observed. 

The SM-like Higgs particle (in Magenta) is almost insensitive on the gaugino mass parameter. Only points of the 
parameter space surviving the LEP2 lower bound are considered. The SUSY lightest scalar is typically the $\tilde \tau_1$ 
(in Green) except for large $m_0$ and small $M_{1/2}$ where the (right-handed) $\tilde{t}_1$ is lighter (in Black). 
In Blue is shown the lightest sneutrino, $\tilde{\nu}_1$, while in Red the lightest sbottom, $\tilde{b}_1$. All the other 
sfermions are not shown in the plot. However all the (almost) left-handed sleptons are practically degenerate with the 
$\tilde{\nu}_1$, while the (almost) right-handed are degenerate with the $\tilde \tau_1$. The first two family squarks 
are heavier than $\tilde{b}_1$ and $\tilde{t}_1$ and nearly degenerate. Finally, the charged Higgs (in Cyan) can be 
the heaviest particle of the spectrum in the large $m_0$ and small $M_{1/2}$ region.

As it can be noticed, comparing the left with the right plot of Fig.~\ref{Fig:scatot}, the SUSY scalar spectra for 
MSUGRA and $T'$ are very similar for small $m_0$ value (upper plots) while two evident differences appear when going 
to heavier scalar masses (lower plots). The first difference can be noticed in the low $M_{1/2}$ region where the $T'$ 
spectrum always extends to lower $M_{1/2}$ compared with the MSUGRA one. This happens because of the ``phenomenological''
requirements that have been imposed on Higgs and SUSY masses\footnote{Specifically, we assumed a Higgs mass $m_H>114$ GeV, 
lightest chargino and slepton masses $m_{\tilde\chi_1^+},m_{\tilde\ell} > 100$ GeV, as well as the neutralino being the LSP.}
In the $T'$ model, due to the freedom in the choice of the (order one) diagonal and off--diagonal parameters, one can 
always have a physically allowed spectrum down $M_{1/2} \sim 100$ GeV. 
The second evident difference is represented by a broadening of all scalar masses, but the lightest Higgs. This effects 
can be already noticed in the $m_0=200$ GeV plot, and then it is strongly enhanced for large values of $m_0$ and $A_0$. 
The reason of such a broadening is related to the strong dependence that these masses have from the order one (randomly 
chosen in the $(1/2,2)$ range) free $T'$ coefficients. Nevertheless, still in this case, the 
``average'' $T'$ scalar masses agree quite well with the MSUGRA predictions. 

\begin{figure}[!th]
\centering
\vspace{-0.5cm}
\subfigure[\it MSUGRA (left) vs. $T'$ (right) spectra for $u=0.01$ and $\tan\beta\approx5$. In the upper (lower) plots 
$A_0 = 2 \, m_{0}=400$ GeV ($A_0 = 2 \, m_{0}=2000$ GeV). \label{Fig:ferRPA}]
{\includegraphics[width=14.5cm,height=10cm]{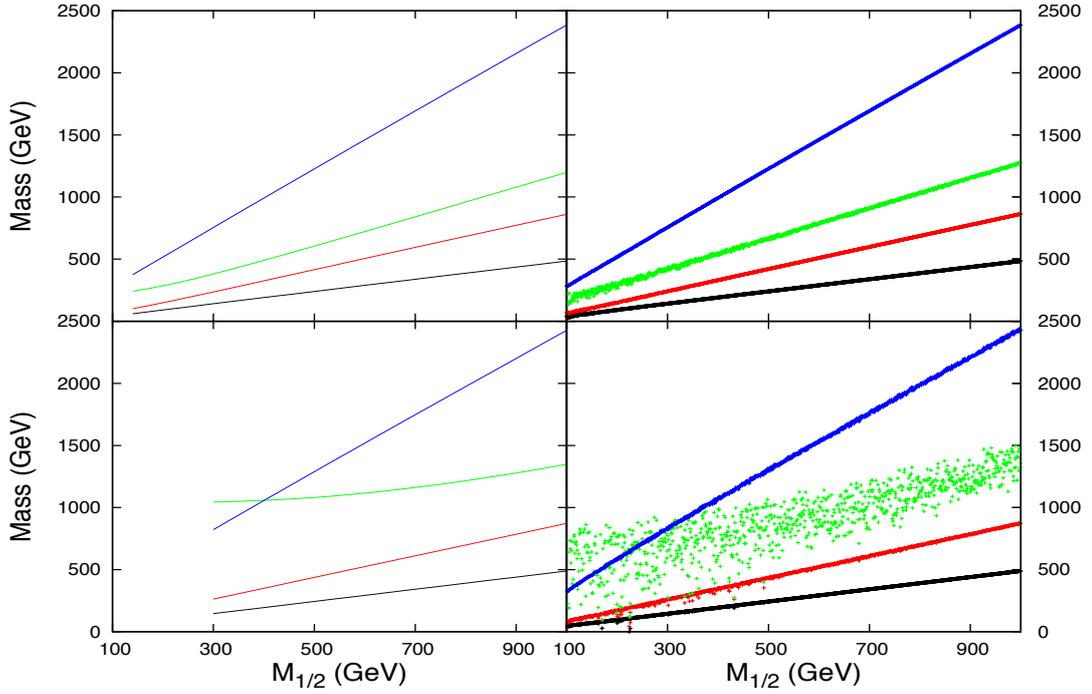}} \\ 
\vspace{-0.5cm}
\subfigure[\it MSUGRA (left) vs. $T'$ (right) spectra for $u=0.05$ and $\tan\beta\approx15$. In the upper (lower) plots 
$A_0 = 2 \, m_{0}=400$ GeV ($A_0 = 2 \, m_{0}=2000$ GeV). \label{Fig:ferRPB}]
{\includegraphics[width=14.5cm,height=10cm]{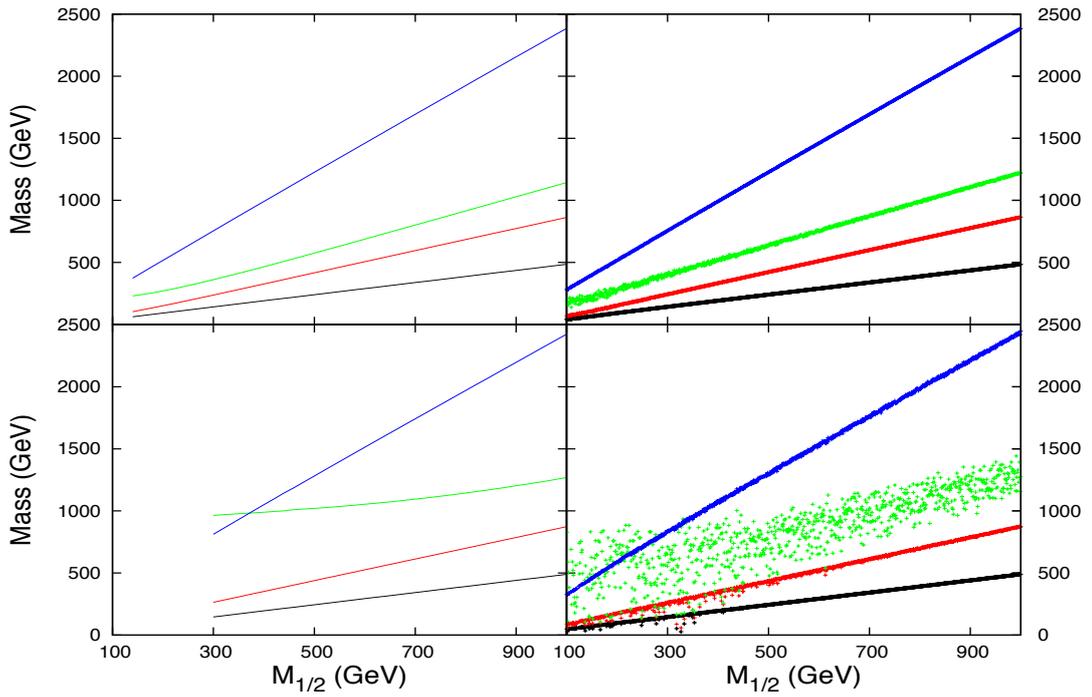}}
\caption{\it SUSY fermion spectra. The colors in the plots refer respectively to: Black for $\tilde{\chi}^0_1$, 
Red for $\tilde{\chi}^0_2$ and $\tilde{\chi}^+_1$, Green for $\tilde{\chi}^0_{3,4}$ and $\tilde{\chi}^+_2$, and 
Blue for $\tilde{g}$.} \label{Fig:fertot}
\end{figure}

In Fig.~\ref{Fig:fertot} we compare the MSUGRA fermionic spectrum (left plots) with the $T'$ model one (right plots), 
as function of common gaugino mass $M_{1/2}$. The results in Fig.~\ref{Fig:ferRPA} refer to the $RP_A$ case, while 
the plots in Fig.~\ref{Fig:ferRPB} cover the $RP_B$ case. In each figure the two upper plots are shown for a common 
bilinear and trilinear scalar SUSY scale $A_0=2\,m_0=400$ GeV, while for the two lower plots $A_0=2\,m_0=2000$ 
GeV has been chosen. For definiteness, again, only the $\sign[\mu]>0$ results have been shown, as no significative 
dependence on it is expected, nor observed. 

As typical in the MSUGRA framework, the lightest supersymmetric fermion is the lightest neutralino $\tilde\chi^0_1$ 
(in Black), while the lightest chargino $\tilde\chi^+_1$ is almost degenerate to the next-to-lightest neutralino 
$\tilde\chi^0_2$ (in Red). The gluino is typically the heaviest fermion (in Blue). As it can be noticed, comparing 
the upper left with the upper right plots of Figs.~\ref{Fig:ferRPA} and ~\ref{Fig:ferRPB}, the SUSY fermionic spectra 
for MSURGRA and $T'$ look very similar for small $m_0$ values, while evident differences appear in the large $m_0$ case.
Again, for $m_0=1000$ GeV one can notice that the $T'$ spectrum is ``phenomenologically'' allowed down to $M_{1/2}\sim 100$ GeV, 
while the MSUGRA one stops to be acceptable at $M_{1/2} \sim 300$ GeV. The second evident difference, as in the scalar 
case, is the presence of the broadening. However, differently from the scalar case, this effect appears only in the mass 
of next-to-heaviest gauginos (in Green), associated to the (mostly) higgsino-like chargino and neutralino, 
$\tilde\chi^+_2$, $\tilde\chi^0_{3,4}$. The reason for such a broadening is due to the way the $\mu$-parameter is derived: 
while in MSUGRA, $|\mu|$ is uniquely fixed once all the common SUSY parameters are defined, in our model $\mu$ indirectly 
depends also from the randomly chosen (order one) parameters that enter in the sfermion mass matrices. This freedom 
largely affects the derived value of $|\mu|$ especially in the large $(m_0,A_0)$ region.

%
%
\subsection{FCNC in the Leptonic Sector}

\begin{figure}
\centering
\vspace{-0.5cm}
\includegraphics[width=16cm]{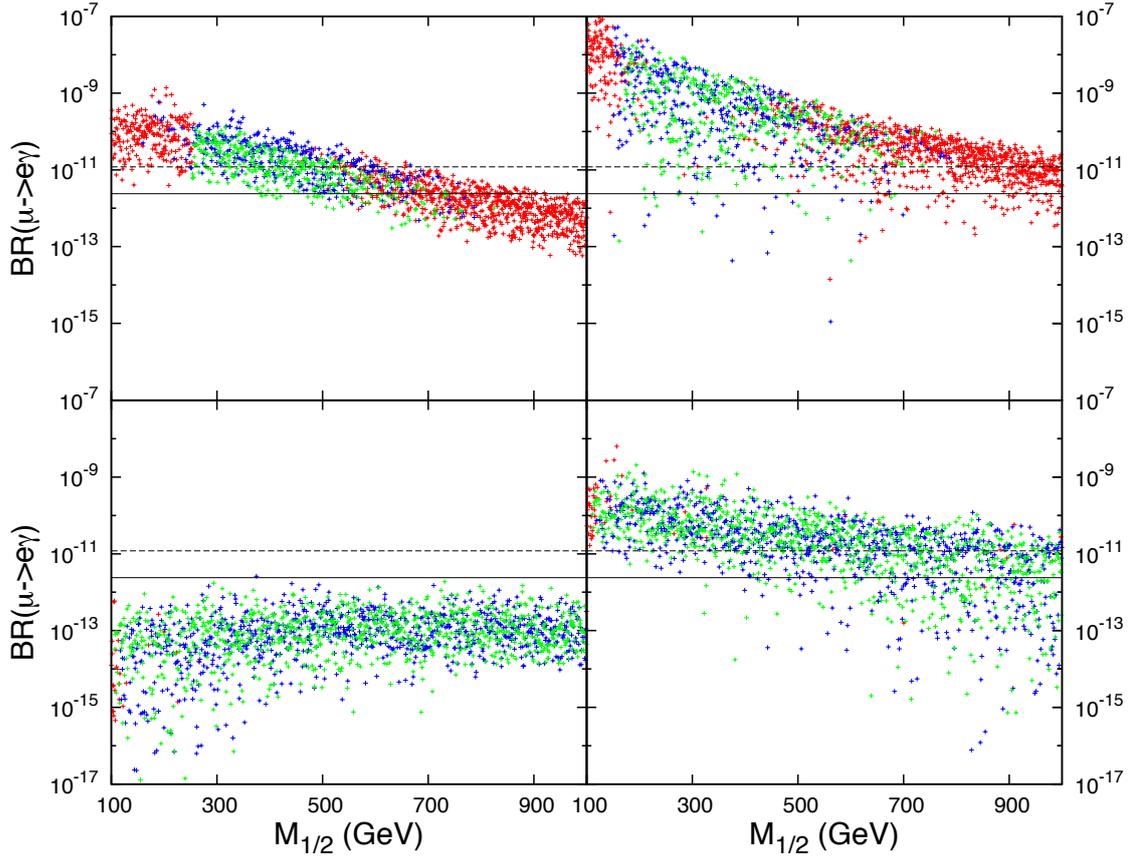} 
\caption{\it Scatter plots for BR(\meg \!\!) for the $T'$ model. Red points are not allowed by ``phenomenological'' 
constraints on Higgs or SUSY masses. Blue (Green) points refer to $\mu > 0$ ($\mu < 0$). The full 
(dashed) line represents MEG (MEGA) bound. The left (right) plots refer to the $RP_A$ ($RP_B$) case for $A_0=2\,m_0=400$ GeV (upper plots) or $A_0=2\,m_0=2000$ GeV (lower plots).}
\label{Fig:meg}
\end{figure}

The LFV analysis of the $T'$ model is going to be a straightforward replica of that performed in Ref.~\cite{FHLM:LFVinSUSYA4} 
for the $A_4$ realisation, as the two models (almost) coincide when restricted to the leptonic sector. As already stated 
in the previous sections, only slight differences are expected due to the presence of the extra flavons $\eta$ and 
$\xi''$ (see Eq.~(\ref{VEVs})) that are introduced when the embedding of $A_4$ in $T'$ is considered.
The only relevant improvement, compared to the $A_4$ analysis of Ref.~\cite{FHLM:LFVinSUSYA4}, is the implementation here of 
the complete (one-loop) RGE running of the high-energy model parameters down to the EW scale, that leads to a somehow 
heavier SUSY spectrum when compared to the un-ran case. 
In order to compare the $T'$ and $A_4$ predictions, the $T'$ branching ratio for \meg is shown in Fig.~\ref{Fig:meg} 
as a function of the common gaugino scale $M_{1/2}$. The full horizontal line represents the recently reported 90\% C.L. 
MEG upper limit Ref.~\cite{Adam:2011ch} of $2.4 \times 10^{-12}$ on the  BR(\meg \!\!), while the dashed line shows,
for comparison, the old MEGA bound of $1.2\times 10^{-11}$.

The two upper plots refer to $A_0=2\,m_0=400$ GeV, respectively for the $RP_A$ case (upper-left) and $RP_B$ case (upper-right). 
Red points are excluded either by the direct experimental requirements on the light Higgs or SUSY masses (mainly in the 
low $M_{1/2}$ region), or by the requirement of a neutral LSP (mainly in the high $M_{1/2}$ region where the $\tilde{\tau}_1$ 
becomes lighter that the $\tilde{\chi}^0_1$). Blue (Green) points refer to $\mu > 0$ ($\mu < 0$) respectively. No 
significative dependence on the $\sign[\mu]$ is observed, consequence of the arbitrary sign with which the $T'$ 
coefficients can enter in most of the off-diagonal entries of the slepton mass matrices. As it can be noticed from the 
two upper plots, the low $(A_0,m_0)$ scenario is almost excluded for both the cases: in the $RP_A$ case mostly due to the 
requirement of neutral LSP, while in the $RP_B$ case because of a too large BR. However, in the $RP_B$ case few points 
are still allowed, independently of the $M_{1/2}$ value, thanks to a fine tuned cancellation between different terms. 
Concerning the $A_0=2\,m_0=2000$ GeV case, the left-lower plot of Fig.~\ref{Fig:meg} shows that no constraints 
comes from the BR(\meg \!\!) for the $RP_A$ scenario, while for large $u$ and $\tan\beta$ (right-lower plot of 
Fig.~\ref{Fig:meg}) our $T'$ model is allowed only for $M_{1/2}\gtrsim 500$ GeV. Our results are quite in agreement 
with the $A_4$ analysis of \cite{FHLM:LFVinSUSYA4}, once a factor 10 suppression in the BR, produced by the heavier spectrum, 
is taken into account. The results for the $\tau$ decays are not shown as no competing results are expected from 
$\tau \raw e \gamma$ and $\tau \raw \mu \gamma$ decays, even assuming Super B factory luminosities.

When studying FCNC, it is quite common to present bounds in terms of Mass Insertion (MI) instead of using directly 
the specific observables. The MI approach has the advantage to provide easy connections between the predictions of 
our $T'$ model with the more general (SUSY) framework. We have not derived explicit formulas in the MI approximation 
as most of them can be found in \cite{FHLM:LFVinSUSYA4}. Instead, a numerical analysis is performed and regions of 
expectation for all the MI have been derived. 

A thoughtful work on leptonic MI in the SUSY context has been performed in \cite{Paradisi:2005fk}, where the upper 
limits, in the single-dominance approximation, for all the relevant MIs are plotted as function of the scalar and 
gaugino common scales, $m_0, M_{1/2}$. To compare the results with the literature we have defined the MIs as the 
corresponding entry in the sfermion mass matrix divided by the ``typical'' SUSY scale that for being conservative 
we have chosen as the mass of the lightest scalar in the corresponding sector.

\begin{figure}
\centering
\vspace{-0.5cm}
\includegraphics[width=15cm,height=10.25cm]{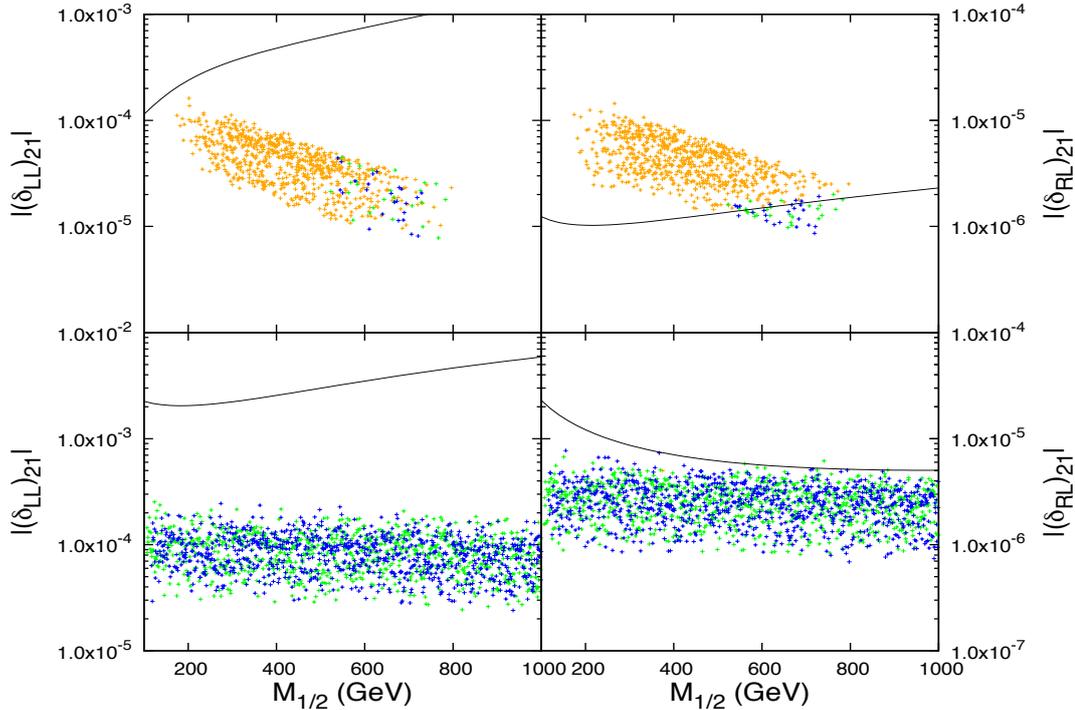}
\caption{\it LL and RL Leptonic Mass insertions $(\delta^{21})_\ell$ for the $RP_A$ case. In the upper (lower) plots 
$A_0=2\,m_0=400$ GeV ($A_0=2\,m_0=2000$ GeV) respectively. Orange points are not allowed by recent MEG data. Blue (Green) 
points refer to $\mu > 0$ ($\mu < 0$). Continuous lines represent MI bounds derived from Ref.~\cite{Paradisi:2005fk}.} 
\label{Fig:MI12RPA}
\end{figure}

In Fig.~\ref{Fig:MI12RPA} the results for the LL and RL leptonic MIs for the 21 sector, $(\delta^{21})_\ell$, are shown 
for the $RP_A$ case and for the two reference values $A_0=2\,m_0=400$ GeV (upper plots) and $A_0=2\,m_0=2000$ GeV 
(lower plots). Dots represent the value of the relevant observable as computed (at one-loop in the NLO 
approximation) by \spheno as function of the SUSY gaugino parameter $M_{1/2}$ and varying randomly all the unknown 
$T'$ coefficients in the $(1/2,2)$ range. Orange points are excluded by present MEG bound on the \meg Branching Ratio, 
while Blue (Green) points represent an allowed region for $\mu > 0$ $(\mu < 0)$ respectively. For all points shown, 
experimental spectrum constraints are satisfied. The continuous black lines represent the MI values, derived from Ref.~\cite{Paradisi:2005fk}, that saturate the experimental data. As it can be noticed from the right-upper plot of  
Fig.~\ref{Fig:MI12RPA}, MEG data severely constrain our model and force the $(\delta^{21}_{RL})_\ell$ MI to be 
below $2\times 10^{-6}$, with a mild dependence on $M_{1/2}$. This is, in fact, expected to be the dominant contribution 
in the low $\tan\beta$ regime. In our model the $(\delta^{21}_{LL})_\ell$ MI can vary in the range $10^{-5}-10^{-4}$, 
roughly a factor 5--10 below the saturation limit. The RR and LR MIs are completely irrelevant in our analysis and 
therefore are not shown. When a higher value for the common scalar mass scale is chosen, $A_0=2\,m_0=2000$ GeV, 
no limits are obtained on the MIs, as shown in Fig.~\ref{Fig:MI12RPA} (lower plots). The main contribution still comes 
from the $(\delta^{21}_{RL})_\ell$ MI, just below the saturation limit.

\begin{figure}
\centering
\vspace{-0.5cm}
\includegraphics[width=15cm,height=10.25cm]{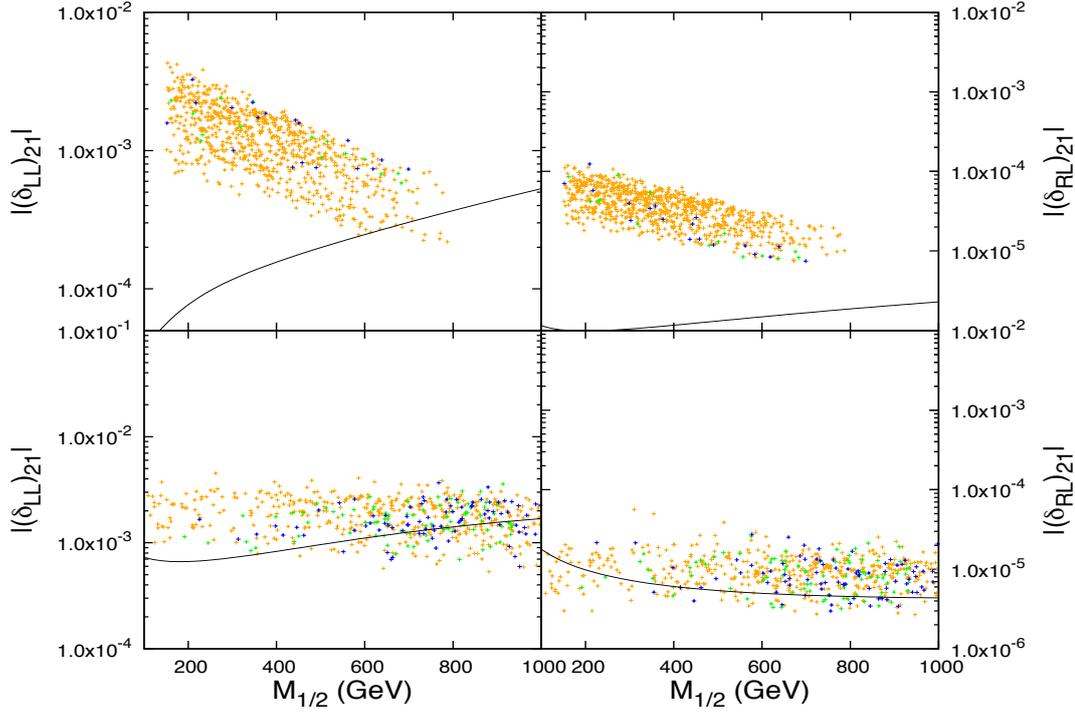}
\caption{\it LL and RL Leptonic Mass insertions $(\delta^{21})_\ell$ for the $RP_B$ case. In the upper (lower) plots 
$A_0=2\,m_0=400$ GeV ($A_0=2\,m_0=2000$ GeV) respectively. Orange points are not allowed by recent MEG data. Blue (Green) 
points refer to $\mu > 0$ ($\mu < 0$). Continuous lines represent MI bounds derived from Ref.~\cite{Paradisi:2005fk}.} 
\label{Fig:MI12RPB}
\end{figure}

In Fig.~\ref{Fig:MI12RPB} the results for the LL and RL leptonic MIs are shown for the $RP_B$ case and for the two 
reference values $A_0=2\,m_0=400$ GeV (upper plots) and $A_0=2\,m_0=2000$ GeV (lower plots). As can be noticed, in this 
case the amount of flavour changing is too high and, at least in the low scalar mass scale case (upper plots of 
Fig.~\ref{Fig:MI12RPB}), almost no points survive the experimental bound. Only at higher scalar scale the model is not 
completely excluded by MEG result, especially at high $M_{1/2}$ values. However, one expects that, for larger values 
of $\tan\beta$ as the one of $RP_B$, both LL and RL MIs can be relevant, opening the possibility of cancellations. 
This indeed happens in both low and high $m_0$ scenario, being particularly evident in the first case where even if 
both the LL and RL MIs are above the saturation limit of \cite{Paradisi:2005fk} by at least a factor 5--10, still few 
points of the model are allowed. Again the RR and LR MIs are completely irrelevant and therefore are not presented here. 

The $T'$ model at hands, predicts on the same footing also FCNC for the 31 and 32 MIs sectors. Bounds on $(\delta^{31})_\ell$ 
and $(\delta^{31})_\ell$ can be mostly obtained by analysing $\tau \raw e \gamma$ and $\tau \raw \mu \gamma$ radiative decays. 
Nevertheless, as already mentioned no relevant information come from these channels and these sectors are not shown here.

%
%
\subsection{FCNC in the hadronic sector}
\label{sec:PhenHad}

Differently from the leptonic case, no previous study to the hadronic FCNC sector for the $T'$ model as ever been 
performed before. Furthermore, for this sector no general MI analysis in all the available SUSY breaking parameters 
($M_{1/2},\, m_0$) has been performed, so we prefer to discuss the phenomenology directly showing the FCNC observables. 

In order to study the hadronic FCNC observables, it has been necessary to modify the corresponding {\tt SPheno} routines, 
and check the agreement with the literature in several ways \cite{Hollik:1997vb,Hollik:1997ph}. In the \bsg Branching 
Ratio, new contributions coming from gluinos and neutralinos, not present in {\tt SPheno-2.2.3}, have also been 
included \cite{bsgstef}. Moreover as \spheno result is not completely up-to-date in reproducing the NNLO SM contribution 
to the \bsg BR, we found more convenient to calculate with \spheno at NLO the ``differential'' Branching Ratio:
\be
\Delta BR(\bar B \raw X_s \gamma) = BR_{SM+NP}(\bar B \raw X_s \gamma) - BR_{SM}(\bar B \raw X_s \gamma)\,. \nn
\ee
We then compare it with the current experimental value \cite{PDG2010}, measured with a photon--energy cut-off 
$E_\gamma > 1.6$ GeV in the $B$-meson rest frame,
\be
BR(\bar B \raw X_s \gamma) = (3.55 \pm 0.24 \pm 0.09) \times 10^{-4} \,, \nn
\ee
subtracted by the SM prediction calculated at NNLO \cite{Misiak:2006zs,Gambino:2001ew,Misiak:2006ab} for the same 
photon energy cut-off,
\be
BR(\bar B \raw X_s \gamma) = (3.15 \pm 0.23) \times 10^{-4} \,. \nn
\ee

\begin{figure}
\centering
\vspace{-0.5cm}
\includegraphics[width=15cm,height=10.25cm]{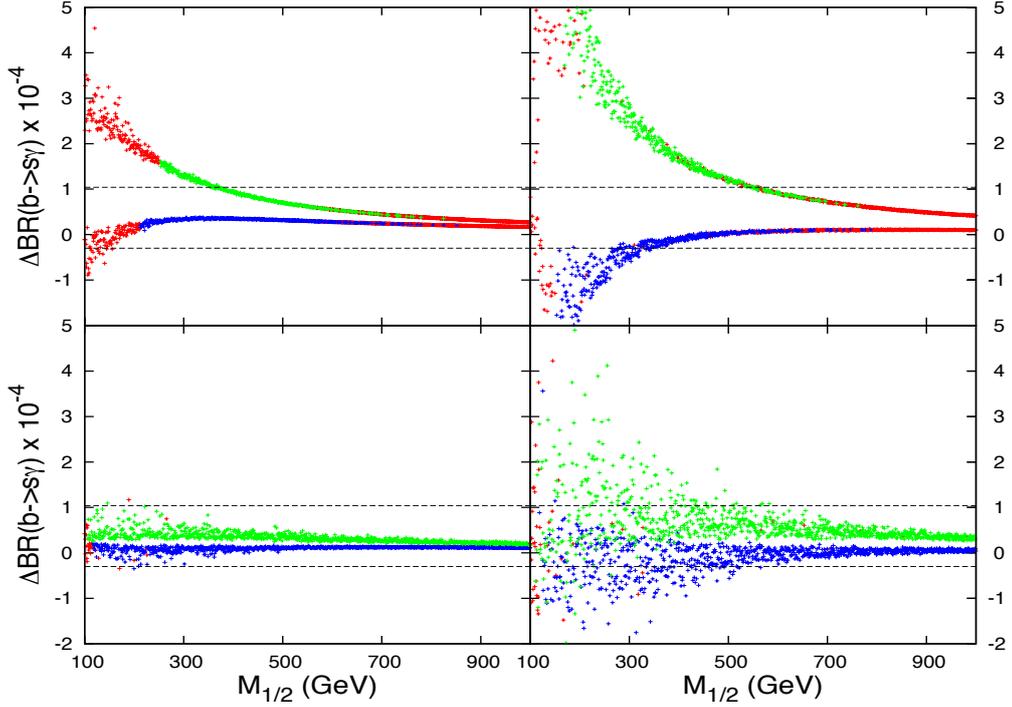}
\caption{\it Scatter plots for the \bsg BR for the $T'$ model. Red points are not allowed by ``phenomenological'' 
constraints on Higgs or SUSY masses. Blue (Green) points refer to $\mu > 0$ ($\mu < 0$). The left (right) plots refer to the $RP_A$ ($RP_B$) case for $A_0=2\,m_0=400$ GeV (upper plots) or $A_0=2\,m_0=2000$ GeV (lower plots).} 
\label{Fig:bsg}
\end{figure}

In Fig.~\ref{Fig:bsg} the $T'$ Branching Ratio for \bsg is shown as function of the common gaugino scale $M_{1/2}$. 
The two upper (lower) plots refer to $A_0=2\,m_0=400$ GeV ($A_0=2\,m_0=2000$ GeV), respectively for the $RP_A$ case 
(left) and $RP_B$ case (right). Red points are excluded by imposing ``phenomenological'' requirements on the Higgs 
and/or SUSY masses. Blue (Green) points refer to $\mu > 0$ ($\mu < 0$) as usual. The horizontal dashed lines represent 
the $2\sigma$ experimental bound from Ref.~\cite{PDG2010}. 
In the small $(A_0,m_0)$ region (upper plots), the dominant $T'$ contribution to the \bsg BR is 
typically the charged Higgs one, due to the fact that the stop is for most of the $M_{1/2}$ range heavier than the 
$H^\pm$ (and the chargino heavier than the top). The Higgs contribution is always concordant in sign with the SM one. 
The second most relevant contribution is the chargino one with a sign depending on $\sign[\mu]$: it tends to enhance 
(cancel) the SM contribution for $\mu < 0$ ($\mu > 0$). Gluino contributions are practically independent 
from $\sign[\mu]$, while neutralino ones are completely negligible. In Appendix~\ref{AppD}, we report the break down 
of the relevant Wilson Coefficient for $BR(\bar B \raw X_s \gamma)$ in terms of the supersymmetric contributions. 
This helps understanding the results showed in Fig.~\ref{Fig:bsg}. As a consequence, experimental constraints on the 
\bsg BR tend to disfavor $\mu < 0$ for $M_{1/2}\lesssim500$ GeV especially for the larger $\tan\beta$ values of the 
$RP_B$ point (right plots). The positive $\mu$ case, instead, turns out to be mostly allowed. In the $A_0=2\,m_0=2000$ 
GeV scenario (lower plots) all new physics contributions get strongly suppressed and no significative limits are 
expected from the \bsg measurement in neither the two cases considered.

\begin{figure}
\centering
\vspace{-0.5cm}
\includegraphics[width=15cm,height=10.25cm]{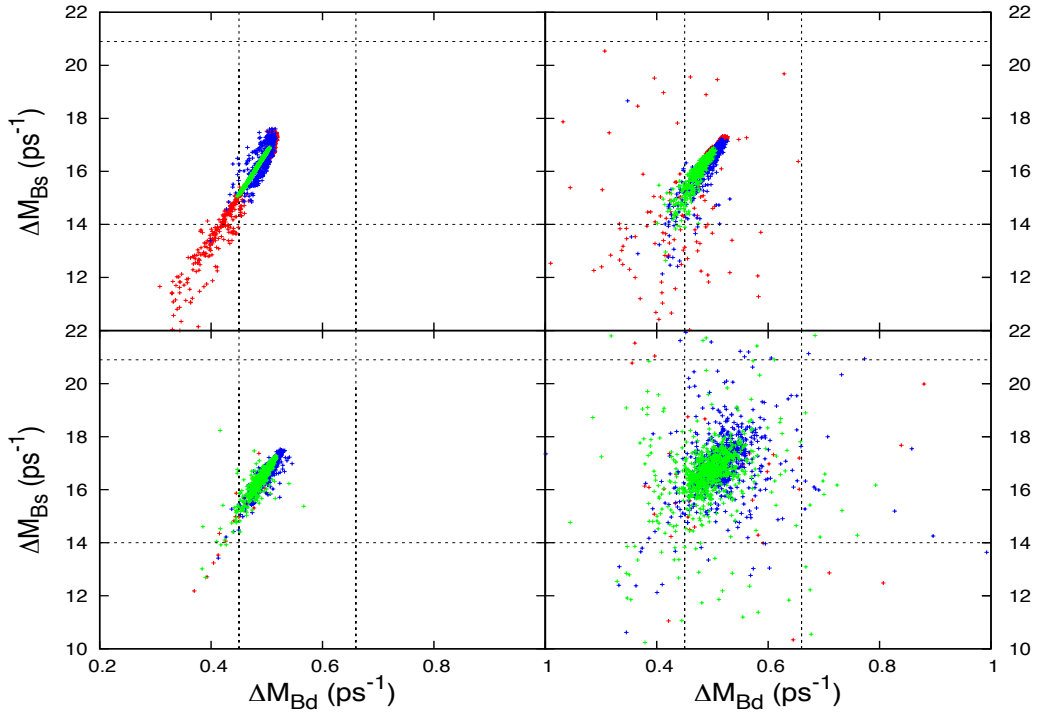}
\caption{\it Scatter plots for $\Delta M_{B_s}$ vs. $\Delta M_{B_d}$ for the $T'$ model. Red points are not allowed by 
``phenomenological'' constraints on Higgs or SUSY masses. Blue (Green) points refer to $\mu > 0$ ($\mu < 0$). 
The dashed lines represent $2\sigma$ experimental bound. The left (right) plots refer to the $RP_A$ ($RP_B$) case for $A_0=2\,m_0=400$ GeV (upper plots) or $A_0=2\,m_0=2000$ GeV (lower plots).} 
\label{Fig:bdvsbs}
\end{figure}

In Fig.~\ref{Fig:bdvsbs} the $T'$ predictions for $\Delta M_{B_s}$ vs. $\Delta M_{B_d}$ are shown as function of 
the common gaugino mass $M_{1/2}$. The two upper (lower) plots refer to $A_0=2\,m_0=400$ GeV ($A_0=2\,m_0=2000$ GeV), respectively for the $RP_A$ case 
(left) and $RP_B$ case (right). Red points, as usual, are excluded by imposing ``phenomenological'' requirements 
on the Higgs and/or SUSY masses while Blue (Green) points refer to $\mu > 0$ ($\mu < 0$). The horizontal and vertical 
dashed lines show the $2\sigma$ bounds on the $B_d$ and $B_s$ mass differences obtained by Ref.~\cite{Lenz:2010gu}: 
\bald
&\Delta M_{B_d} =  0.55^{+0.11}_{-0.10} \quad {\rm ps}^{-1} \\ 
&\Delta M_{B_s} = 16.8^{+4.1}_{-2.8} \quad {\rm ps}^{-1} \,,\nn
\eald
perfectly compatible with present experimental bounds from Ref.~\cite{PDG2010}:
\bald
&\Delta M_{B_d} = 0.507 \pm 0.005 \quad {\rm ps}^{-1} \\
&\Delta M_{B_s} = 17.77 \pm 0.12 \quad {\rm ps}^{-1} \,.\nn
\eald
The $T'$ SUSY contribution to $\Delta M_{B_{d,s}}$ has always opposite sign compared to the SM one, independently 
on the $\sign[\mu]$. As evident from the plots of Fig.~\ref{Fig:bdvsbs} this fact always leads to a suppression 
of the neutral meson mass differences, where the ``comet'' tail is obtained for large values of $M_{1/2}$, 
while for small $M_{1/2}$ the points addensate in the ``comet'' head region. However, both $B_d$ and $B_s$ bounds 
are still too loose to provide any useful constraints especially in the $RP_A$ scenario (left plots). 
Nevertheless it is interesting to note that, for the higher $\tan\beta$ of the $RP_B$ scenario (right plots) the low 
$M_{1/2}$, tails start getting excluded by the present value of $\Delta M_{B_d}$ both in the low and in the high $m_0$ 
case. One can expect, that with a future improvement in the hadronic flavour parameters at LHCb or at Super B 
factories, constraints from the hadronic FCNC will start being competitive with the leptonic experiments.

Finally, no relevant bounds from $K^0-\bar{K}^0$ oscillations are obtained in our model as the first two family squarks 
are nearly degenerate, producing no significant deviations from the SM (though poorly known) reference value.

%
%
\subsection{Cross-correlations between leptonic and hadronic observables}

The main feature of the SUSY $T'$ model presented in this paper is the simultaneous prediction of the SM fermion 
mass sector together with the sfermion mass matrix structures. Furthermore, the amount of FCNC in the leptonic and hadronic 
sectors are tightly connected, providing very peculiar signatures. 

\begin{figure}
\centering
\vspace{-0.5cm}
\subfigure[Scatter plots for \meg vs. \bsg for the $T'$ model. The left (right) plots refer to the $RP_A$ ($RP_B$) case 
respectively for $A_0=2\,m_0=400$ GeV (upper plots) or $A_0=2\,m_0=2000$ GeV (bottom plots).
\label{Fig:bsgvsmeg}]
{\includegraphics[width=15cm,height=10cm]{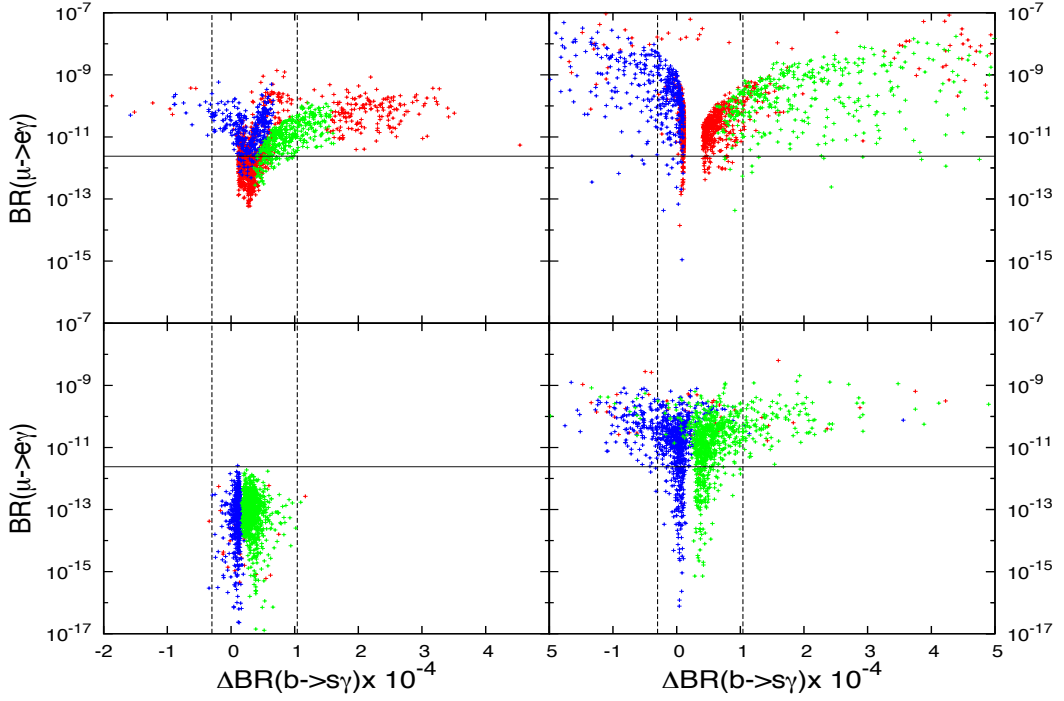}} \\ 
\vspace{-0.5cm}
\subfigure[Scatter plots for \meg vs. $\Delta M_{B_d}$ for the $T'$ model. The left (right) plots refer to the $RP_A$ ($RP_B$) 
case respectively for $A_0=2\,m_0=400$ GeV (upper plots) or $A_0=2\,m_0=2000$ GeV (bottom plots).
\label{Fig:bdvsmeg}]
{\includegraphics[width=14.5cm,height=10cm]{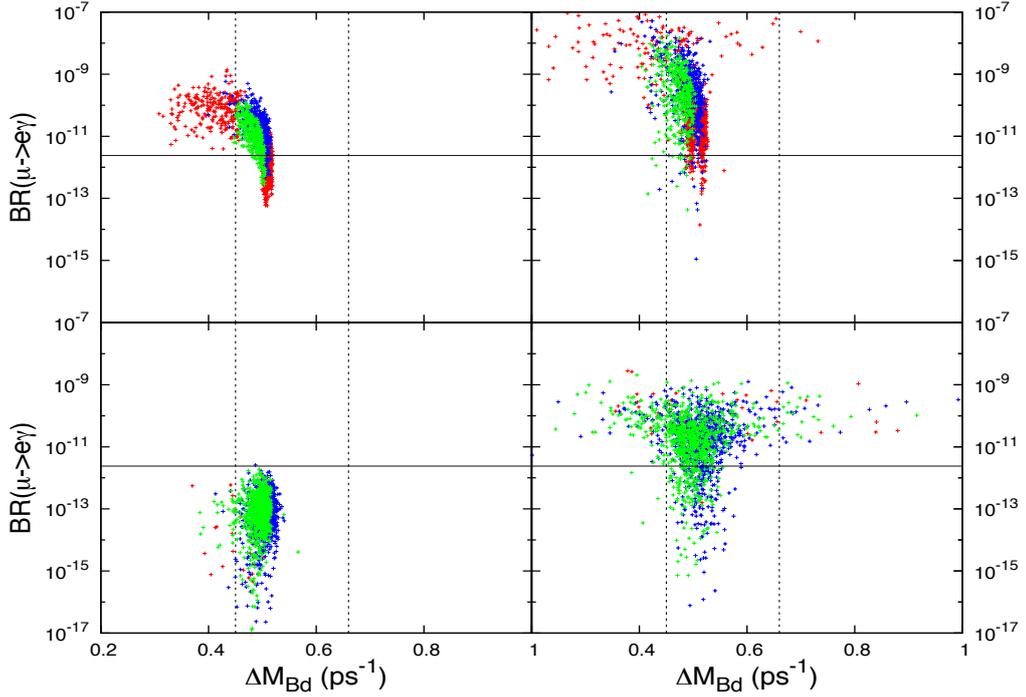}}
\caption{\it Cross correlations between hadronic and leptonic observables. Red points are not allowed by ``phenomenological'' 
requirements on Higgs and/or SUSY masses. Blue (Green) points refer to $\mu > 0$ ($\mu < 0$).} \label{Fig:adrlep}
\end{figure}

In Fig.~\ref{Fig:adrlep} the $T'$ cross--correlations between \meg vs. \bsg (Fig.~\ref{Fig:bsgvsmeg}) and \meg vs. 
$\Delta M_{B_d}$ (Fig.~\ref{Fig:bdvsmeg}) are shown varying the common gaugino mass $M_{1/2}$ in the range 
$(100,1000)$ GeV. The left (right) plots refer to the $RP_A$ ($RP_B$) case, while in the upper (lower) plots of 
each figure the $A_0=2\,m_0=400$ GeV ($A_0=2\,m_0=2000$ GeV) case is presented. Red points are excluded by 
imposing ``phenomenological'' requirements on the Higgs and/or SUSY masses. Blue (Green) points refer to $\mu > 0$ 
($\mu < 0$) as usual. The horizontal dashed lines indicate the $2\sigma$ experimental bounds as reported in the 
previous subsections.

It is evident from Fig.~\ref{Fig:bsgvsmeg} that the strongest constraint still comes from the \meg BR, which almost excludes 
the small $m_0$ region. However, when larger values of $m_0$ are considered (lower plots), an improvement in the 
\meg bound will not impose any severe exclusion on the model, as BR values as small as $10^{-15}$ are still acceptable. 
Some help can come, instead, from the hadronic sector. For the moment the precision in the \bsg BR measurement is not 
sufficiently good to impose further bounds on the $T'$ model, when larger values of $m_0$ are considered (lower plots). 
However, an improved measure of \bsg, can eventually leads in the large $\tan\beta$ scenario to a discrimination of 
the $\sign[\mu]$. An improvement in the knowledge of the meson mass differences at future facilities can help in further 
constraining the model in the large $\tan\beta$ regime (see lower--right plot of Fig.~\ref{Fig:bdvsmeg}), while the large 
$m_0$, small $\tan\beta$ case will be almost impossible to exclude.

%
%
\section{Conclusions}
\label{sec:Conclusions}

In the Standard Model (SM) of particle physics an explanation of the origin of the fermionic mass and mixing patterns 
is missing. A complete description of Nature should include a comprehensive description of the flavour sector. 
In this paper, a SUSY model based on the discrete flavour group $T'$, mainly based on Ref.~\cite{FHLM:Tp}, has been analysed. 
This model accounts both for leptons, predicting the TB mixing pattern, and for quarks, providing a realistic CKM matrix. 

While several flavour models can accomplish these goals, however, a discrimination between 
them can only be obtained through a detailed analysis of the associated phenomenology. In this paper we have  
studied the $T'$ model predictions to the most relevant, leptonic and hadronic, FCNC observables. Concerning the 
leptonic FCNC we have essentially confirmed the \meg results obtained in Ref.~\cite{FHLM:LFVinSUSYA4} for 
the $A_4$ model, as the two realisations (almost) coincide when restricted to leptons. The main difference 
is related to the full implementation in our study of the RGE for the low energy SUSY spectrum, by means of the use of 
the \spheno routines. In addition we performed, for the first time, a detailed analysis of the $T'$ predictions 
for FCNC in the hadronic sector. It turns out that the amount of FCNC in the leptonic and hadronic sectors are 
tightly connected.

The strongest bounds on the FCNC sector still come from the \meg Branching Ratio. Making use of the latest available 
MEG data, supplemented by ``phenomenological'' constraints on the Higgs and SUSY spectrum, almost all the available 
parameter space is excluded in the low $(A_0,m_0)$ region. No bounds can be set when the common soft scalar breaking 
scale is increased to the TeV range and small values of the $T'$ symmetry breaking parameter $u$ are selected, while 
the low $M_{1/2}$ region is excluded for larger $u$ and $\tan\beta$. Hadronic FCNC observables like \bsg Branching 
Ratio or the neutral $B$ meson mass differences, $\Delta M_{B_d}$ and $\Delta M_{B_s}$ can provide independent 
information, even if they are not yet enough precise to severely constrain the model. A factor 2 improvement in their 
experimental and theoretical determinations could however be enough for making the corresponding constraints 
quite competitive. The neutral $K$ meson mass difference, $\Delta M_{K}$ turns out to almost coincide with SM prediction, 
due to the quasi degeneracy in the first two squark families.

The forthcoming results from LHCb and Super B factories will provide stronger constraints in the hadronic sector 
and allow a better study of the parameter space for the $T'$ model.

%
%
\section*{Acknowledgments}
The authors thanks A.~Casas, M.~E.~Cabrera, F.~Feruglio and P.~Paradisi for very useful discussions.
L.~Merlo and B.~Zald\'ivar Montero thank the Dipartimento di Fisica ``Galileo Galilei" of the Universit\`a degli 
Studi di Padova for hospitality during the development of this project. L.~Merlo and S.~Rigolin thank the Departamento 
de F\'isica Te\'orica of the Universidad Aut\'onoma de Madrid for hospitality during the development of this project. 
L.~Merlo acknowledges the German Bundesministerium f\"ur Bildung und Forschung' under contract 05H09WOE. 
S.~Rigolin acknowledges the partial support of an Excellence Grant of Fondazione Cariparo and of the 
European Program‚ Unification in the LHC era‚ under the contract PITN- GA-2009-237920 (UNILHC).
B. Zald\'ivar Montero acknowledges the financial support of the FPI (MICINN) grant BES-2008-004688, and the contracts 
FPA2010-17747 and PITN-GA-2009-237920 (UNILHC) of the European Commission. 

%
%

%
%
\appendix

\mathversion{bold}
\section{The group $T'$}
\label{AppA}
\mathversion{normal}

The group $T'$ has $24$ elements and $7$ irreducible representations: one triplet $\bf3$, three doublets $\bf2$, $\bf2'$ and $\bf2''$ and three singlets $\bf1$, $\bf1'$ and $\bf1''$. It is generated by two elements $S$ and $T$ fulfilling the relations
\begin{equation}
S^{2}=\mathbb{R}\,, \;\; T^{3}=\mathbb{1}\,, \;\; (S T)^{3}=\mathbb{1}\,, \;\; \mathbb{R}^{2}=\mathbb{1}\,,
\end{equation}
where $\mathbb{R}=\mathbb{1}$ in case of the odd-dimensional representation and $\mathbb{R}=-\mathbb{1}$ for $\bf2$, $\bf2'$ and $\bf2''$ such that $\mathbb{R}$ commutes with all elements of the group. Beyond the center of the group, generated by the elements $\unity$ and $\mathbb{R}$, there are other Abelian subgroups: $Z_3$, $Z_4$ and $Z_6$. In particular, there is a $Z_4$ subgroup here denoted by $G_S$, generated by the element $TST^2$ and a $Z_3$ subgroup here called $G_T$, generated by the element $T$. $G_S$ and $G_T$ are of great importance because they represent the low-energy flavour structures of the fermion masses.

The multiplication rules of the representations are as follows:
\be
\begin{array}{l}
{\bf1}^a \times {\bf r}^b = {\bf r}^b\times {\bf1}^a={\bf r}^{a+b}\qquad\qquad\text{for}\;{\bf r}={\bf1},\,{\bf2}\\
{\bf1}^a \times {\bf3} = {\bf3} + {\bf1}^a = {\bf3}\\
{\bf2}^a \times {\bf2}^b = {\bf3} + {\bf1}^{a+b}\\
{\bf2}^a \times {\bf3} = {\bf3}\times {\bf2}^a = {\bf2} + {\bf2'} + {\bf2''}\\
{\bf3} \times {\bf3} = {\bf3} + {\bf3} + {\bf1} + {\bf1'} + {\bf1''}
\end{array}
\label{TpTBM:mult}
\ee
where $a,\,b=0,\pm1$ and we have denoted ${\bf1}^0\equiv{\bf1}$, ${\bf1}^{1}\equiv{\bf1'}$, ${\bf1}^{-1}\equiv{\bf1''}$ and similarly for the doublet representations. On the right-hand side the sum $a+b$ is modulo 3. The Clebsch-Gordan coefficients for the decomposition of product representations can be found in Ref.~\cite{FHLM:Tp}.

%
%
\mathversion{bold}
\section{Details on the sfermion mass matrices}
\label{AppB}
\mathversion{normal}

In this Appendix we give more details on the computation of the sfermion mass matrices listed in the Sects. \ref{sec:SfermionMasses} and \ref{sec:PhysicalBasis}.

The contributions to the sfermion masses from the SUSY breaking terms in the K\"ahler potential are given by two distinct terms: $(m^2_{fLL})_K$ of Eqs.~(\ref{me2LL})--(\ref{md2RR}) and those contributions related to the non-vanishing VEV of the auxiliary fields of the flavon supermultiplets. Denoting these second ones as $(m^2_{fLL})_{K2}$, we find for the slepton sector:
\be
(m^2_{(e,\nu)LL})_{K2}=
\left(
	\begin{array}{ccc}
		t^\ell_{F1}~ u^2 & t^\ell_{F4}~ u^2 & t^\ell_{F5}~ u^2\\
		\ov{t}^\ell_{F4}~ u^2 & t^\ell_{F2}~ u^2 & t^\ell_{F6}~ u^2\\
		\ov{t}^\ell_{F5}~u^2 & \ov{t}^\ell_{F6}~ u^2 & t^\ell_{F3}~ u^2 
	\end{array}
\right)\,,
\label{me2LL2}
\ee
\be
(m^2_{eRR})_{K2}=
\left(
	\begin{array}{ccc}
		t^e_{F1}~ u^2 & t^e_{F4}~ u^2 t& t^e_{F5}~ u^2\, t^2\\
		\ov{t^e}_{F4}~ u^2 t&  t^e_{F2}~ u^2 & t^e_{F6}~ u^2\, t\\
		\ov{t^e}_{F5}~ u^2\, t^2& \ov{t^e}_{F6}~ u^2\, t &  t^e_{F3}~ u^2 
	\end{array}
\right)\,,
\label{me2RR2}
\ee
where the coefficients $t^{\ell,e}_{F}$ are combinations of $p_I$ in Eq. (\ref{deck}) and the distinct $c_F$ in Eqs. (\ref{VEVsFTerms}). Such relations are not particularly significant and we avoid to report them here. Notice that these contributions are absent in Ref.~\cite{FHLM:LFVinSUSYA4}, but do not change those results: indeed $(m^2_{eLL})_{K2}$, $(m^2_{\nu LL})_{K2}$ and $(m^2_{eRR})_{K2}$ can be safely absorbed in a redefinition of the parameters of $(m^2_{eLL})_{K}$, $(m^2_{\nu LL})_{K}$ and $(m^2_{eRR})_{K}$.

For the same contributions in the quark sector we get:
\be
(m^2_{(u,d)LL})_{K2}=
\left(
	\begin{array}{ccc}
		t^q_{F1}~ u^2 & t^q_{F4}~ u^2 & t^q_{F5}~ u^2\\
		\ov{t}^q_{F4}~ u^2 & t^q_{F2}~ u^2 & t^q_{F6}~ u^2\\
		\ov{t}^q_{F5}~u^2 & \ov{t}^q_{F6}~ u^2 & t^q_{F3}~ u^2 	
	\end{array}
\right)\,,
\label{mq2LL2}
\ee
\be
(m^2_{uRR})_{K2}=
\left(
	\begin{array}{ccc}
		t^u_{F1}~ u^2 & t^u_{F5}~ u^2 & t^u_{F6}~ u^2\,t \\
		\ov{t}^u_{F5}~ u^2 &t^u_{F2}~ u^2  & t^u_{F4}~ u^2\,t \\
		\ov{t}^u_{F6}~u^2\,t & \ov{t}^u_{F4}~ u^2\,t & t^u_{F2}~u^2 \\ 
	\end{array}
\right)\,,
\label{mu2RR2}
\ee
\be
(m^2_{dRR})_{K2}=
\left(
	\begin{array}{ccc}
		t^d_{F1}~ u^2 & t^d_{F4}~ u^2 & t^d_{F5}~ u^2\\
		\ov{t}^d_{F4}~ u^2 & t^d_{F2}~ u^2 & t^d_{F6}~ u^2\\
		\ov{t}^d_{F5}~u^2 & \ov{t}^d_{F6}~ u^2 & t^d_{F3}~ u^2  
	\end{array}
\right)\,,
\label{md2RR2}
\ee
where the coefficients $t^{q,u,d}_{F}$ are combinations of $p_I$ in Eq. (\ref{deck}) and the distinct $c_F$ in Eqs. (\ref{VEVsFTerms}). Such relations are not particularly significant. Furthermore notice that $(m^2_{uLL})_{K2}$, $(m^2_{dLL})_{K2}$, $(m^2_{uRR})_{K2}$ and $(m^2_{dRR})_{K2}$ can be safely absorbed in a redefinition of the parameters in $(m^2_{uLL})_{K}$, $(m^2_{dLL})_{K}$, $(m^2_{uRR})_{K}$ and $(m^2_{dRR})_{K}$.\\

We then comment on additional SUSY contributions to the $m^2_{fRR}$ mass matrices coming form the $D$ terms. The relevant factors in the scalar potential (through a $D$ term) are the following:
\be
\begin{split}
V_{D,FN}=&
\dfrac{1}{2} \,\left(M_{FI}^2+ g_{FN} \,Q_{FN}^i \dfrac{\partial \cK}{\partial \phi_i} \phi_i\right)^2+ 
          q_{FN}\, m_0^2 \vert \theta_{FN}
 \vert^2\\
 =& g_{FN}^2\, c_\theta\, m_0^2\Big[ 2|\tilde{e}^c|^2+|\tilde{\mu}^c|^2+|\tilde{u}^c|^2+|\tilde{c}^c|^2+
       |\tilde{d}^c|^2+|\tilde{s}^c|^2+|\tilde{b}^c|^2+\\
 &\hspace{2.7cm}+\left(t_4^e \,u^2 t\, {\ov{\tilde{e}}}^c \tilde{\mu}^c+ 
       t_6^e \,u^2 t\, {\ov{\tilde{\mu}}}^c \tilde{\tau}^c+ t^u_6\,ut\,\ov{\tilde{u}}^c \tilde{t}^c+ \hc\right) \Big]+...
\end{split}
\ee
where $Q_{FN}^i$ stands for the FN charge of the scalar field $\phi_i$, and in the second line we have displayed 
only the leading contributions to the terms quadratic in the matter fields for the slepton and the squark sectors. 
This result has been recovered considering the FN field vevs in Eq. (\ref{VEVFN}), but in the SUSY broken phase:
\be
\dfrac{M_{FI}^2}{g_{FN}}-|\mean{\theta_{FN}}|^2=c_\theta\, \msusy^2\,.
\ee
As commented in Sect.~\ref{sec:SfermionMasses}, these contributions can be simply reabsorbed in $(m^2_{fRR})_K$ through a field redefinition.\\

Once in the physical basis and assuming for simplicity that all the parameters of the model are real, we find the following results for the sfermion mass matrices.

We start with the LL block. For the sleptons we find
\be
(\hat{m}_{(e,\nu)LL}^2)_K=
	\left( \begin{array}{ccc}
                n^\ell_0 + 2 \, \hat{n}^\ell_1 \, u
                                & \hat{n}^\ell_4 \, u^2
                                & (\hat{n}^\ell_5 + (3 \, \hat{n}^\ell_1 - \hat{n}^\ell_2) \, c_{T3}) \, u^2 \\[1mm]
                 \hat{n}^\ell_4 \, u^2
                                & n^\ell_0 - (\hat{n}^\ell_1 + \hat{n}^\ell_2) \, u
                                & \hat{n}^\ell_6 \, u^2 \\[1mm]
                (\hat{n}^\ell_5 + (3 \, \hat{n}^\ell_1 - \hat{n}^\ell_2) \, c_{T3}) \, u^2
                                & \hat{n}^\ell_6 \, u^2
                                & n^\ell_0 - (\hat{n}^\ell_1 - \hat{n}^\ell_2) \, u \\
\end{array}\right) \, m_0^2\,,
\label{App:meLL_hat}
\ee
where the coefficients are defined by
\be
\hat{n}^\ell_i = n^\ell_i - t^\ell_i n^\ell_0\qquad( i=1,2,4,5,6)\,.
\label{App:meLL_hat_coefficients}
\ee
For the squarks we have
\be
(\hat{m}_{(u,d)LL}^2)_K =
\left( \begin{array}{ccc}
                \hat{n}^q_0 - \hat{n}^q_2  \, u & 2\,c''\,\hat{n}^q_2\,\dfrac{\tilde{y}_6}{y_5}u\,\sqrt{t} & c''\,\hat{n}^q_3\,\dfrac{\tilde{y}_6}{y_5}u\,\sqrt{t}\\[3mm]
                2\,c''\,\hat{n}^q_2\,\dfrac{\tilde{y}_6}{y_5}u\,\sqrt{t} & \hat{n}^q_0 + \hat{n}^q_2  \, u & \hat{n}^q_3 \, u \\[3mm]
                c''\,\hat{n}^q_3\,\dfrac{\tilde{y}_6}{y_5}u\,\sqrt{t} & \hat{n}^q_3 \, u & \hat{n}^q_1 \\
\end{array}\right) \, m_0^2\,,
\label{App:mqLL_hat}
\ee
where the coefficients are defined by
\be
\begin{gathered}
\hat{n}^q_0 = \dfrac{n^q_0}{t^q_d}\,,\qquad \qquad
\hat{n}^q_1 = \dfrac{n^q_1}{t^q_s}\,,\qquad \qquad
\hat{n}^q_2 =\dfrac{1}{t_d^q}\left(n_2^q-t_1^q\,\hat{n}^q_0\right)\,,\\
\hat{n}^q_3 =\dfrac{1}{\sqrt{t_s^q\,t_d^q}}\left(n^q_4-t_4^q\,\hat{n}^q_0\right)+ c'\dfrac{y_7}{y_b}\sqrt{\dfrac{t_s^q}{t_d^q}}\left(\hat{n}^q_0-\hat{n}^q_1\right)\,.
\end{gathered}
\label{App:mqLL_hat_coefficients}
\ee

For the RR block we find that $(\hat{m}_{fRR}^2)_K$ for the sleptons is given by
\be
(\hat{m}_{eRR}^2)_K = \left( \begin{array}{ccc}
                 n_1^c & 2 \, c_{T3} \,  (n_1^c - n_2^c) \, \dfrac{m_e}{m_\mu} u & 2 \, c_{T3} \, (n_1^c - n_3^c) \, \dfrac{m_e}{m_\tau} u\\[2mm]
                2 \, c_{T3}  \, (n_1^c - n_2^c) \, \dfrac{m_e}{m_\mu} u & n_2^c & 2 \, c_{T3} \, (n_2^c - n_3^c) \, \dfrac{m_\mu}{m_\tau} u\\[2mm]
                2 \, c_{T3}  \, (n_1^c - n_3^c) \, \dfrac{m_e}{m_\tau} u & 2 \, c_{T3}  \, (n_2^c - n_3^c) \,\dd \frac{m_\mu}{m_\tau} u & n_3^c\\
	\end{array}\right) \, m_0^2\,,
\label{App:meRR_hat}
\ee
while for the up squarks we have
\be
(\hat{m}_{uRR}^2)_K =\left(
        \begin{array}{ccc}
            \hat{n}^u_0-\hat{n}^u_2\,u & \hat{n}^u_5\,u^2 & \hat{n}^u_6\,t\,u^2 \\
            \hat{n}^u_5\,u^2  & \hat{n}^u_0+\hat{n}^u_2\,u & \hat{n}^u_4\,t\,u \\
            \hat{n}^u_6\,t\,u^2 & \hat{n}^u_4\,t\,u & \hat{n}^u_1 \\
        \end{array}
\right)\,\msusy^2\,,
\label{App:muRR_hat}
\ee
where the coefficients are defined by
\be
\begin{gathered}
\hat{n}^u_0 = \dfrac{n^u_0}{t^u_d}\,,\qquad \qquad
\hat{n}^u_1 = \dfrac{n^u_1}{t^u_s}\,,\qquad \qquad
\hat{n}^u_2 = \dfrac{1}{t_d^u}\left(n_2^u-t_1^u\,\hat{n}^u_0\right)\,,\\
\hat{n}^u_4 = \dfrac{1}{\sqrt{t_d^ut_s^u}}\left(n_4^u-t_4^u\dfrac{n^u_0-n_1^u}{t_d^u-t_s^u}\right)\,,\qquad
\hat{n}^u_5 = \dfrac{1}{t_d^u}\left(n_5^u-\dfrac{t_5^u}{t_1^u}\,n^u_2\right)\,,\\
\hat{n}^u_6 = \dfrac{1}{\sqrt{t_d^ut_s^u}}\left(n_6^u-t_6^u\dfrac{n^u_0-n_1^u}{t_d^u-t_s^u}\right)- \dfrac{t_5^u}{2\,t_1^u}\hat{n}^u_4\,,
\end{gathered}
\label{App:muRR_hat_coefficients}
\ee
and finally for the down squarks we get
\be
(\hat{m}_{dRR}^2)_K =\left(
        \begin{array}{ccc}
            n^d_0 - \hat{n}^d_2  \, u & -2\,c''\,\hat{n}^d_2\,\dfrac{\tilde{y}_6}{y_5}u\,\sqrt{t} & -c''\,\hat{n}^d_3\,\dfrac{\tilde{y}_6}{y_5}u\,\sqrt{t}\\[3mm]
            -2\,c''\,\hat{n}^d_2\,\dfrac{\tilde{y}_6}{y_5}u\,\sqrt{t} & n^d_0 + \hat{n}^d_2  \, u & \hat{n}^d_3 \, u \\[3mm]
            -c''\,\hat{n}^d_3\,\dfrac{\tilde{y}_6}{y_5}u\,\sqrt{t}& \hat{n}^d_3 \, u & n^d_1 \\
        \end{array}
\right)\,\msusy^2\,,
\label{App:mdRR_hat}
\ee
with the coefficients given by
\be
\begin{gathered}
\hat{n}^d_0 = \dfrac{n^d_0}{t^d_d}\,,\qquad \qquad
\hat{n}^d_1 = \dfrac{n^d_1}{t^d_s}\,,\qquad \qquad
\hat{n}^d_2 =\dfrac{1}{t_d^q}\left(n_2^d-t_s^d\,\hat{n}^d_0\right)\,,\\
\hat{n}^d_3 =\dfrac{1}{\sqrt{t_s^d\,t_d^d}}\left(n^d_4-t_4^d\,\hat{n}^d_0\right)+ c'\dfrac{y_8}{y_b}\sqrt{\dfrac{t_s^d}{t_d^d}}\left(\hat{n}^d_0-\hat{n}^d_1\right)\,.
\end{gathered}
\label{App:mdRR_hat_coefficients}
\ee

Finally, we report the results for the RL block. For the charged sleptons, the contributions are the following:
\be
(\hat{m}^2_{eRL})_1 = \hat{A}^e_1 \dfrac{v\, \cos\beta}{\sqrt{2}} A_0\,,\qquad\qquad
(\hat{m}^2_{eRL})_2 = \hat{A}^e_2 \dfrac{v\, \cos\beta}{\sqrt{2}} A_0\,,
\label{App:meRL_hat}
\ee
where $\hat{A}^e_1$ is given by
\be
\begin{aligned}
&\left[\hat{A}^e_1\right]_{11}=\dfrac{z_e}{y_e} \, m_e\,\dfrac{\sqrt2}{v\, \cos\beta}\,,\\
&\left[\hat{A}^e_1\right]_{12}=c_{T3}\,\dfrac{(z_e y_\mu - z_\mu y_e)} {y_e y_\mu}\, m_e u\,\dfrac{\sqrt2}{v\, \cos\beta}\,,\\
&\left[\hat{A}^e_1\right]_{13}=c_{T3}\,\dfrac{(z_e y_\tau - z_\tau y_e)}{y_e y_\tau}  \, m_e u\,\dfrac{\sqrt2}{v\, \cos\beta}\,,\\
&\begin{split}
\left[\hat{A}^e_1\right]_{21}=& \left[c_{T2} \, \dfrac{(z_\mu y_\mu^\prime - z_\mu^\prime y_\mu)}{y_\mu^2} \, m_\mu tu^2+ c_{T3}\,\dfrac{(z_e y_\mu -  z_\mu y_e)}{y_\mu^2} m_e tu+\right.\\ 
&\quad\left.-c_{T3}\left(\dfrac{(z_\mu y_\mu^\prime - z_\mu^\prime y_\mu)}{y_\mu^2} +\dfrac{(z_\tau y'_\tau-z'_\tau y_\tau)}{y_\tau^2}\right)m_\mu u^3 \right]\dfrac{\sqrt2}{v\, \cos\beta}\,,
\end{split}\\
&\left[\hat{A}^e_1\right]_{22}=\dfrac{z_\mu}{y_\mu} m_\mu\,\dfrac{\sqrt2}{v\, \cos\beta}\,,\\
&\left[\hat{A}^e_1\right]_{23}=c_{T3}\,\dfrac{(z_\mu y_\tau -  z_\tau y_\mu)}{y_\mu y_\tau} m_\mu u\,\dfrac{\sqrt2}{v\, \cos\beta}\,,\\
&\left[\hat{A}^e_1\right]_{31}=c_{T3} \,\dfrac{(z_\tau y_\tau ^\prime - z_\tau ^\prime y_\tau)}{y_\tau^2} \, m_\tau u^2\,\dfrac{\sqrt2}{v\, \cos\beta}\,,\\
&\left[\hat{A}^e_1\right]_{32}=\left[c_{T2} \,\dfrac{(z_\tau y_\tau ^\prime - z_\tau ^\prime y_\tau)}{y_\tau^2} \, m_\tau tu^2+ c_{T3}\,\dfrac{(z_\mu y_\tau -  z_\tau y_\mu)}{y_\tau^2} m_\mu tu\right]\,\dfrac{\sqrt2}{v\, \cos\beta}\,,\\
&\left[\hat{A}^e_1\right]_{33}=\dfrac{z_\tau}{y_\tau} \, m_\tau\,\dfrac{\sqrt2}{v\, \cos\beta}\,,\\
\end{aligned}
\label{App:Ae1_hat}
\ee
while for $\hat{A}^e_2$ we have
\begin{equation}
\hat{A}^e_2 =\left( \begin{array}{ccc}
               c^F\,m_e & (c^F_{T3} -c^Fc_{T3})  \, m_e\, u & c^F_{T2}\, m_e\, u\\[1mm]
               c^F_{T2} \, m_\mu\, u & c^F\, m_\mu  & (c^F_{T3} -c^Fc_{T3}) \, m_\mu\, u\\[1mm]
               (c^F_{T3} -c^F c_{T3}) \, m_\tau\, u & c^F_{T2} \, m_\tau\, u & c^F\, m_\tau \\
\end{array}
\right) \,\dfrac{\sqrt2}{v\, \cos\beta} \,,
\label{App:Ae2_hat}
\end{equation}

Moving to the up squarks, the contributions are the following:
\be
(\hat{m}^2_{uRL})_1 = \hat{A}^u_1 \dfrac{v \sin\beta}{\sqrt{2}} A_0\,,\qquad\qquad
(\hat{m}^2_{uRL})_2 = \hat{A}^u_2 \dfrac{v \sin\beta}{\sqrt{2}} A_0\,,
\label{App:muRL_hat}
\ee
where $\hat{A}^u_1$ is given by
\be
\begin{aligned}
&\begin{split}
\left[\hat{A}^u_1\right]_{11}=&\dfrac{c''\,\tilde{y}_6}{y_5}\left[\left(\dfrac{c_{T3}\,z_1+2c''\,z_2}{2\sqrt{t^q_d\,t^u_d}}- \dfrac{t^u_5\,z_1}{2\sqrt{t^q_d\,t^u_d}\,t^u_1}+ \dfrac{c'\,t^u_6(z_3\,y_b-z_t\,y_7)}{\sqrt{t^q_d\,t^u_d}\,(t^u_d-t^u_s)\,y_b}\right)\,\sqrt{t}u^3+\right.\\
&\qquad\quad\left.-\dfrac{c'\,t^u_4\,t^u_5\,(z_3\,y_b-z_t\,y_7)}{2\sqrt{t^q_d\,t^u_d}\,t^u_1\,(t^u_d-t^u_s)\,y_b}\,\dfrac{u^4}{\sqrt{t}} \right]\,,
\end{split}\\
&\begin{split}
\left[\hat{A}^u_1\right]_{12}=&\left(\dfrac{c_{T3}\,z_1+2c''\,z_2}{2\sqrt{t^q_d\,t^u_d}}- \dfrac{t^u_5\,z_1}{2\sqrt{t^q_d\,t^u_d}t^u_1}+ \dfrac{c'\,t^u_6(z_3\,y_b-z_t\,y_7)}{\sqrt{t^q_d\,t^u_d}\,(t^u_d-t^u_s)\,y_b}\right)\,tu^2+\\
&\quad-\dfrac{c'\,t^u_4\,t^u_5(z_3\,y_b-z_t\,y_7)}{2\sqrt{t^q_d\,t^u_d}\,t^u_1\,(t^u_d-t^u_s)\,y_b}\,u^3 \,,
\end{split}\\
&\left[\hat{A}^u_1\right]_{13}=\dfrac{t^u_6\,z_t}{\sqrt{t^q_st^u_d}(t^u_d-t^u_s)}\,tu-\dfrac{t^u_4\,t^u_5\,z_t}{2\sqrt{t^q_s\,t^u_d}\,t^u_1\,(t^u_d-t^u_s)}\,u^2\,,\\
&\left[\hat{A}^u_1\right]_{21}=\dfrac{c''\,z_1\,\tilde{y}_6}{\sqrt{t^q_d\,t^u_d}\,y_5}\,\sqrt{t}u^2+ \dfrac{c'\,t^u_4(z_3\,y_b-z_t\,y_7)}{\sqrt{t^q_d\,t^u_d}\,(t^u_d-t^u_s)\,y_5\,y_b}\,\dfrac{u^3}{\sqrt{t}}\,\\
&\left[\hat{A}^u_1\right]_{22}=\dfrac{z_1}{\sqrt{t^q_d\,t^u_d}}\,tu+ \dfrac{c'\,t^u_4(z_3\,y_b-z_t\,y_7)}{\sqrt{t^q_d\,t^u_d}\,(t^u_d-t^u_s)\,y_b}\,u^2 \,,\\
&\left[\hat{A}^u_1\right]_{23}=\dfrac{t^u_4\,z_t}{\sqrt{t^q_s\,t^u_d}\,(t^u_d-t^u_s)}\,u\,,\\
&\left[\hat{A}^u_1\right]_{31}=\dfrac{c'\,\,c''\,\tilde{y}_6\,(z_3\,y_b-z_t\,y_7)}{\sqrt{t^q_d\, t^u_s}\,y_5\,y_b}\,\dfrac{u^2}{\sqrt{t}}\,,\\
&\left[\hat{A}^u_1\right]_{32}=\dfrac{c'\,(z_3\,y_b-z_t\,y_7)}{\sqrt{t^q_d\, t^u_s\, y_b}}\,u\,,\\
&\left[\hat{A}^u_1\right]_{33}=\dfrac{z_t}{\sqrt{t^q_s\,t^u_s}}\,,
\end{aligned}
\label{App:Au1_hat}
\ee
while for $\hat{A}^u_2$ we have
\be
\begin{aligned}
&\left[\hat{A}^u_2\right]_{11}=\dfrac{c^F_{T2}\,y_1}{\sqrt{t^q_d\,t^u_d}}\,tu^2\,,\\
&\left[\hat{A}^u_2\right]_{12}=\left(\dfrac{c_{T3}\,y_1+c^{F\prime\prime}\,y_2}{2\sqrt{t^q_d\,t^u_d}}- \dfrac{c^F\,t^u_5\,y_1}{2\sqrt{t^q_d\,t^u_d}\,t^u_1}+ \dfrac{c^{F\prime}\,t^u_6\,y_3}{\sqrt{t^q_d\,t^u_d}\,(t^u_d-t^u_s)}\right)\,tu^2-\dfrac{c^{F\prime}\,t^u_4\,t^u_5\,y_3}{2\sqrt{t^q_d\,t^u_d}\,t^u_1\,(t^u_d-t^u_s)}\,u^3 \,,\\
&\left[\hat{A}^u_2\right]_{13}=-\dfrac{(c^{F\prime}\,t^u_5+2\,c^F_{\eta_2}\,t^u_1)y_4}{2\sqrt{t^q_s\,t^u_d}\,t^u_1}\,tu^2\,,\\
&\left[\hat{A}^u_2\right]_{21}=\dfrac{c^F\,c''\,y_1\,\tilde{y}_6}{\sqrt{t^q_d\,t^u_d}\,y_5}\,\sqrt{t}u^2+ \dfrac{c^{F\prime}\,c''\,t^u_4\,y_3\,\tilde{y}_6}{\sqrt{t^q_d\,t^u_d}\,(t^u_d-t^u_s)\,y_5}\,\dfrac{u^3}{\sqrt{t}}\,\\
&\left[\hat{A}^u_2\right]_{22}=\dfrac{c^F\,y_1}{\sqrt{t^q_d\,t^u_d}}\,tu+ \dfrac{c^{F\prime}\,t^u_4\,y_3}{\sqrt{t^q_d\,t^u_d}\,(t^u_d-t^u_s)}\,u^2 \,,\\
&\left[\hat{A}^u_2\right]_{23}=\dfrac{c^{F\prime}\,y_4}{\sqrt{t^q_s\,t^u_d}}\,tu\,,\\
&\left[\hat{A}^u_2\right]_{31}=\dfrac{c^{F\prime}\,c''\,y_3\,\tilde{y}^6}{\sqrt{t^q_d\, t^u_s}\,y_5}\,\dfrac{u^2}{\sqrt{t}}\,,\\
&\left[\hat{A}^u_2\right]_{32}=\dfrac{c^{F\prime}\,y_3}{\sqrt{t^q_d\, t^u_s}}\,u\,,\\
&\left[\hat{A}^u_2\right]_{33}=\dfrac{c^{F\prime}\,y_3(c'\,t^q_s\,y_7-t^q_4\,y_b)}{\sqrt{t^q_s\,t^u_s}\,t^q_d\,y_b}\,u^2\,.
\end{aligned}
\label{App:Au2_hat}
\ee

Finally we deal with the  down squarks and the contributions are the following:
\be
(\hat{m}^2_{dRL})_1 = \hat{A}^d_1 \dfrac{v \cos\beta}{\sqrt{2}} A_0\,,\qquad\qquad
(\hat{m}^2_{dRL})_2 = \hat{A}^d_2 \dfrac{v \cos\beta}{\sqrt{2}} A_0\,,
\label{App:mdRL_hat}
\ee
where $\hat{A}^d_1$ is given by
\be
\hat{A}^d_1=\left(
        \begin{array}{ccc}
             -\dfrac{c^{\prime\prime2}\,z_5\,\tilde{y}_6^2}{\sqrt{t^q_d\,t^d_d}\,y_5^2}\,u^3
             & -\dfrac{c''\,z_5\,\tilde{y}_6}{\sqrt{t^q_d\,t^d_d}\,y_5}\,\sqrt{t}u^2
             & \dfrac{c'\,c''\,\tilde{y}_6\,(z_8\,y_b-z_b\,y_8)}{\sqrt{t^q_s\,t^d_d}\,y_5\,y_b}\,\sqrt{t}u^2 \\[5mm]
             \dfrac{c''\,z_5\,\tilde{y}_6}{\sqrt{t^q_d\,t^d_d}\,y_5}\,\sqrt{t}u^2
             & \dfrac{z_5}{\sqrt{t^q_d\,t^d_d}}\,tu 
             & \dfrac{c'(z_8\,y_b-z_b\,y_8)}{\sqrt{t^q_s\,t^d_d}\,y_b}\, tu \\[5mm]
             \dfrac{c'\,c''\,\tilde{y}_6\,(z_7\,y_b-z_b\,y_7)}{\sqrt{t^q_d\,t^d_s}\,y_5\,y_b}\,\sqrt{t}u^2
             & \dfrac{c'(z_8\,y_b-z_b\,y_8)}{\sqrt{t^q_d\,t^d_s}\,y_b}\,t u 
             & \dfrac{z_b}{\sqrt{t^q_s\,t^d_s}}\,t \\
        \end{array}
\right)
\label{App:Ad1_hat}
\ee
while for $\hat{A}^d_2$ we have
\be
\hat{A}^d_2=\left(
        \begin{array}{ccc}
             \dfrac{c''(2c^{F\prime\prime}-c^F\,c'')\,\tilde{y}^2_6}{\sqrt{t^q_d\,t^d_d}\,y_5}\,u^3
             & \dfrac{(c^{F\prime\prime}-c^F\,c'')\,\tilde{y}_6}{\sqrt{t^q_d\,t^d_d}}\,\sqrt{t}u^2
             & -\dfrac{c^{F\prime}\,c''\,\tilde{y}_6\,y_8}{\sqrt{t^q_s\,t^d_d}\,y_5}\,\sqrt{t}u^2 \\[5mm]
             -\dfrac{(c^{F\prime\prime}-c^F\,c'')\,\tilde{y}_6}{\sqrt{t^q_d\,t^d_d}}\,\sqrt{t}u^2
             & \dfrac{c^F\,y_5}{\sqrt{t^q_d\,t^d_d}}\,tu 
             & \dfrac{c^{F\prime}\,y_8}{\sqrt{t^q_s\,t^d_d}}\, tu \\[5mm]
             \dfrac{c^{F\prime}\,c''\,\tilde{y}_6\,y_7}{\sqrt{t^q_d\,t^d_s}\,y_5}\,\sqrt{t}u^2
             & \dfrac{c^{F\prime}\,y_7}{\sqrt{t^q_d\,t^d_s}}\,t u 
             & \left[\hat{A}^d_2\right]_{33}\\
        \end{array}
\right)
\label{App:Ad2_hat}
\ee
where
\be
\left[\hat{A}^d_2\right]_{33}= \left(\dfrac{y_8\,(c'\,t^d_s\,y_8-t^d_4\,y_b)}{\sqrt{t^q_s\,t^d_s}\,t^d_d} + 
\dfrac{y_7\,(c'\,t^d_s\,y_7-t^q_4\,y_b)}{\sqrt{t^q_s\,t^d_s}\,t^q_d}\right)\,\dfrac{c^{F\prime}\,tu^2}{y_b}\,.
\ee
 
%
%
\mathversion{bold}
\section{Canonical normalisation of the kinetic terms and diagonalisation of $M_{\ell,d}$}
\label{AppC}
\mathversion{normal}

We first perform the transformations to go to the basis in which the kinetic terms are canonically normalised and 
after we diagonalise the charged lepton and the down quark mass matrices. We perform 
these transformations not only on the fermions, but also on the sfermions in order to ensure that the 
gaugino-fermion-sfermion vertices do not violate flavour at this stage.

To diagonalise the hermitian matrices $K^f$, $f=\ell,e,q,u,d$, we apply the unitary transformations $W^f$: 
\be
W^{f\dag} K^f W^f = \diag\,.
\ee
Normalizing $K^f$ requires a rescaling of the fields via the real (diagonal) matrices $R^f$:
\be
R^f W^{f\dag} K^f W^f R^f = \unity\,.
\ee
The superfields $\Psi_f=\{\ell,\,\ell^c,\,q,\,q^c\}$ are expressed as
\be
\Psi_f =W^f R^f \, \Psi_f ^{\prime}\,,
\ee
so that the kinetic terms are in their the canonical form
\be
i \, \ov{\Psi}^{\prime}_{f,i} \ov{\sigma}^{\mu} \DD_\mu \Psi^{\prime}_{f,i} + |\DD_\mu \tilde{\Psi}^{\prime}_{f,i}|^2\,.
\ee
The mass matrices for fermions in this basis read as
\be
\begin{aligned}
\ell^c\, M_\ell \, \ell &= \ell^{c \, \prime}\, R^e\, (W^e)^{T}\, M_\ell\, W^\ell\, R^\ell \, \ell^{\prime}\, \equiv\, \ell^{c \, \prime}\, M_\ell^\prime\, \ell^{\prime}\,,\\
U^c\, M_u \, Q &= U^{c \, \prime}\, R^u\, (W^u)^{T}\, M_u\, W^q\, R^q \, Q^{\prime}\, \equiv\, U^{c \, \prime}\, M_u^\prime\, Q^{\prime}\,,\\
D^c\, M_d \, Q &= D^{c \, \prime}\, R^d\, (W^d)^{T}\, M_d\, W^q\, R^q \, Q^{\prime}\, \equiv\, D^{c \, \prime}\, M_d^\prime\, Q^{\prime}\,,
\end{aligned}
\ee
while for sleptons as
\be
\begin{aligned} 
\ov{\tilde{\ell}} \, m_{eLL}^{2} \, \tilde{\ell} &= \ov{\tilde{\ell}}^{\prime} \, R^\ell \,(W^\ell)^\dag  \, m_{eLL}^2 \, W^\ell R^\ell \, \tilde{\ell}'\,,\\
\tilde{\ell}^c \, m_{eRR}^{2} \, \ov{\tilde{\ell}}^c &= \tilde{\ell}^{c \, \prime} \,  R^e\, (W^e)^T\, m_{eRR}^2 \, W^{e\,*}\, R^e \, \ov{\tilde{\ell}}^{c\,\prime}\,,\\
\tilde{\ell}^c m_{eRL}^{2} \, \tilde{\ell} &= \tilde{\ell}^{c\,\prime}\, R^e\, (W^e)^T\, m_{eRL}^2\, W^\ell\, R^\ell \, \tilde{\ell}'\,,
\end{aligned}
\ee
and for quarks as
\be
\begin{aligned} 
\ov{\tilde{Q}} \, m_{qLL}^{2} \, \tilde{Q} &= \ov{\tilde{Q}}^{\prime} \, R^q \,(W^q)^\dag  \, m_{qLL}^2 \, W^q R^q \, \tilde{Q}'\,,\\
\tilde{U}^c \, m_{uRR}^{2} \, \ov{\tilde{U}}^c &= \tilde{U}^{c \, \prime} \,  R^u\, (W^u)^T\, m_{uRR}^2 \, W^{u\,*}\, R^u \, \ov{\tilde{U}}^{c\,\prime}\,,\\
\tilde{D}^c \, m_{dRR}^{2} \, \ov{\tilde{D}}^c &= \tilde{D}^{c \, \prime} \,  R^d\, (W^d)^T\, m_{dRR}^2 \, W^{d\,*}\, R^d \, \ov{\tilde{D}}^{c\,\prime}\,,\\
\tilde{U}^c m_{uRL}^{2} \, \tilde{Q} &= \tilde{U}^{c\,\prime}\, R^u\, (W^u)^T\, m_{uRL}^2\, W^q\, R^q \, \tilde{Q}'\,,\\
\tilde{D}^c m_{dRL}^{2} \, \tilde{Q} &= \tilde{D}^{c\,\prime}\, R^d\, (W^d)^T\, m_{dRL}^2\, W^q\, R^q \, \tilde{Q}'\,.
\end{aligned}
\ee

We diagonalise the resulting mass matrices $m'_\ell$ and $m'_d$ of the charged leptons and down quarks, respectively, by the usual bi-unitary transformations:
\be
\begin{aligned}
U_e^{T} M_\ell^\prime V_e &=\diag (m_e,\,m_\mu,\,m_\tau)\,,\\
U_d^{T} M_d^\prime V_d &=\diag (m_d,\,m_s,\,m_b)\,.
\end{aligned}
\ee
In this way we can define the mass eigenbasis $\ell'',\,\ell^{c\,\prime\prime},\,Q'',D^{c\,\prime\prime}$:
\be
\ell^{\prime} = V_e\, \ell^{\prime\prime}\,,\qquad \ell^{c \, \prime} = U_e\, \ell^{c \, \prime\prime}\,, \qquad 
 Q^{\prime} = V_d\, Q^{\prime\prime}\,,\qquad D^{c \, \prime} = U_d\, D^{c \, \prime\prime}\,.
\ee
Finally, the slepton mass matrices are given by
\be
\begin{aligned}
\ov{\tilde{\ell}} \, m_{eLL}^{2} \, \tilde{\ell} &=\ov{\tilde{\ell}}^{\prime\prime} \left[V_e^\dag R^\ell W^{\ell\dag} \, m_{eLL}^2 \, W^\ell R^\ell V_e\right] \, \tilde{\ell}^{\prime\prime} \equiv \ov{\tilde{\ell}}^{\prime\prime} \hat{m}_{eLL}^2 \,\tilde{\ell}^{\prime\prime}\,,\\
\tilde{\ell}^c \, m_{eRR}^{2} \, \ov{\tilde{\ell}}^c &=\tilde{\ell}^{c\,\prime\prime}\left[U_e^T R^e (W^e)^T m_{eRR}^2\,W^{e\,*} R^e \, U_e^*\right] \,\ov{\tilde{\ell}}^{c\,\prime\prime} \equiv \tilde{\ell}^{c \, \prime\prime} \hat{m}_{eRR}^2 \,\ov{\tilde{\ell}}^{c\,\prime\prime}\,,\\
\tilde{\ell}^c m_{eRL}^{2} \, \tilde{\ell} &= \tilde{\ell}^{c\,\prime\prime}\left[U_e^T R^e (W^e)^T m_{eRL}^2 W^\ell R^\ell V_e\right] \, \tilde{\ell}^{\prime\prime} \equiv \tilde{\ell}^{c\,\prime\prime} \hat{m}_{eRL}^2 \, \tilde{\ell}^{\prime\prime}\,,
\end{aligned}
\ee
and the squark ones by
\be
\begin{aligned} 
\ov{\tilde{Q}} \, m_{qLL}^{2} \, \tilde{Q} &= \ov{\tilde{Q}}'' \,\left[V_d^\dag R^q \,(W^q)^\dag  \, m_{qLL}^2 \, W^q R^q V_d\right] \, \tilde{Q}''\,,\\
\tilde{D}^c \, m_{dRR}^{2} \, \ov{\tilde{D}}^c &= \tilde{D}^{c \, \prime\prime} \, \left[U_d^T R^d\, (W^d)^T\, m_{dRR}^2 \, W^{d\,*}\, R^d U_d^*\right] \, \ov{\tilde{D}}^{c\,\prime\prime}\,,\\
\tilde{U}^c m_{uRL}^{2} \, \tilde{Q} &= \tilde{U}^{c\,\prime}\,\left[R^u\, (W^u)^T\, m_{uRL}^2\, W^q\, R^q V_d\right] \, \tilde{Q}''\,,\\
\tilde{D}^c m_{dRL}^{2} \, \tilde{Q} &= \tilde{D}^{c\,\prime\prime}\,\left[U_d^T R^d\, (W^d)^T\, m_{dRL}^2\, W^q\, R^q V_d\right] \, \tilde{Q}''\,.
\end{aligned}
\ee
Note that the $m_{uRR}^{2}$ has been not affected by these last transformations.

To arrive at this result we assume that all couplings involved are real and as a result the matrices 
$W^f$, $U_{e,d}$ and $V_{e,d}$ turn out to be orthogonal instead of unitary. 

%
%
\mathversion{bold}
\section{Wilson coefficient contributions for $BR(\bar{B}\to X_s\gamma)$}
\label{AppD}
\mathversion{normal}

\begin{figure}
\centering
\vspace{-0.5cm}
\includegraphics[width=16cm]{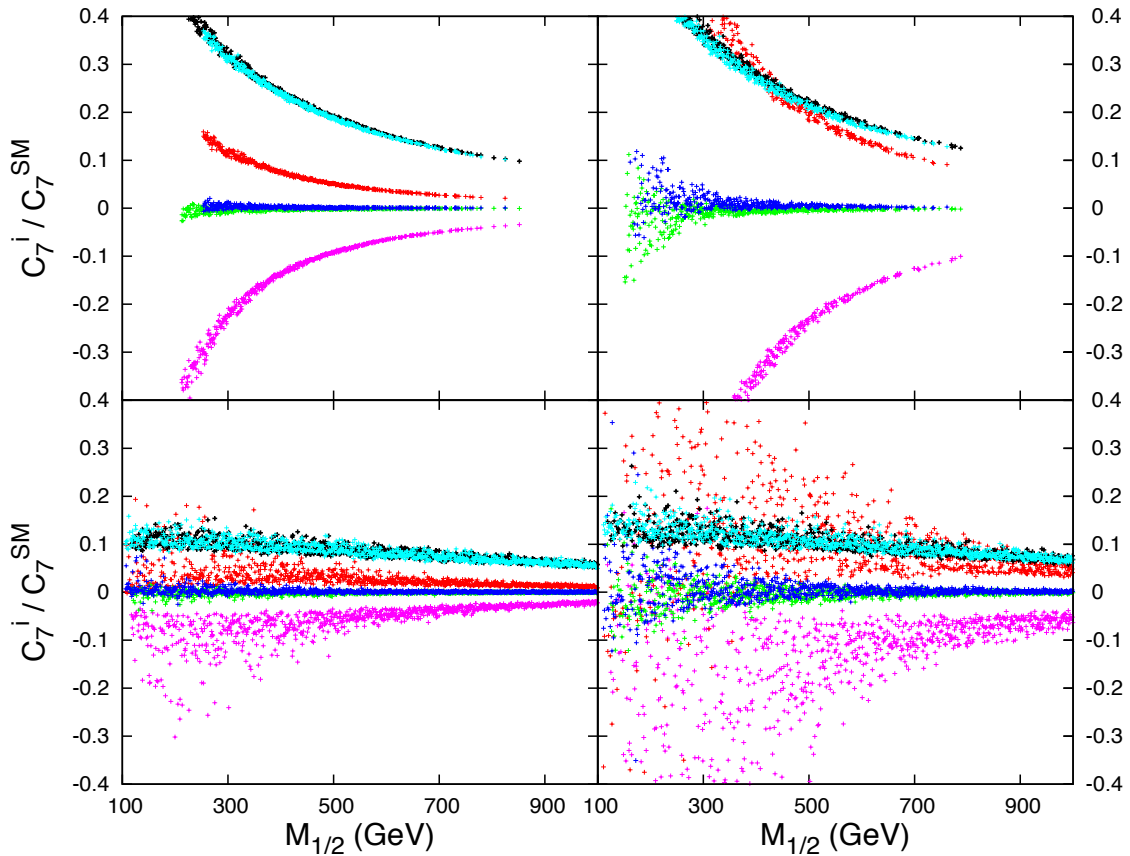} 
\caption{\it Wilson coefficient contributions for $BR(\bar{B}\to X_s\gamma)$. The left (right) plots refer to the $RP_A$ ($RP_B$) case for $A_0=2\,m_0=400$ GeV (upper plots) or $A_0=2\,m_0=2000$ GeV (lower plots). See the text for details.}
\label{Fig:Ci}
\end{figure}

In this Appendix we report the break down of the $C_{7\gamma}$ Wilson coefficient for $BR(\bar{B}\to X_s\gamma)$ in terms of the distinct supersymmetric contributions. In Fig.~\ref{Fig:Ci}, the two upper (lower) plots refer to $A_0=2\,m_0=400$ GeV ($A_0=2\,m_0=2000$ GeV), respectively for the $RP_A$ case (left) and $RP_B$ case (right). The different colors refer to distinct contributions: Magenta (Red) for the chargino contribution with positive (negative) $\sign[\mu]$; Black (Cyan) for the charged Higgs contribution with positive (negative) $\sign[\mu]$; Green (Blue) for the gluino contribution with positive (negative) $\sign[\mu]$.

Fig.~\ref{Fig:Ci} confirms the analysis reported in Sect.~\ref{sec:PhenHad}: in the small $(A_0,m_0)$ region (upper plots) the dominant $T'$ contribution to the \bsg BR is typically the charged Higgs one, that is always concordant in sign with the SM one. 
The second most relevant contribution is the chargino one with a sign depending on $\sign[\mu]$: it tends to enhance 
(cancel) the SM contribution for $\mu < 0$ ($\mu > 0$). Gluino contributions are practically independent 
from $\sign[\mu]$. In Fig.~\ref{Fig:Ci} we do not show the neutralino contribution, since it turns out to be completely negligible.

%
%

\providecommand{\href}[2]{#2}\begingroup\raggedright\endgroup

\end{document}